\documentclass[twoside,12pt]{article}
\pdfoutput=1
\usepackage{graphicx}
\usepackage{epsfig}
\usepackage{amssymb}
\usepackage{amsmath}
\usepackage{mathtools}

\newcommand{\be}{\begin{equation}}
\newcommand{\ee}{\end{equation}}
\newcommand{\bea}{\begin{eqnarray}}
\newcommand{\eea}{\end{eqnarray}}

\begin{document}

\title{ \vspace{1cm}  Geometry Optimization \\ of a Muon-Electron Scattering Detector}
\author{Tommaso Dorigo \\
INFN, Sezione di Padova, Italy}
\maketitle

\begin{abstract} 
A high-statistics determination of the differential cross section of elastic muon-electron scattering as a function of the transferred four-momentum squared, $d \sigma_{el}(\mu e \to \mu e)/dq^2$, has been argued to provide an effective constraint to the hadronic contribution to the running of the fine-structure constant, $\Delta \alpha_{had}$, a crucial input for precise theoretical predictions of the anomalous magnetic moment of the muon. An experiment called ``MUonE'' is being planned at the north area of CERN for that purpose. We consider the geometry of the detector proposed by the MUonE collaboration and offer a few suggestions on the layout of the passive target material and on the placement of silicon strip sensors, based on a fast simulation of elastic muon-electron scattering events and the investigation of a number of possible solutions for the detector geometry.
\end{abstract}



\section{Introduction}
\label{s:intro}

A clear picture of fundamental physics emerges at the dawn of the third millennium, after Run 2 of the Large Hadron Collider delivered over $150/fb$ of integrated luminosity of 13 TeV proton-proton collisions. The detailed studies of particle phenomenology at high energy by the CMS and ATLAS experiments, together with the high-intensity and high-precision studies of heavy quark properties offered by the LHCb and Belle experiments, and the wealth of additional information collected by a number of other dedicated facilities, all show that the Standard Model of electroweak interactions and the theory of Quantum Chromodynamics jointly provide a completely successful description of the phenomenology of elementary fermions and hadrons down to length scales of $10^{-18}$m. 

While from a theoretical standpoint the Standard Model is considered incomplete, and at most an effective theory which is bound to break down at as of yet untested energy scales, there is no experimental evidence that the theory may eventually fail to describe any of the phenomena we will test with present or future facilities, with one notable exception. 

\subsection{ The muon anomaly and its uncertanties}

At the time of writing, one observable quantity stands out as the only systematical, persistent discrepancy of theory and experiment in particle phenomenology: the anomalous magnetic moment of the muon. The precise determination of the muon g-2, or specifically $a_{\mu} = (g_{\mu}-2)/2$~\cite{amuexp1, amuexp2,amuexp3,amuexp4}, performed at the Brookhaven laboratories, has shown a disagreement with its theoretical prediction~\cite{amuth} $a_{\mu}^{th}$, at a significance level ($3.7 \sigma$) that deserves serious consideration: \par

\begin{equation}
a_{\mu}^{meas} - a_{\mu}^{th} = (116592091 \pm 63 - 116591820 \pm 36) \times 10^{-11}.
\end{equation}

\noindent
The experiment is being repeated with a more intense muon source at Fermilab by the E989 group, where it is foreseen that the total uncertainty on $a_{\mu}$ will eventually be brought down by a further factor of four~\cite{e989_1,e989_2,e989_3}. Such a result has the potential of offering a conclusive proof that new physical phenomena need to be accounted for in the calculation of quantum loop diagrams affecting the muon-photon vertex; however, uncertainties in the calculation of $a_\mu^{th}$ do limit the severity of the hypothesis test. 

A limiting factor in the theoretical calculation of $a_\mu$ is the precise evaluation of hadronic loop contributions at the muon vertex. Until recently, those contributions were estimated through the calculation of a dispersion integral of the hadronic production cross section for s-channel electron-positron annihilation. That reaction includes the same loop contributions that affect the $a_{\mu}$ calculation, but is complicated by several resonant processes, to which correspond poles whose integration limits the overall theoretical precision. 

It has been recently noted~\cite{originalidea} that the hadronic term could alternatively be computed by integration over the space-like muon-electron elastic scattering process:\par

\begin{equation*}
\begin{aligned}
a_\mu^{HLO} = \frac{\alpha}{\pi} \int_{0}^{1} dx (1-x) \Delta \alpha_{had}(t), \\
t = \frac{x^2 m_\mu^2}{x-1}.
\end{aligned}
\end{equation*}

\noindent
In the above formula $\Delta \alpha_{had}$ is the hadronic contribution to the running of $\alpha$, which can be determined without the need of complex integration over resonant states if one is able to measure the differential cross section of elastic muon scattering on electrons as a function of four-momentum squared. An experimental determination of the hadronic loops contribution to that reaction relies on the subtraction of the theoretically-computed electroweak contributions to the differential cross section, which are known over the full kinematical range to three-loop accuracy~\cite{daewk}. As the size of the hadronic contribution is of only a few percent at most, concentrated in the region of large four-momentum transfer, from an experimental standpoint one needs to envision a very precise measurement of the differential cross section as a function of $q^2$. A shape fit to the distribution, where the electroweak component constitutes a template with free normalization (the normalization of the electroweak contribution is in fact less precisely known than its shape) may then enable the extraction of the wanted parameter. 

\begin{figure}[h!]
\begin{center}
\includegraphics[scale=0.6]{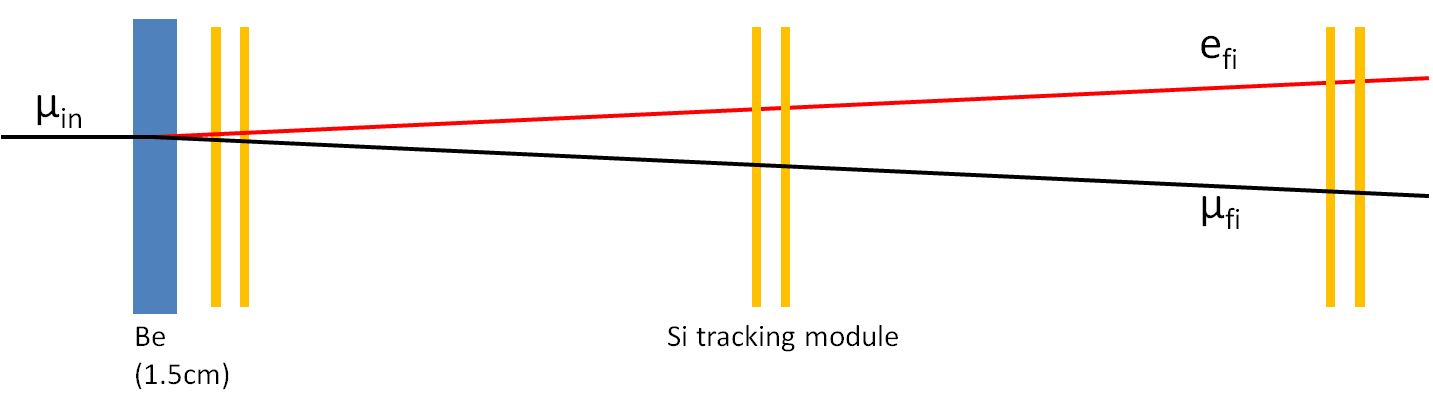}
\includegraphics[scale=0.7]{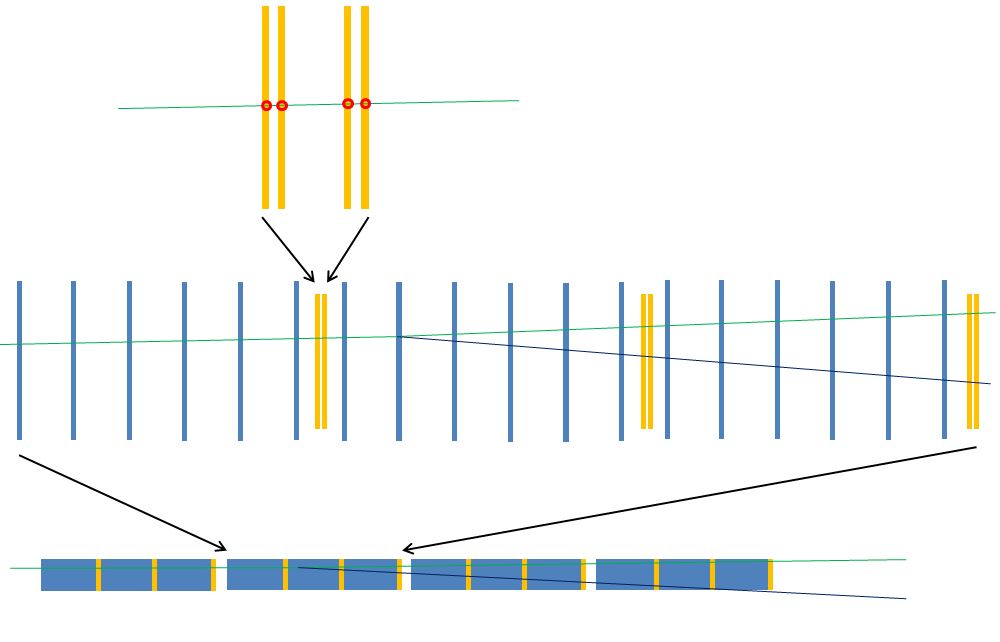}
\caption{\em Top: Proposed layout of a 1m-long tracking station the MUonE detector. Bottom: layout resulting from a distributed target scheme with 18 equally-spaced layers per station. From top to bottom is shown the arrangement of a detection module into two double-sided silicon strip sensors, the layout of a 1m-long detection station, and the full apparatus.}
\label{f:layoutmuone}
\end{center}
\end{figure}

In order to be able to produce a significant decrease of the total uncertainty on $a_{\mu}^{th}$, the total hadronic contribution to the scattering cross section must be evaluated with a relative uncertainty of the order of a percent or less. This poses demanding requirements on a successful experimental campaign: very high statistics, as well as extreme care in beating down systematic uncertainties. An intense beam of muons, well suited for the task at hand, is available at the CERN north area. The muon beam, originated by secondary decays of hadrons produced by fixed target collisions of the SpS beam, at an energy of about 150 GeV has a root-mean-square (RMS) cross section downstream of the COMPASS experiment of about 2.6 by 2.7 cm, with small angular divergence (RMS of 2.0 by 2.7 milliradians in the vertical $y$ direction and the horizontal $x$ direction transverse to the beam, respectively). The MUonE collaboration plans to instrument 40 meters of available space downstream of COMPASS with 40 1-m-long measuring stations, each composed of a relatively thin beryllium target followed by three tracking modules. The latter are each made by coupling two double-sided silicon strip sensors, respectively reading the $x$ and $y$ coordinates of incoming charged particles; the proposed arrangement is shown in Fig.~\ref{f:layoutmuone}a. 

The envisioned modular arrangement of the detection system enables a straightforward triggering strategy for the scattering events, as well as simplicity of assembly and independence of the measurement from systematic effects arising from the imprecise relative positioning of the stations along the beam axis. An electromagnetic calorimeter located at the end of the array of stations might complement the system, providing redundancy in the measurement of the final state electron, as well as reduction of beam-induced and physics backgrounds and a removal of the ambiguity in the signal kinematics for configurations in which the muon and electron emerge from the interaction with similar divergence.

As already pointed out, the success of the proposed measurement rests on the control of a number of subtle systematic uncertainties. In this respect, the resolution (and its uncertainty) with which the parameters of electron and muon trajectories can be determined, once experimental biases are accounted for, is the crucial ingredient of the measurement: the large sample statistics then allow for a precise {\em in-situ} calibration and inter-alignment of the detector components. The choice of silicon strip modules for the tracking of incoming muon and outgoing muon and electron is certainly sound and cost-effective, in particular in view of the good properties of appropriately-sized sensors that are being developed for the much more massive task of instrumenting the CMS tracker for its Phase 2 upgrade~\cite{cms}, and which the MUonE collaboration plans to employ in their detector construction.

\subsection{ Goal of this study and plan of the document}

In this document we consider the issue of what could be the optimal arrangement of detection elements and target material for the final goal of a precision measurement of the hadronic contribution to the $\Delta \alpha$ parameter, the running of the EM coupling constant. Indeed, the choice of the position  of passive and active material along the beam axis will be shown to have a significant effect in the precision with which the event kinematics can be reconstructed. That this is the case can be appreciated intuitively by considering that, for a given beam energy, the scattering kinematics are essentially determined by the knowledge of the incident muon direction and by the angles $\theta_e$, $\theta_\mu$ at which electron and muon scatter off it. A concentrated target (a 1.5 cm-thick layer of beryllium is envisioned for each detection station in the submitted design of the MUonE detector~\cite{muonedoc}) will cause a small amount of multiple scattering to incoming and outgoing particles before the incoming muon interacts and the outgoing pair exits the target. This small smearing in the particles' directions corresponds to a loss of information on the event kinematics that is irrecoverable, regardless of the precision of the trajectory measurements upstream and downstream. A distribution of that 1.5-cm Be-equivalent material into three layers of a third of that thickness, each one alternating with a tracking module, would already allow to obtain for each track at least two pairs of measurement points ``closer'' (in radiation length metric $X_0$) to the interaction point, with a reduction of the uncertainty on their angles.

In addition, the careful positioning of a large number of thinner target layers, which could be precisely spaced from one another to uniformly populate the space between the tracking modules (a spacing by 3mm would {\em e.g.} do the job if 300 $50 \mu$m-thick layers per station were used) if stacks of target layers interleaved by proper spacing frames were constructed, would yield a great benefit through the constraining power of the scattering position along the beam axis (which we will denote by {\em z axis} in the following) in the fit to the particle trajectories. In fact, since the scattering takes place only within the layers of target material or silicon sensors~\footnote{We neglect interactions with electrons from nitrogen or oxygen in the air, which contribute to the total material budget by up to 3.9\% at STP. More discussion on this detail is provided in Sec.~\ref{s:generation} {\em infra}.}, the knowledge of where the layers are placed becomes a powerful constraint on the $z$ position of the interaction vertex, which in turn can be used to constrain the event kinematics. We will return to this important point in Sec.~\ref{s:zvertex}.

In our study we consider a number of possible arrangements of the target material, with the goal of identifying the design choices minimizing the uncertainty with which $\Delta \alpha_{had}$ can be extracted from a sample of interactions. Of course, an accurate assessment of the overall uncertainty requires in principle a complete model of the detector, of the physics of the scattering and of electron and muon radiation losses, of the detection of particle hits in the silicon sensors, and of all relevant backgrounds. Such a task can only be achieved by a full simulation in GEANT4. For a quick study, however, which could more nimbly explore the space of alternative design choices, we produced a simplified description of the above elements with $C^{++}$ code. We attempted to limit the modeling to the essential ingredients, creating a fast custom simulation which we believe is still sufficiently accurate to provide the answers we are looking for. Those answers restrict the space of advantageous geometries to a subset on which a full simulation can more narrowly focus, to fine tune the desired answers. We leave this optional investigation task to the MUonE collaborators.

The contents of this article are as follows. In Section 2 we offer a quick reminder of the main aspects of the theory of muon-electron scattering and its relevance for the measurement of the hadronic contribution to $g-2$. In Section 3 we describe the simulation code used for the optimization studies. Section 4 is devoted to describing the event reconstruction and the likelihood function. In Section 5 we show how a distributed target is capable of significantly increasing the precision of the event reconstruction, and we quantify the potential gain in the achievable precision on the $\Delta \alpha$ parameter. In Section 6 we further the studies of Section 5 by considering the effect of additional variations that concern the placement of the detection modules, the offset of the placement of strips in the two sides of double-sided sensors, and the angle of stereo strip sensors. In Section 7 we discuss how the uncertainty in the longitudinal positioning of sensors as well as their tilt or bow off the plane orthogonal to the $z$ axis, whose value may affect the precision with which the incoming muon momentum is determined, can be constrained to arbitrary precision by a large statistics sample of scatterings. This offers a powerful complement, or even a cheap alternative, to the laser-based holographic system envisioned by the MUonE collaboration to constrain those parameters. In Section 8 we summarize our findings in a set of recommendations on the most favourable detector geometries and design choices.
 
\section{ Elastic muon-electron scattering \label{s:scattering}}

The interaction of energetic muons with electrons in a fixed target is dominated by its elastic scattering part, which at leading order proceeds through the t-channel exchange of a single virtual photon: indeed, the determination of the differential rate of that process as a function of $q^2$ is what motivates the measurement, due to the contributions that the leading electromagnetic process receives from hard-to-calculate hadronic loops. Electromagnetic and weak contributions to the running of $\alpha$ are calculated to very good precision; granted that, one can subtract off the measured differential cross section the calculated electroweak part, obtaining an estimate of the hadronic part.

 From a purely experimental point of view, elastic scattering $\mu e $ events are quite easy to distinguish from anything else, thanks to stringent kinematic relations binding the scattering angles at which the two bodies emerge in the final state~\footnote{Of course, in a strict sense elastic scattering is an idealization of the physics of muon-electron interactions, as the emission of arbitrarily soft photons, {\em e.g.}, has to be considered beyond leading order; we neglect this aspect in what follows, although we do note its power to slightly modify some of the conclusions of this study.}. We will briefly review those relations in what follows. 

In the laboratory frame~\footnote{The frames of reference used in this document are described {\em infra}, Sec.~\ref{s:coordsystem}.} we call $p_{\mu}$ and $E_{\mu}$ the four-momentum and energy of the incoming muon and $p_e$, $E_e$ the four-momentum and energy of the scattered electron. In that frame the target electron can be considered to be at rest to good approximation. The following relations allow to compute the variables $s$ and $q^2=-t$:\par

\begin{equation*}
\begin{aligned}
s = m_\mu^2 + m_e^2 + 2 m_e E_{\mu}, \\
q^2 = -t = 2 m_e E_e - 2 m_e^2.
\end{aligned}
\end{equation*}

\noindent
For any given value of the incoming muon momentum, there exists a maximum four-momentum transfer, $-t_{min} = q^2_{max}$. This can be obtained as \par

\begin{equation}
  q^2_{max} = \frac{s^2+m_{\mu}^2+m_e^2 -2s m_{\mu} -2s m_e -2 m_{\mu} m_e}{s}.
\end{equation}

\noindent
Since we will consider, in the rest of this document, the specific experimental conditions of muon-electron scattering produced by the beam of muons available at the CERN north area, which offers muons of energies in the ballpark of $E_\mu=150 \div 160$ GeV at high intensities (with a nominal average rate of $1.3 \, 10^{7}$Hz), it is useful to quote in passing the maximum four-momentum transfer that can be produced in those conditions: taking the reference value $E_\mu = 150$ GeV, we find $q^2_{max}=0.143$ GeV$^2$. 

\begin{figure}[h!]
\begin{center}
\includegraphics[scale=0.6]{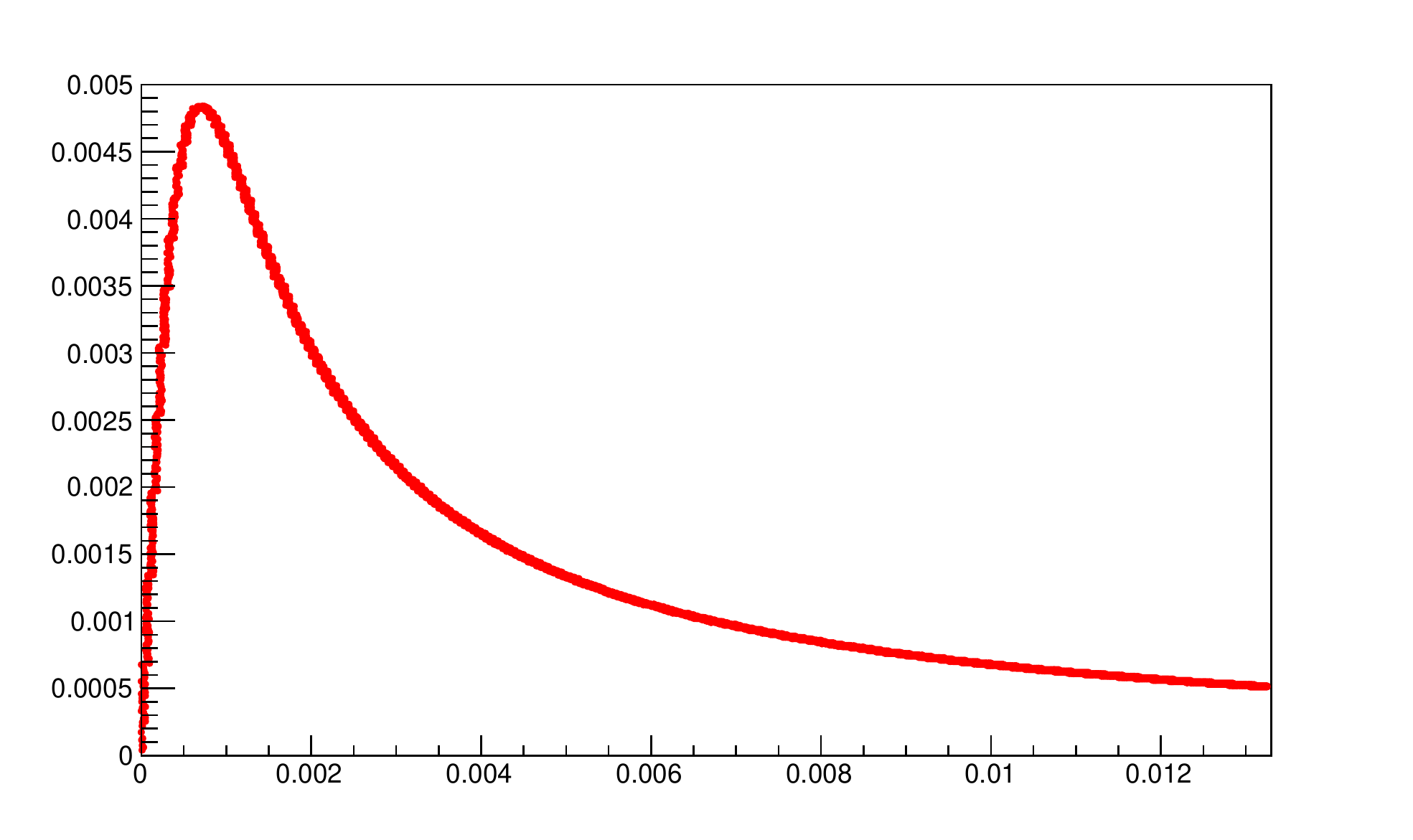} 
\caption{{\em Relation between the divergence of final state muon (}$\theta_{\mu}${\em, on the y axis) and the divergence of final state electron (}$\theta_e${\em, on the x axis) for elastic muon scattering of 150 GeV on a fixed target.}}
\label{f:thetas}
\end{center}
\end{figure}

In the considered frame, where the initial state electron is at rest, the elasticity condition provides a relation between the polar angles $\theta_{\mu}$, $\theta_e$ of the final-state bodies, measured relative to the direction of the incoming muon:\par

\begin{equation}
tan \theta_\mu = \frac {2 tan \theta_e} {(1+\gamma^2 tan^2 \theta_e)(1+\frac{E_\mu m_e+m^2_\mu}{E_\mu m_e + m^2_e})-2}.
\end{equation}

\noindent
where we have set $\gamma = \frac{E_\mu + m_e }{ \sqrt{s}}$. The relation corresponds to a characteristic curve in the $\theta_{\mu} \div \theta_e$ plane, which has a maximum value of $\theta_{\mu} \simeq 0.005$ (see Fig.~\ref{f:thetas}) for an incident energy $E_\mu=150$ GeV. Since the kinematical region where the hadronic contribution to $\Delta \alpha$ is the highest corresponds to the largest values of four-momentum transfer, where $\theta_e$ is of the same order of magnitude of the muon scattering angle $\theta_{\mu}$, it is clear that the measurement is quite challenging, since the determination of the $q^2$ of the corresponding elastic scattering events will have to rely on estimating track angles to an absolute precision in the $10^{-4}$ radians ballpark. This can, however, be achieved with silicon tracking detectors, as will be described in Sec.~\ref{s:detector}.

\section{Fast simulation of elastic muon-electron scattering and event reconstruction}

\subsection{Generalities and generation of the scattering}

For our study we produced a software description of the physics and of the detector, as well as a reconstruction of the event kinematics, based on $C^{++}$ code, wherein we made use of several libraries from the ROOT analysis software~\cite{root}.  ROOT offers a random number generator of good quality, TRandom3~\cite {trandom3}, which is based on the Mersenne Twister Generator; its periodicity is of about $10^{6000}$. We use four different sequences of random numbers: one to simulate the scattering kinematics, one to simulate the multiple scattering effects to particles propagation in the material, one to deal with noise in the silicon strips of the tracking modules, and one to simulate a non-perfect efficiency of the sensors. In this way, by properly reusing the same random sequences we can subject the very same scattering events to different detector geometries~\footnote{ Of course final state particles with identical initial direction will undergo different scattering, even in an average sense, if different detector assemblies are considered, hence the correlation is imperfect.}, minimizing the effect of random sampling of their physical phase space, as will be clarified in the following.

The formulas of the previous section allow a complete description of the scattering kinematics. However, we need to define a range of $q^2$ for the events we wish to simulate, since from an experimental standpoint the extraction of the value of $\Delta \alpha_{had}$ requires to consider events in a restricted kinematical region. The MUonE collaboration suggests that the region $\theta_e <0.02$ radians be used; in any case, larger scattering angles for the electron correspond to very small values of $q^2$, where the hadronic contribution to the running of $\alpha$ is completely negligible. Since the cross section falls very steeply with $q^2$, and we wish to consider unweighted events in our study to simplify the statistical treatment, the setting of a lower threshold on the four-momentum transfer of simulated scatterings speeds up all calculations. As is shown below, the experimental resolution on $q^2$ is of about $0.0002$ at its low end (see Sec.~\ref{s:baselinegeometry}), so the simulation must extend to slightly smaller values than those we aim to study, in order to correctly model the shape of the measured distribution after accounting for experimental smearing. We found that generating interactions in the $q^2/GeV^2$ range $[0.006,0.143]$ is appropriate for our study. This corresponds to $\theta_{el}$ values up to 0.0132 radians.

\subsubsection{Incoming muon beam}

In the following we discuss the generation of the scattering kinematics and the propagation of particles through the material. First an incoming muon is generated, sampling from a Gaussian bivariate distribution in $x,y$ the particle position at the $z=0$ coordinate we take as the origin of the detector along the beam line (see Sec.~\ref{s:coordsystem}, {\em infra}), and sampling another Gaussian bivariate distribution in $\theta_x, \theta_y$ to model its initial direction, where the two angles correspond to the particle divergence from the z axis.  We consider the following nominal parameters of the CERN muon beam, assumed to operate at an energy of 150 GeV:\par

\begin{itemize}
\item average muon energy  $\overline{E_\mu} = 150$ GeV;
\item energy spread 3.5\% (assumed Gaussian), so $\sigma_{E_\mu} = 5.25$ GeV;
\item beam transverse cross section: $\sigma_x = 2.6 cm$, $\sigma_y = 2.7 cm$ (profile assumed Gaussian);
\item beam divergence: $\sigma (\theta_x) = 0.00027$ rad, $\sigma (\theta_y) = 0.00020$ rad (profile assumed Gaussian).
\end{itemize}

\noindent
In this work we assume that it is possible to extract an arbitrarily precise measurement of the average beam energy~\footnote {The procedure requires limited statistics to be carried out, so even relatively unstable beam conditions can be coped with.} by inverting the kinematics for scattering events where final state muon and electron emerge with the same angle; this has been demonstrated by the proponents of the MUonE experiment~\cite{muonedoc}. We do not assign any uncertainty to the average muon beam energy $\overline{E_\mu}$; the same is done with the above parameters, which model much less crucial aspects of the incoming muons kinematics. For studies of the effects of different detector geometries on the resolution achievable on the scattering kinematics we set to zero the energy spread $\sigma (E_\mu)$, which eliminates that nuisance parameter from the point estimate problem. This corresponds, in statistical terms, to the factoring out of that ancillary statistic, effectively conditioning to a subspace of the measurement space where the statistical inference is more precise. 

\subsubsection {Modeling of multiple scattering in the material \label{s:msc}}

The incoming muon is propagated through the material of the detector apparatus (whose description is given below, Sec.~\ref{s:detector}) as a straight line in regions devoid of material, broken by deviations and shifts due to the multiple scattering effects that the particle undergoes in crossing each material layer. To model the latter we follow the description proposed by the PDG~\cite{pdg}. The model suggests that the crossing of a layer of thickness $\Delta z$ and of radiation length $X_0$ (both properly modified by the factor $1/cos(\theta_\mu)$ to account for the divergence of the incident particle off the z axis) by a particle of momentum $p$ produces an angular deviation $\Delta \theta_{MS}$ and a transverse offset $\Delta h_{MS}$, the latter distributed uniformly in $[0, 2 \pi]$ around the original particle trajectory. Following~\cite{pdg} we sample from a Normal distribution two numbers $g_1$ and $g_2$, and then we compute: \par

\begin{equation*}
\begin{aligned}
\theta_0 = \frac{0.0136}{p} \sqrt X_0 (1+0.038 \log X_0), \\
\Delta \theta_{MS} = \sqrt 2 g_2 \theta_0, \\
\Delta h_{MS} = \frac{1}{\sqrt 12} g_1 \Delta z \sin \theta_0 + 0.5 g_2 \Delta z \sin \theta_0, \\
\Delta x_{MS} = \sqrt 2  \Delta h_{MS} \cos \Delta \phi_{MS}, \\
\Delta y_{MS} = \sqrt 2  \Delta h_{MS} \sin \Delta \phi_{MS}, 
\end{aligned}
\end{equation*}

\noindent
where $\Delta x_{MS}$ and $\Delta y_{MS}$ above are the resulting offsets from the $x,y$ position where the particle hits the layer. $\Delta \theta_{MS}$ and $\Delta \phi_{MS}$ are then combined with the incident particle direction to obtain the emerging particle direction. 

\subsubsection{Generation of elastic $\mu e$ scattering \label{s:generation}}

The fast simulation we produced is unsuitable to handle the generation of backgrounds and their effect on tracking resolution and other beam-related effects. We note that these degradation effects have arguably no large impact on the determination of the relative merits of different geometries. In our study all of the simulated incoming muons undergo elastic scattering with an electron, and therefore constitute our ``signal''~\footnote {The one neglected effect that has a potentially large impact in a geometry optimization is constituted by inelastic scattering events, which ``thicken'' the curve describing the functional $\theta_{\mu} (\theta_e)$ relation of Fig.~\ref{f:thetas} and thus potentially contaminate the cross section determination if the $q^2$ resolution is not very high. We believe that our focus on the precise determination of that parameter in our optimization study does indirectly account for it, although indeed more studies are necessary of this ingredient.}. In other words, no simulation of beam backgrounds or of muons not undergoing elastic scattering is attempted. The $q^2=-t$ value of the scattering interaction is sampled from the formula\par

\begin{equation}
\frac {d \sigma} {d t} = 4 \pi \alpha^2 \frac{(m_\mu^2+m_e^2)^2-su+t^2/2}{\lambda t^2} [1+2 \Delta \alpha(t)],
\end{equation}

\noindent
where $\lambda$ and $s, u$ are defined by \par

\begin{equation*}
\begin{aligned}
s = m_{\mu}^2 + m_e^2 +2 m_e E_{beam}, \\
u = 2 m_\mu^2 + 2m_e^2 -s - t, \\
\lambda = s^2 + m_\mu^4 + m_e^4 -2sm_\mu^2 -2sm_e^2 -2m_\mu^2 m_e^2.
\end{aligned}
\end{equation*}

\noindent
The function $\Delta \alpha(t)$ is modeled by the following two-parameter ``fermion-like'' form~\footnote{The functional form and the fitted parameter values were provided by C. Carloni Calame, to whom we are indebted.}: \par \par

\begin{equation*}
\begin{aligned}
p_1 = 0.00239479, \\
p_2 = 0.0523448, \\
\end{aligned}
\end{equation*}
\begin{equation}
\Delta \alpha (t) = \frac{p_1}{3} \left(-\frac{5}{3} -4 \frac{p_2}{t} + \frac{8 \frac{p_2^2}{t^2} + 2\frac{p_2}{t}-1}{\sqrt{1-4 \frac{p_2}{t}}} \log | \frac{1-\sqrt{1-4 p_2 t}}{1+\sqrt{1-4 p_2 t}}| \right).
\label{eq:hadcontrib}
\end{equation}

\noindent
The scattering is generated in the second section of an array of four 1-meter-long sections of equal geometry. This arrangement allows the modeling of the interaction of the incoming muon with at least one full station, and its detection in the corresponding silicon modules; as well as the study of the effect of at least two full stations and their material distribution on the reconstruction of the final state kinematics. 

The position along $z$ at which the interaction is simulated to occur is sampled from a uniform distribution in radiation length metric within each of the layers of beryllium-equivalent or silicon material of which the station is chosen to be composed (see {\em infra}), in such a way that the total number of simulated interactions distributes exactly evenly along the material depth, but retains stochasticity within the thickness of each material element. So, for instance, if the station is composed of a single 1.5cm thick beryllium layer (corresponding to $0.0425 X_0$) followed by twelve 320-$\mu$m silicon sensors (corresponding to $0.003415 X_0$ each) arranged in three modules of two double-sided detection units, the total thickness of the station is of $0.08348 X_0$. At the start of the simulation we determine how many interactions to generate in each layer from the total number of requested scatterings $N_{tot}$, enforcing that all of them take place within the same station (see {\em infra}, Sec.~\ref{s:detector}, for a description of the simulated detector): $0.509 N_{tot}$ of them evenly distributed among the beryllium targets, and $0.0409 N_{tot}$ of them in each of the 12 silicon sensor layers. Within each layer, the actual $z$ position of each of the required interactions is generated at random from a uniform distribution. This arrangement allows to reduce the stochasticity of the simulated dataset, in the sense that the z-vertex distribution of the dataset is the same as those of all other datasets produced to test alternative geometries; the same random number sequences are also used for the other stochastic parameters in these scattering events, for the same reason. The randomness within each layer is necessary to avoid annoying discreteness effects which would occur for a completely fixed spacing of the interactions along $z$, because the interplay of a uniform spacing of the scatterings with the discrete placement of silicon detector strips would produce non-smooth resolution maps as a function of the muon and electron angles. 

The attentive reader will no doubt have noticed that above we have neglected to discuss the presence of a medium between the layers of beryllium and silicon along the particles' paths. At standard pressure and temperature, air filling the 98.118cm of empty space within each station corresponds to a non-negligible addition of $0.03229 X_0$, {\em i.e.} an increase of about $3.9\%$ of the material thickness provided by target and detection layers. Due to its distribution along the station, the effect of air goes in the same direction we are advocating in this article --that of distributing the scattering interactions along the stations width. However, it also worsens the power of the $z$ vertex constraint, as one must account for the possibility of scatterings taking place where there is no solid material of exactly known position. Simulating interactions in air requires a doubling of the layers described in the code, both in the propagation of particles (with the resulting need to model multiple scattering in air) ad in the likelihood fit; we found this too taxing for the CPU consumption of our studies, so we omitted the description of air in our fast simulation. We believe that for the scope of this document the approximation of neglecting the effect of scatterings in air can be accepted, although it should be kept in mind as an improvement for a more precise study. Here we limit ourselves to point out that if the target layers of each station are assembled into three rigid 31.5-cm-long structures, as seems opportune (see {\em infra}), these can easily be filled with low-pressure helium and sealed. The resulting layout of a station then consists of 94.5cm of target blocks containing in total 1.5cm of Be and 93cm of gaseous He, plus 0.384cm of Si in 12 layers, plus a remaining 5.116cm of air. The equivalent $X_0$ of such a setup is of $0.8348 X_0 (Be+Si) + 0.00163 X_0 (He) + 0.00168 X_0 (air) = 0.8528 X_0$ for standard pressure filling, {\em i.e.} an increase of less than $0.4\%$ of total radiation length from that due to beryllium and silicon alone. If straightforward to implement, this is a simple and advantageous remedy to the worsening effect of scatterings with no $z$-vertex constraint. 

\subsubsection { Rate of scattering events}

A calculation of the rate of the interactions in the station, and a corresponding determination of the equivalent integrated luminosity and run time of a simulated data set of $N_{tot}$ scatterings, is not necessary for a study focusing on relative differences in the resolution of the measurable quantities. In any case, given a total width $W_{Be}=1.5$ cm of beryllium and $W_{Si}=0.384$cm of silicon material per station, the following calculation provides those numbers:\par

\begin{equation*}
\begin{aligned}
U_{Si} = 28.0855 \, gr/mol \\
U_{Be} = 9.0122  \, gr/mol \\
\rho_{Si} = 2.33 \, gr/cm^3 \\
\rho_{Be} = 1.85 \, gr/cm^3 \\
N_{e^-, Si} = \rho_{Si} Z_{Si} N_A/U_{Si} = 6.994 \, 10^{23} \, cm^{-3} \\
N_{e^-, Be} = \rho_{Si} Z_{Si} N_A/U_{Si} = 4.945 \, 10^{23} \, cm^{-3} \\
\sigma_{\mu e, el.} = 245 \mu b = 2.45 \, 10^{-28} \, cm^2 \\
N_{e^- tot} = W_{Be} N_{e^-, Be} + W_{Si} N_{e^-, Si} = 10.103 \, 10^{23} \, cm^{-2}  \\
N_{el. \, sc.}/\mu = \sigma_{\mu e, el} N_{e^- tot} = 2.475 \, 10^{-4}, \\
\end{aligned}
\end{equation*}

\noindent
from which one obtains \par

\begin{equation}
N_{el. \, sc. tot} = R_{\mu} N_{el. \, sc.}/\mu  = 3217.87 \, Hz,
\end{equation}

\noindent
where we have used the nominal average rate of muons of the CERN beam at 150 GeV running energy, $R_{\mu} = 1.3 \, 10^7$ Hz. 
Therefore a simulation of $10^6$ incident muons, all of which are forced to produce an elastic scattering interaction within a station, corresponds to a running time of about five minutes for the considered station.

The scattering $q^2$ determines uniquely the emerging angles of the electron and muon. In a reference system where the incoming muon travels exactly along the z axis, the divergences from the z axis of the two outgoing particles, $\theta_e$ and $\theta_{\mu}$,  follow the distributions determined by the equations of the previous section. The azimuthal angles of the two particles in the orthogonal plane $xy$ are generated such that $\theta_e$ has a uniform distribution in $[0, 2 \pi]$ and, of course, $\phi_{\mu} = \phi_e + \pi$. The generated three-vectors of electron and muon are then rotated to obtain their value in the laboratory frame, accounting for the incoming muon direction; for the transformation of coordinates see {\em infra}, Sec.~\ref{s:coordsystem}.

Following the interaction, the two final state particles are propagated through the detector. Besides the multiple scattering effects already mentioned above, we account for radiative losses of the electron momentum, so that the description of the electron trajectory correctly accounts for that effect (the multiple scattering formula (see Sec.~\ref{s:msc}) of course includes the dependence on momentum for the tracked particles).

\subsection {Detector description \label{s:detector}}

In order to study the effect of different design choices for the MUonE apparatus, we decided to simulate a set of four contiguous stations --four meters of apparatus, {\em i.e.} a tenth of its full length. The description of four stations in series allows to fully simulate the relevant inputs to a full-blown reconstruction of the event kinematics. In particular, by enforcing that all scatterings take place in the second station, we allow for the complete measurement of incoming and outgoing particles for the simulated events: the measurement of an incoming muon in three to six silicon modules ({\em i.e.}, in up to two contiguous stations) and the accounting for multiple scattering effects on the measurement precision due to the chosen material configuration (which is always assumed here to be identical in all stations); the interaction of the incoming muon with material in any one of the beryllium targets or within each of the silicon sensors of the second station of the set of four; and the tracking of the outgoing muon and electron trajectories in six to nine silicon modules downstream of the interaction. 

\subsubsection {General considerations}

Since each tracking module is composed of two adjoined double-sided strip sensors, the first one providing two readings of a coordinate transverse to the $z$ axis (chosen to be the $x$ coordinate here) and the second two measurements of the other ($y$), each module nominally provides up to four independent measurement points along the trajectory. Following the MUonE reconstruction logic, the two $x$ and the two $y$ measurements are combined to create $x$ and $y$ ``stubs''. Then, particle tracks considered in this study are constructed from triplets of $x$ and $y$ stubs in three consecutive modules; one of them is then by construction a stereo module, where the stubs are created in the rotated $x'$, $y'$ local coordinate set (see {\em infra}, Sec.~\ref{s:coordsystem} for the rotation relations). 

Here we recall that the original design choice of the MUonE detector envisions the repeated scheme of 40 independent tracking stations, each offering a limited $X_0$ thickness, as a way to acquire enough statistics of muon-electron scatterings (which are foreseen to allow to carry out the desired measurement in two years of data taking, if a target of 60cm equivalent of beryllium is employed) without suffering from large resolution losses due to the resulting multiple scattering effects that incoming and outgoing particles would undergo in a single thick target block. The modularity of the system is also a way to avoid systematic uncertainties related to the interalignment and positioning of the stations with respect to one another, as well as to provide for the ideal granularity of a triggering logic. This implicitly assumes that three tracking modules are sufficient for an effective reconstruction of the trajectory of the incoming and outgoing particles. We will see that this assumption is well borne by simulation studies; on the other hand, the taming of systematic uncertainties due to longitudinal misplacements, which may come into play in case one wishes to combine measurements in adjoining stations, is a very complex issue. We discuss some aspects of the general problem in Sec.~\ref{s:systs}. As for the triggering strategy, in this study we waive the constraint that each station be endowed with self-triggering capabilities. This allows to be free to consider the combination of measurements in different stations, and to investigate advantageous alternatives to the original design; of course, the price to pay is that the resulting trigger logic to be constructed becomes slightly more complicated. In addition, one has to renounce to the freedom of a non-calibrated positioning of the stations next to each other, as the tracking becomes sensitive to it.

As an additional point to be noted, we ignore in this study the effect of the possible addition, at the end of the set of 40 stations, of an electromagnetic calorimeter. The calorimeter may be useful for the identification of electron and muon signals when the two particles emerge with similar angles from the scattering, breaking the kinematic ambiguity; it may also provide for a stand-alone determination of the electron energy, which helps the determination of the scattering $q^2$; and it may help distinguish inelastic scatterings producing photons, $\mu e \to \mu e \gamma$. However, strictly speaking the calorimetric measurement is not mandatory to carry out a measurement, as for a well-known incoming muon momentum a full closure of the scattering kinematics only requires the determination of the trajectories of incoming muon and outgoing electron --the outgoing muon direction is already redundant if one aims to determine just the event $q^2$. The inclusion in this study of an electron energy measurement, riddled as it is with complex issues related to the description of the radiative losses of electrons produced far upstream, as well as with the effect of beam-induced backgrounds, would make significantly more complex an already quite extended and multi-parametric problem, and would ultimately prevent the formulation of very specific optimization questions, as other considerations --relative cost of the two sub-detectors being one of them-- would then come into play. We leave the study of combining an optimized tracking system with the most appropriate calorimetric design to future work.

\subsubsection {Description of the stations layout}

\begin{figure}[h!]
\begin{center}
\includegraphics[scale=0.8]{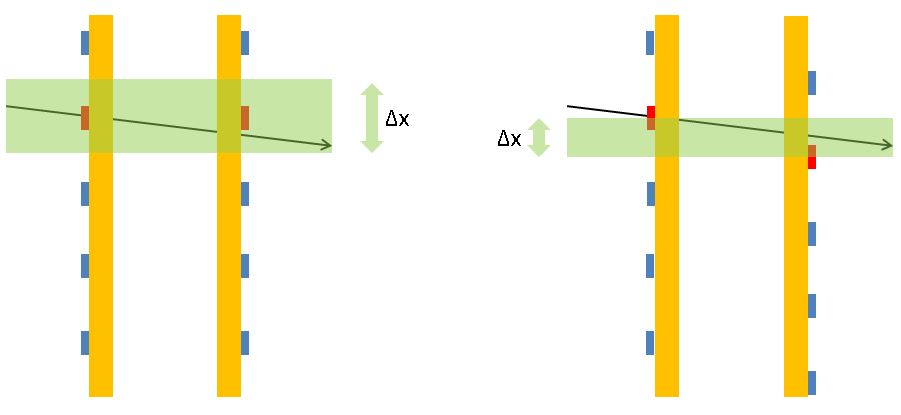}
\caption {\em Effect of vertical staggering of one of the two arrays of silicon strips of a double-sided sensor with respect to the other. On the left is shown the case of no staggering, when a low-divergence track produces a single-strip hit on both sides. The green band shows the inferred precision of the vertical position of passage of the track, which equals the strip pitch. On the right, a staggering equal to half the strip pitch produces a reduction by a factor of two of the combined uncertainty in the vertical position of the track crossing.}
\label{f:stripoffset}
\end{center}
\end{figure}

For each station we retain the general MUonE scheme of three silicon modules, each comprised of two double-sided silicon sensors (the two sides separated by a gap ranging from 1.8mm to 5.8mm~\footnote {The CMS modules will be built with a 1.8mm spacing between the two sides of double-sided sensors; however, {\em infra} (Sec.~\ref{s:relspacingsisi}) we entertain the possibility that the sensors be glued together with a wider gap, while keeping the total width of a module fixed.}, each pair reading out one coordinate ($x$ the first one, and $y$ the second one, moving from smaller to larger $z$ values along the beam); the two double-sided elements are mounted together such that the assembly has a total width of 1.5 cm.  We also keep the proposed scheme of having the second of the three modules measuring coordinates at a stereo angle, but the rotation angle is treated as a changeable parameter in our model, such that we may be sensitive to the effect of a different angular rotation from the default one of 45 degrees (although we do not expect any, due to approximate azimuthal symmetry considerations~\footnote {In truth, a small acceptance loss results from the rotation of one of the tracking modules, if stubs are requested to be recorded there, as there is then an imperfect overlap of coverage of the transverse plane; the acceptance loss is correctly factored in by the quantitative figures of merit discussed in Sec.~\ref{s:fom}.}). We assume, as MUonE does, that these are 1016-strip sensors with a pitch of 90 $\mu m$ and 320 $\mu m$ of thickness, arranged in a $10 \times 10$ $cm^2$ layout: as already mentioned, this conforms to the assumption that the experiment to be built with the same sensors used for the Phase-2 upgrade of the CMS tracker, with very considerable savings of time and money. We do allow for a transverse staggering (in the [0-45] $\mu m$ range) in the placement of the strips on the two sides of a double-sided module, to study what relative offset of the strips guarantees optimality of the resulting tracking. Intuitively, a 45-micron offset of the strip of one of the two sides reduces by a factor of two (hence from $90/\sqrt{12}=26 \mu m$ to $13 \mu m$) the position uncertainty for orthogonally incident particles which leave a signal in only one silicon strip on each side (see Fig.~\ref{f:stripoffset}). Charge sharing in more than one strip, with a resulting multi-strip cluster, further decrease the position uncertainty, but this is a rare occurrence for most of the tracks of interest, which travel with very small divergence (see Fig.~\ref{f:baseline1}, Sec.~\ref{s:baselinegeometry}). In any case, since the fraction of multi-strip clusters increases in a non-trivial way with the angle of incidence of the particles, we leave it to our simulation to determine the optimal configuration~\footnote {We note here that our modeling of hit generation in the sensors is rather crude (see Sec.~\ref{s:silicon}), hence this parameter should be subjected to studies using a full GEANT4 description.}. 

Wee keep the equivalent radiation length of the target material in each station fixed to the value chosen in the original MUonE design, {\em i.e.} 1.5cm of beryllium. This allows for an apples-to-apples comparison which factors out possible differences in the statistical uncertainty resulting from changes in the total radiation length, a parameter with which the number of useful scatterings scales linearly. However, there ends our set of assumptions for the layout of the target material. In fact we aim to study, with the definition of appropriate parameters, the performance of the measurement resulting from the following choices:\par

\begin{itemize}
\item the number of layers into which an equivalent 1.5cm Be thickness is divided;
\item their relative placement ({\em i.e.}, the inter-layer spacing);
\item the distance of the set of layers from the closest silicon module downstram;
\item the stereo angle of rotation of the middle module of each station;
\item the staggering of strips between the left and right sensor in each double-sided sensor;
\item the spacing between the two double-sided sensors in each tracking module.
\end{itemize}

\noindent 
As already noted, it is impractical to construct a device with a large number of thin beryllium layers. Other materials provide for easier handling and machining, and offer better rigidity. One such material is graphene, but there are other possible candidates. As the exact choice of target material has little or no effect on the measurable features of muon-electron scattering, in this study we stick with a description which uses layers of beryllium-equivalent material. If the study should evidence advantages of some geometrical layout with respect to others, it would have to be complemented with a more precise study of similar solutions employing different materials. The subtlety which requires this additional step lays in the non-negligible effect on the angular and $q^2$ resolution of varying even by small amounts the thickness (as measured in length units, by keeping the equivalent $X_0$ fixed and changing the material) of thin layers of target material, due to the constraining effect of the prior on scattering vertex z position that one can impose in a multi-track fit to the event kinematics. We will discuss this point in detail in Sec.~\ref{s:likelihood}.

In our study we assume that the relative precision with which thin layers of target material can be placed is of 10 $\mu m$. While this is also a parameter in our detector description, whose effect is duly studied, we believe the quoted figure is a reasonable assumption. In fact, it seems feasible to construct a stack of thin layers of, {\em e.g.}, graphene (say, 50 $\mu m$ thick each) alternated with spacing ``frames'' (which keep the target layers in place while providing no impedment to the passage of particles in a $10 \times 10$ $cm^2$ fiducial transverse area) of, say, $3$mm of thickness. A stack of 100 such layers would form a $30.5$cm long distributed target, which could be placed with high longitudinal accuracy between two silicon modules using a laser alignment system such as the one currently under development by the MUonE collaboration, or by other methods discussed below (see Sec.~\ref{s:systs}). A well-built distributed target would guarantee a very precise relative placement of each of the thin layers, offering a very tight constraint on the z position of the scattering vertex, provided that the structure retained sufficient rigidity. While the above is only a preliminary consideration, we indeed show {\em infra} (Sec.~\ref{s:geometries}) similar arrangements of different thicknesses of target material, in a number of spacing configurations.

\subsubsection{Parametrization of the stations layout \label{s:defaultpars}}

The layout of each station is specified by choosing the following parameters relative to the placement of the target and sensor layers.\par
\par

\vskip .2cm
\noindent
{\bf Fixed parameters:}\par
\begin{itemize}
\item Number of detection modules per station: 3;
\item Pitch $= 0.009$ cm: strip pitch in silicon sensors;
\item $W_{Si} = 0.032$ cm: width of each element of a double-sided silicon sensor;
\item $W_{mod} = 1.5$ cm: total module width;
\item $\Delta z_{mod} = 0.2$ cm: space left and right of tracking modules;
\item $\Delta z_{st} = 0.0$ cm: space between stations;
\item $W_{Be}^{tot} = 1.5$ cm: total width of beryllium per station;
\item Station length: 100 cm.
\end{itemize}

\vskip .2cm
\noindent
{\bf Variable parameters:}\par
\begin{itemize}
\item $\Delta z_{Si}$ (default $1.8$ mm): spacing between silicon layers in double-sided sensors;
\item $Z_{mod0}$: z position of left edge of first detection module in a station; 
\item $Z_{mod1}$: z position of left edge of second detection module in a station;
\item $Z_{mod2}$: z position of left edge of third detection module in a station;
\item $N_{Be0}$ (default 1): number of target layers to the left of the first detection module;
\item $N_{Be1}$ (default 0): number of target layers to the left of the second detection module;
\item $N_{Be2}$ (default 0): number of target layers to the left of the third detection module;
\item $N_{Be} = N_{Be0}+N_{Be1}+N_{Be2}$ (default 1): total number of target layers per station;
\item $\Delta z_{BeSi}$ (default 3.5cm): spacing between right edge of rightmost target and left edge of subsequent detection module;
\item $\Delta h_{stag}$ (default $0 \mu$m): staggering of strips on right side of double-sided sensor with respect to strips on left side of same sensor;
\item $\Phi_{stereo}$ (default $\pi/4$): angle of rotation of stereo strips in middle detection module.
\end{itemize}

\noindent
Once a value is defined for the above parameters, the uniform spacing between the $N_{Be}$ target layers then results to be
$\Delta z_{Be} = (100 -\Delta z_{st} -3 W_{mod} - 1.5 - 3 \Delta z_{BeSi})$ cm. A possible layout with eighteen uniformly spaced target layers was shown {\em supra}, in the bottom panel of Fig.~\ref{f:layoutmuone}.

\subsubsection{Coordinate systems \label{s:coordsystem}}

In the laboratory system we consider the $z$ axis as oriented along the nominal center of the muon beam direction. The $y$ coordinate points upward, and the $x$ coordinate points horizontally and is oriented to make a right-handed $(xyz)$ system. When orienting positive $z$ directions toward the right (as we do in all sketches of the detector and in all discussions of the event topologies in this document), we take the origin of the $z$ axis at the left edge of the first of the simulated detection stations.
Particle trajectories in this reference system are described by their divergence off the $z$ axis, $\theta$, and by their azimuthal angle in the $xy$ plane, $\phi$. Throughout the document we distinguish angles of incoming muons, outgoing muons, and electrons by using the subscripts $in$, $\mu$, and $e$ respectively. 

Tracking modules have strips oriented along the $y$ axis in the sensor positioned at smaller $z$ coordinate and along the $x$ axis in the sensor at larger $z$, hence these sensors respectively read out $x$ and $y$ positions for crossing particles. Strip positions and hits on these sensors are measured in cm from the center of the sensor, where the $z$ axis lays; hence local module coordinates in the transverse plane coincide with laboratory coordinates. An exception is the center module of each sensor, which is rotated by a stereo angle $\phi_{stereo}$ around the $z$ axis with respect to the other two modules. In this case the $x$, $y$ coordinates of hits in the laboratory system are derived from the local $x'$, $y'$ coordinates of the module rotated by an angle $\phi_{stereo}$ through the following rotation relations:\par

\begin{equation*}
\begin{aligned}
x = x' \cos \phi_{stereo} - y' \sin \phi_{stereo} \\
y = x' \sin \phi_{stereo} + y' \cos \phi_{stereo} .
\end{aligned}
\end{equation*}

\noindent
The elastic scattering reaction results from the incidence along the direction $\vec{u}=(u_x,u_y,u_z)$ of a muon from the beam on an electron considered with very good approximation at rest in the laboratory. In that frame of reference the incoming muon has a divergence $\theta_{in} = \arccos{u_z}$ and an azimuthal angle $\phi_{in} = \arcsin{u_y/u_x}$. It is advantageous to initially describe the scattering kinematics in a system $(x^{SC},y^{SC},z^{SC})$ rotated such that the incoming muon direction coincides with the $z^{SC}$ axis: one may define the direction of $x^{SC}$ and $y^{SC}$ by performing a rotation of the laboratory frame $(x,y,z)$ by an angle $\theta_{in}$ around the axis defined by the vector product of the beam axis versor $\hat{z}=(0,0,1)$ (the z axis in the laboratory system) with the versor $\vec{u}$:\par

\begin{equation*}
\begin{aligned}
\vec{R} = \hat{z} \times \vec {u}.
\end{aligned}
\end{equation*}

\noindent
The rotation is undefined if the incoming muon has zero divergence; in that case, the two systems coincide. It is worth noting here that numerical instabilities may arise in the calculation of derivatives of the likelihood function (see Sec.~\ref{s:likelihood}, {\em infra}) in the case of extremely small incidence angles. These have no effects on the results presented here, but should be considered with care if more accurate studies are performed.

\subsection{Reconstruction of hits and stubs in silicon sensors \label{s:silicon}}

\begin{figure}[h!]
\begin{minipage}{0.49\linewidth}
\begin{center}
\includegraphics[scale=0.5]{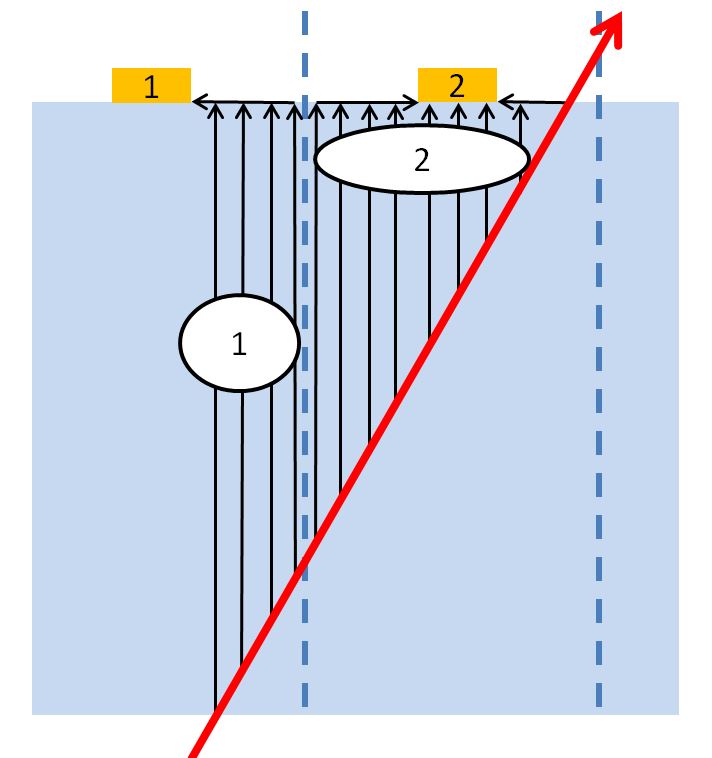}
\end{center}
\end{minipage}
\begin{minipage}{0.49\linewidth}
\begin{center}
\includegraphics[scale=0.65]{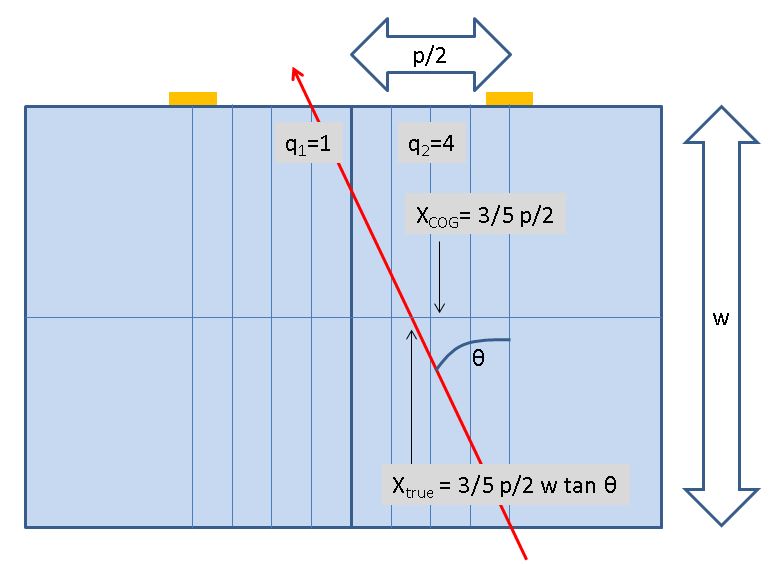}
\end{center}
\end{minipage}
\caption {\em Left: model of charge migration in the silicon sensors. Charge is generated in proportion to the particle path in the silicon bulk, and migrates to the strip plane, where it is then collected by the closest strip. Right: graphical demonstration of the bias of the center-of-gravity calculation for tracks with small divergence from normal incidence. The sensor is shown in section, with two neighboring readout strips of pitch $p$ on the top surface. The charge read out by the left strip, for the track shown in red, is one fifth of the total charge, and the charge read out by the right strip is four fifths. The COG calculation produces the position estimate shown in the figure for the crossing point of the center of the silicon layer by the track, which is displaced to the right of the true point, whose coordinate is correctly computed only using Eq.~\ref{eq:xi}. }
\label{f:cogbias}
\end{figure}

As we discussed {\em supra}, the silicon sensors considered in this work are those designed for the inner tracker of the CMS Phase-2 detector upgrade. These are double-sided, $w = 320 \mu m$-thick silicon layers, of approximately $10 \times 10$ cm$^2$ in size, instrumented with 1024 readout strips separated by $p = 90 \mu m$~\footnote {In the designed CMS Phase-2 tracker modules, the strips are broken into two 5-cm-long segments with separate readout. This detail has not been simulated, as it has no relevance to the resolution of the tracker, but only on the noise in the sensors and in background rejection, which are not treated here.}. In order to appreciate the effect of a discrete layout of silicon micro-strips in the detection elements, the fast simulation must account for the different resolution that results when ionizing particles deposit a signal above threshold in only one strip or in two or more adjoining strips. The crude model we constructed involves an evaluation of the total charge that would be read out by the electronics if charge migrated along straight paths orthogonally to the silicon surface, and were entirely collected by the closest strip at the surface (see Fig.~\ref{f:cogbias}, left).

In the above model, orthogonally incident particles producing ionization charge at less than $45 \mu m$ from a strip center in the direction orthogonal to the strips will yield signal (nominally $21,120$ electrons) only in that strip. Conversely, particles that cross the inter-strip boundary during their propagation in the silicon material, or particles hitting the silicon with a large incidence angle, will instead see their produced ionization charge split into two or more adjacent strips according to their incidence angle and crossing position. We generate noise in the strips reading out the ionization charge, as well as in their two closest neighbor strips on each side, as a Gaussian distribution of zero mean and sigma of 1000 electrons~\footnote {With the use of these approximated parameter values (courtesy N. Bacchetta, private communication) we have chosen to include for the sake of completeness a rough description of effects of electronic noise in our simulation; their values have however practically no effect on all the results discussed in this work.}; then we assume a threshold charge of 3000 electrons for the readout, dropping from consideration strips that collected a total charge below that value~\footnote{Here we have assumed that the MUonE electronics will be able to read out analog information on the deposited charge in the strips. This is however not granted, due to difficulties connected to reading out the strips in an asynchronous way (as the timing of arrival of muons is not fixed). The absence of information of the deposited charge reduces the resolution of multiple-strip hits, but does not substantially modify the conclusions of our study, due to the small fraction of multi-strip clusters.}.

Once the total charge above threshold is known, the position of the track along the local strip coordinate
, at a $z$ coordinate corresponding to half the width of the silicon sensor, is calculated as follows. For one-strip clusters, the position is defined to coincide with the position of the strip center, and its uncertainty is given by $90/\sqrt{12} \mu m$. For two-strip clusters, we indicate as $q_1$, $q_2$ the charge in the two adjacent strips, assuming $q_2>q_1$, and we compute the coordinate as \par

\begin{equation}
\xi_{COG} = 0.5 p \frac{q_2-q_1}{q_1+q_2} = 0.5 p (2\rho -1)
\end{equation}

\noindent
where $p$ is the strip pitch, and where we have defined $\rho = q_2/(q_1+q_2)$. We set the local coordinate $\xi=0$ at the interstrip boundary, and $\xi>0 (<0)$ around the strip reading more (respectively, less) charge. The position along $z$ is always calculated as the center of the silicon layer, {\em i.e.} at a $z$ distance of $160 \mu$m from the smaller-$z$ edge of the layer.

In passing we note that, as shown geometrically in Fig.~\ref{f:cogbias}(right), the COG as defined above is an unbiased estimator of the center of the trajectory (given a sharing of charge in two neighboring strips) only if the angle of track incidence $\theta$ is equal to or larger than the geometry ratio $\tan \theta = p/w$, hence for very large angles, $\theta>0.274$, which never arise in the considered setup. For smaller angles, the unbiased position estimator would rather be \par

\begin{equation}
\xi_{unb} = w \tan \theta \left( \frac{q_2-q_1}{q_1+q_2} \right)
\label{eq:xi}
\end{equation}

\noindent
Equation~\ref{eq:xi} requires a knowledge of the track angle of incidence $\theta$ on the sensor, which is not available at the time of hit finding. It could still be used in a more refined likelihood definition than the one we have adopted here~\footnote { The hit positions, which are the data upon which the likelihood definition relies, may be made themselves a function of the polar angles of the tracks. This in practice means incorporating the hit positions calculation inside the likelihood function, which therefore moves from being defined by hit positions data to being defined by charge depositions data. We believe this approach should be investigated for the MUonE experiment. }; we believe the effect of this improvement would be small, again due to the fact that the vast majority of the tracks of relevance to this study produce single-strip clusters in the modules; on the other hand, it is a fact that their angles are in all cases very small, such that the COG definition is, indeed, a biased one in an idealized sensor where charge drifts in the silicon bulk perfectly orthogonally to the strips.

We assign an uncertainty to the measured position $\xi$ of two-strip clusters as the propagation of the above-mentioned noise level ($\sigma_q = 1000$ electron charges) on the center of gravity calculation, \par

\begin{equation}
\sigma_{\xi}^2 = p^2 \frac{(q_1^2 +q_2^2) \sigma_q^2}{(q_1+q_2)^4}, 
\end{equation}

\noindent
which reduces to \par

\begin{equation}
\sigma_{\xi} = p \rho (1 - \rho) \sigma_q \sqrt{1/q_1^2 + 1/q_2^2}.
\end{equation}

\noindent

For clusters of larger multiplicity (which, because of the small angles of the involved tracks, may only result from the effect of noise above threshold in strips adjacent to those receiving ionization charge), we keep for simplicity the uncertainty calculation above, using the two strips with highest collected charge. This approximation has no effect, as we have practically no such cases even considering the largest datasets we simulated.

Finally, we note that for simulated events where the scattering interaction takes place within a silicon sensor we do consider the combined effect of ionization by the incoming track and by the two outgoing tracks, properly accounting for the charge deposition of each track segment, albeit by applying the same crude charge transport model described above. In those cases, the likelihood definition includes the hit produced by the three tracks as a shared hit of the three trajectories; $60\%$ of the resulting charge clusters are multiple-strip ones. The scattering position is thus in general better known than the position of any other hit, if it takes place inside the silicon. 

\clearpage

\section{ Reconstruction of the event kinematics}

\subsection { The likelihood function \label{s:likelihood}}

When dealing with event reconstruction based on hits in tracking detectors, one usually starts by defining a criterion to construct track segments from a restricted set of detector components, then iteratively associates other hits to those segments, and finally refits the full trajectory of charged particles; high-level information on the event characteristics can then be constructed with the latter. Such a bottom-up strategy works very well even in the most complex environments, and is robust to noise and other experimental effects. As noted elsewhere, however, in this work we aim at estimating the best possible performance that the experimental setup can provide; therefore we want to decouple from noise and combinatorial effects that, while always present, can be tamed with tools we have no chance to study with a fast simulation. The straightforward way to reconstruct the elastic scattering events would therefore be to fit to straight lines the hit collection of each track, and then derive, from their relative angles, information on the event $q^2$. In so doing we would however encounter the issue of having to combine the information provided by each of the two final state particles: electron and muon divergences from the incoming muon direction both offer in principle independent estimates of the event $q^2$, albeit with significantly different precision (in most of the phase space, in fact, the electron is measured with a much higher relative precision). 

Combining electron and muon post-fit information is possible but not optimal, as linear approximations to the covariance terms must be used. A very attractive alternative is offered by the simplicity of the topology we aim to reconstruct. We can directly fit the event $q^2$ starting from a univocal association of the hits to the three involved particle tracks. In so doing, the full information is exploited more effectively and precisely. Such a procedure, in a real experimental situation, would have to be preceded by the identification of the signal hits for each track; here its optimality indicates that we must use it as our baseline $q^2$ determination in our study.

The likelihood function we aim to define depends on the following parameters: \par

\begin{enumerate}
\item $p_0 = q^2$, the squared four-momentum transfer;
\item $p_1 = x_0$, the $x$ coordinate of the scattering interaction;
\item $p_2 = y_0$, the $y$ coordinate of the scattering interaction;
\item $p_3 = z_0$, the $z$ coordinate of the scattering interaction;
\item $p_4 = \phi_{e}^{SC}$, the azimuthal angle of the final-state electron in the scattering frame;
\item $p_5 = \theta_{in}$, the divergence of the incoming muon with respect to the $z$ axis, as measured in the laboratory frame;
\item $p_6 = \phi_{in}$, the azimuthal angle of the incoming muon in the $xy$ plane orthogonal to the $z$ axis, in the laboratory frame.
\end{enumerate}

\noindent
As mentioned in Sec.~\ref{s:coordsystem} the scattering frame, in which the scattering is generated, is defined such that the incoming muon travels aligned with the positive verse of the $z^{SC}$ axis. In that system the azimuthal angles $\phi_e^{SC}$ and $\phi_{\mu}^{SC}$ are related by $\phi_e^{SC} = \phi_{\mu}^{SC}+ \pi$, {\em i.e.} they are back-to-back. 
From the parameters defined above one may compute the final state electron energy $E_e$, the final state muon energy $E_{\mu}$, and the other angles of the outgoing particles ($\phi_e^{SC}$, $\phi_{\mu}^{SC}$, $\theta_{\mu}^{SC}$, $\theta_e^{SC}$) in the scattering frame, using the formulas of Sec.~\ref{s:scattering}.

The likelihood can only be defined once we have experimental (simulated, in our case) data. These come as measurement pairs $(x_j \pm \sigma (x_j), z_j \pm \sigma (z_j))$ for the left double-sided sensor of non-rotated modules, pairs $(y \pm \sigma (y_j), z \pm \sigma (z_j)$ for the corresponding right sensor, and corresponding pairs $(x_j^{'},z_j)$, $(y_j^{'},z_j)$ in left and right sensors of stereo modules. Uncertainties are computed as discussed in Sec.~\ref{s:silicon}. For all results discussed in this document we only consider events for which we have reconstructable tracks with at least three $x$ and three $y$ stubs for each of the particles involved in the scattering, and we use the three of them closest to the scattering position in the likelihood calculation; therefore, the likelihood includes 36 distinct coordinate pairs in its definition. The rationale of this is to emulate the original choices of the experiment --in particular, those relative to the idea of triggering on stub triplets. However, we did study the effect of relaxing the above conditions, finding that besides a general worsening of the fit quality when larger number of hits along the tracks are considered~\footnote{ We warn the reader here that this conclusion only refers to a fit that combines the three tracks in a global determination of the $q^2$; fits that determine separately the trajectories of each of the three particles may instead benefit from using information from a larger number of hits for each track; yet what really matters is the uncertainty at the scattering position along the z axis, once a constraint of single origin is applied to the three tracks.}, there is little wisdom to obtain as far as geometry optimization is concerned. Of course, this effect should be explored in more detail with a more precise simulation, once an optimized fit strategy for the particle trajectories is devised ({\em e.g.}, one which includes in the likelihood definition the modeling of the multiple scattering on particle trajectories with its non-Gaussian distributions, as well as background effects causing hit precision degradation, a more precise modeling of charge deposition in the silicon sensors, knowledge of the tracks incident angle in the hit position determination, and so on).

We may define our likelihood in a concise form as follows: \par

\begin{equation*}
\begin{aligned}
\log L(\vec{p}) = - \sum_{i=1}^{N_{SC}} [  
   \sum_{j=1}^{n_{x hit,in}} 0.5 \frac{[x_j^{(')}-x_{in}^{(')}(z_j)]^2}{\sigma_{x_j^{(')}}^2} +  
   \sum_{j=1}^{n_{y hit,in}} 0.5 \frac{[y_j^{(')}-y_{in}^{(')}(z_j)]^2}{\sigma_{y_j^{(')}}^2} +  \\
   \sum_{j=1}^{n_{x hit,\mu}} 0.5 \frac{[x_j^{(')}-x_{\mu}^{(')}(z_j)]^2}{\sigma_{x_j^{(')}}^2} +  
   \sum_{j=1}^{n_{y hit,\mu}} 0.5 \frac{[y_j^{(')}-y_{\mu}^{(')}(z_j)]^2}{\sigma_{y_j^{(')}}^2} + \\
   \sum_{j=1}^{n_{x hit,e}} 0.5 \frac{[x_j^{(')}-x_e^{(')}(z_j)]^2}{\sigma_{x_j^{(')}}^2} +  
   \sum_{j=1}^{n_{y hit,e}} 0.5 \frac{[y_j^{(')}-y_e^{(')}(z_j)]^2}{\sigma_{y_j^{(')}}^2} ]  + 
\log P(p_3) 
\end{aligned}
\label{eq:likelihood}
\end{equation*}

\noindent
Above, true particle coordinates are obtained by propagating as straight lines the trajectories of the three particles from the scattering position (defined by $x_0=p_1,y_0=p_2,z_0=p_3$) to the nominal measurement coordinates $z_j$. The propagation uses the angles parametrized by $p_4, p_5, p_6$ as well as those derived from combining the kinematical constraints of Sec.~2 and Sec.~3.1.3 and using $p_0=q^2$. Also, in the stereo layers the $x^{'}$, $y^{'}$ coordinates are of course determined by rotating the laboratory ones by the appropriate stereo angle $\phi_{stereo}$. So, for instance, for a hit in the $j$-th x-measurement layer assigned to the incoming muon, we compute the expected particle position as\par

\begin{equation*}
\begin{aligned}
x_{in} = (z_j-z_0) \tan{\theta_{in}} \cos{\phi_{in}} + x_0,
\end{aligned}
\end{equation*}

\noindent 
while for a hit in a stereo layer measuring the $y^{'}$ coordinate assigned to the outgoing muon, we compute\par 

\begin{equation*}
\begin{aligned}
x^{'}_{\mu} = (z^{'}_j-z_0) \tan{\theta_{\mu}} \sin{(\phi_\mu-\phi_{Stereo})} + (-x_0 \sin{\phi_{Stereo}}+y_0 \sin{\phi_{Stereo}}).
\end{aligned}
\end{equation*}

\noindent
As for the measurement uncertainties $\sigma(x_j^{(')}), \sigma(y_j^{(')})$, in the likelihood model they result from the combination of two contributions: the uncertainty from the strip cluster position reconstruction, and the estimated uncertainty in the trajectory resulting from the amount of crossed material from the interaction point. For the first contribution, we assume that single-strip clusters have a nominal uncertainty of $90/\sqrt{12}=26 \mu m$, and for multiple-strip clusters the position uncertainty along the measurement coordinate is instead determined by propagating the uncertainties on the center-of-gravity calculation of the deposited charge, as discussed {\em supra}(Sec.~\ref{s:silicon}. We do not attempt to model the non-Gaussianity of the sampling distribution of the position uncertainty which results from the discreteness of the position measurement for single-strip clusters, as our studies indicate that it makes no practical difference on the value of the parameters at the likelihood maximum, nor on their uncertainty, while it considerably increases the CPU load for the event reconstruction. 

For the second contribution to the position uncertainties in x and y we proceed as follows. We first evaluate the expected divergence of a particle from its initial trajectory caused by multiple scattering in the total amount of traversed material, also accounting for its estimated momentum as described in Sec.~\ref{s:msc}. The calculation differs for initial and final state particles, as for the incoming muon the traversed material is computed as the sum of contributions of all layers from the considered measurement layer to the scattering position, and the assumed momentum is the beam momentum; while for final state muon and electron the traversed material is computed as the sum of contributions of all layers from the scattering position to the considered measurement layer downstream it, and the assumed particle momentum is derived from the $q^2$ using the formulas of Sec.~\ref{s:scattering}. 

We then compute, for {\em e.g.} a measurement of the x coordinate of a final state muon of momentum $p_{\mu}(q^2)$, \par

\begin{equation*}
\begin{aligned}
\theta_{ms} = \sqrt{2} \frac{0.0136}{p_{\mu}(q^2)} \sqrt {\Delta X_0(z_j-z_0)} (1+0.038 \log [\Delta X_0(z_j,z_0)]), \\
\sigma(x_j)^2 = \sigma(x_j, hit)^2 + ((z_j-z_0)\frac{\tan{(\theta+\theta_{ms})}-\tan{(\theta-\theta_{ms})}}{2} \cos{\phi_{\mu}})^2.
\end{aligned}
\end{equation*}

\noindent
Above, $\Delta X_0(z_j,z_0)$ is the estimated radiation length traversed by the particle in traveling from $z_0$ to $z_j$; it is a function of both coordinates as different positions along the detector will correspond to different material thicknesses for a given $\Delta z$. The formula above correctly models the smearing effect on the particle trajectories due to angular variations. We instead ignore the less important contribution to the hit position uncertainty of the position shifts $\Delta x_{MS}$, $\Delta y_{MS}$ as modeled in Sec.~\ref{s:msc}. Due to the inclusion of the $\theta_{ms}$ effect, the position uncertainty of each hit is itself indirectly a function of the $p_3=z_0$ and $p_0=q^2$ parameters in the fit, and duly varies during maximization along with them.

Uncertainties in the z position of the hits are considered only when studying systematical effects resulting from the precision of the placement of detection and target layers (see {\em infra}, Sec.~\ref{s:systs}); they are instead ignored ({\em i.e.} $\sigma(z_j)=0$ for hit measurements) in the studies of relative merits of the different geometries, conforming to the general methodology adopted in the present optimization study.

In the likelihood definition the hits associated to incoming and outgoing particles are univocally assigned to each of the true particles that produced them, speeding up the calculation. While this simplifying assumption looks like some sort of cheating at first sight (as it equates to assuming, in addition to the absence of background hits, that a perfect identification of the particle species is available prior to the kinematic fit), we trust it does not affect the conclusions we can draw in our study, as we take the ansatz that the relatively rare ambiguous kinematic configurations may be resolved by considering the signal left in the calorimeter by the two particles. In Sec.~\ref{s:switch} we briefly study the level of degradation to the resolution in $q^2$ and other measured quantities caused by a complete ignorance on the identity of the two outgoing particles. 

\subsubsection {The $z$-vertex constraint \label{s:zvertex}}

The last term in the likelihood function above is an important ingredient. The $P(p_3=z)$ function can be defined as the probability distribution of the possible $z$ positions of the scattering vertex. It should be intuitively evident that a precise knowledge of the interaction point benefits the correct reconstruction of the event kinematics, but it is hard to gauge by back-of-the-envelope calculations how much do angular measurements depend on applying a constraint that the vertex $z$ must lay where there are electrons along the path.

The function $P(z)$ must be defined in a way that accounts for measurement precision of the layers placement along the $z$ axis. We take this number to be $10 \mu$m, as that is the specification originally required, and considered achievable, by the MUonE collaboration for the placement of the detection sensors. In truth, we will show in Sec.~\ref{s:biases} how elastic scattering data may be used to constrain with much higher accuracy the placement of target and detection layers along $z$, but we keep the $10 \mu$m precision as a baseline in the definition of $P(z)$. Of course, while the probability should decrease to a negligible value when the scattering $z$ position falls away from the nominal position of the closest layer, within the material it should be constant. We chose therefore to model it by using two back-to-back $Erf(z)$ functions, as follows:\par

\begin{equation*}
\begin{aligned}
P(z) = \sum_{i=i^*-3,i^*+3} \frac{X_0^i}{X_0^{tot}} \left( 0.5+0.5 Erf(\frac{z+w^i/2-\bar{z}^i}{\sigma_z} \right) S(\bar{z_{i}}-z) \\
     = \sum_{i=i^*-3,i^*+3} \frac{X_0^i}{X_0^{tot}} \left( 0.5+0.5 Erf(\frac{\bar{z}^i+w^i/2-z}{\sigma_z} \right) S(z-\bar{z_{i}})
\end{aligned}
\end{equation*}

\noindent
where $S(z)$ is a step function ($S(z>0)=1$, $S(z<0)=0$), $X_0^i$ is the total width in radiation lengths of the considered layer $i$, and $X_0^{tot}$ is the sum of the radiation lengths of the considered adjacent layers, such that $p(z)$ is correctly normalized. The sum over nearby layers allows for very large occasional deviations from the true $z$ value to correctly contribute to the total probability~\footnote {Due to numerical precision issues, for large absolute values of the argument of the $Erf(z)$ functions their value is set to $10^{-16}$ in the code (the smallest value returned by the function TMath::Erf() from the used mathematical functions library); this apparently creates no convergence issues to the likelihood maximization, provided that the initial step in the related variable is set to a large enough value.}.

\subsubsection { Likelihood maximization }

We use Minuit~\cite{minuit} for the search of the likelihood maximum in the 7-parameter space defining the kinematics of every elastic scattering interaction, and minimize the $-\log L$ value computed as discussed {\em supra}. In the initialization phase, Minuit requires the user to specify a range for every parameter, as well as an initial guess of the steps to be taken in each direction in search for the minimum. We use the following range and step values:\par

\begin{enumerate}
\item $q^2$: $[0., 0.15]$ GeV, step = 0.00001 GeV;
\item $x_0$: $[-10., 10.]$ cm, step = 0.001 cm;
\item $y_0$: $[-10., 10.]$ cm, step = 0.001 cm;
\item $z_0$: $[-10., 410.]$ cm, step = 0.001 cm;
\item $\phi_{el}^{SC}$: $[-2 \pi, 4 \pi]$ rad, step = 0.001 rad;
\item $\theta_{in}$: $[0., 0.1]$ rad, step = 0.00001 rad;
\item $\phi_{in}$: $[-2 \pi, 4 \pi]$ rad, step = 0.01 rad.
\end{enumerate}

\noindent
To make Minuit work, a starting value for each of the parameters must also be provided by the user. Although the minimization usually converges regardless of what initial values are given, CPU consumption is significantly reduced if we give as starting parameters the true ones --the true generated $q^2$, the real $x,y,z$ coordinates of the generated scattering interaction, the true value of the $\phi_{el}^{SC}$ angle in the scattering frame, and the true incoming muon beam divergence and azimuthal angle $\theta_{in}$, $\phi_{in}$. This ``illegal'' procedure --one which we may not apply to real data-- does not invalidate our results, as a careful minimization that considered in turn the different possible configurations of free parameters would allow to find the same global minimum. Again, our focus here is to compare different geometrical configurations of the detector, and we do it by voluntarily choosing an idealized situation. In this case, the benefit is in the speed of the minimization, which translates in the chance of analyzing larger simulated datasets, obtaining more precise information on the relative merits of the different considered choices.

One further note has to be made concerning the minimization strategy. We found that in some event configurations and for values of $q^2$ around 0.084 and 0.13, the standard minimization strategy invoked by the ``migrad'' command sometimes fails to provide the true likelihood minimum, being affected by numerical precision issues connected with the vanishing gradients of the trigonometric functions used in the transformation of coordinates from the scattering to the laboratory frame. The use of the less rigorous ``simplex'' strategy instead is unaffected by those peculiarities. As the performance of the two routines is otherwise undistinguishable in our case, we use the latter.

\subsection{ What should we optimize on? \label{s:optimize}} 

In general, the approximations adopted in the present study all go in the direction of producing an idealized situation. In particular, no background hits worsen the resolution of track reconstruction; inelastic scatterings are ignored; no ambiguity is introduced in the identification of the scatterings (although we do study the issue in Sec.~\ref{s:switch}); no delta rays affect hit resolutions; no non-Gaussian tails affect the propagation of particles in the material. Careful studies of the real detector which will hopefully be built, and analysis of the resulting real data, will no doubt allow the production of a reconstruction software capable of minimizing the deteriorating effects of those approximations. Here, on the other hand, we believe that their consideration would confuse the issue of pinpointing the relative merits of the different geometry options under study. 

What we believe must be the focus here is to discuss {\em what it is that we want to measure as precisely as possible}, given the experimental situation we model and regardless of its approximate nature and its simplifications. There is no doubt on what a principled answer should be: for an end-to-end optimization we should aim for the smallest possible uncertainty on the value of $\Delta \alpha_{had}$ obtained from a given integrated luminosity collected by the apparatus, such as the one corresponding to two years of data-taking ({\em e.g.}, the number used by the MUonE collaboration in their studies, $L = 1.5 \times 10^7 nb^{-1}$) once the most effective reconstruction of the events is carried out, and once all systematic uncertainties are considered. {\em That} parameter is indeed the one we ultimately need to determine with precision in a self-respected optimization study. Of course, the above is a really tall order, for reasons which should be obvious: we do not have an optimal reconstruction software handy (while, in fact, we do offer our global likelihood as a bid for the general direction to take in an optimized reconstruction here, numerous improvements should be considered in its definition), nor can we model all systematic sources in a credible way before real data are collected~\footnote{We nonetheless stress here, in passing, that in our experience most instrumental systematic uncertainties can usually be beaten down to smaller values than originally believed, by careful studies of real data and using techniques not evident at a design stage.}. Hence, we need to consider the various elements separately below, to try and simplify our task.

$\Delta \alpha_{had}$ may be extracted from a shape fit to the distribution of the differential cross section for elastic scattering, $d \sigma /d q^2$,  \par

\begin{equation}
\Delta \alpha_{had}(q^2) = 0.5 \left( \frac{\int \frac{d \sigma}{d q^2} dq^2}{\int \frac{d \sigma_{LO}}{ dq^2}dq^2} - 1 \right) 
\end{equation}

\noindent
by integration over $q^2$. The precise determination of the differential cross section as a function of $q^2$ required to perform that calculation rests on a determination of the $q^2$ of each scattering event with the smallest possible uncertainty, particularly in the region of high $q^2$ where the hadronic contribution (and thus the integrand at the numerator in the equation above) reaches its largest relative value. This is because, even in presence of a very accurate model of experimental effects, the worsening of $q^2$ resolution amounts to an irrecoverable loss of information. In addition, the extraction of $\Delta \alpha_{had}$ from such a shape fit is riddled with very hard to control systematic effects that modify the shapes of the electroweak and hadronic contributions from their calculated values. A detector offering the highest resolution on elastic scattering parameters will improve the constraining power of the data on the values of the parameters describing those effects. In this document we do provide, for some sample geometry choices, the variation of the relative statistical uncertainty on $\Delta \alpha_{had}$ resulting from a template fit as a function of the studied parameters; the fit methodology is described in Appendix B. In general, those results confirm the results of the more straightforward optimization measures discussed below, but they are not as precise, as they are much more affected by stochastic noise.

A different approach to estimate the hadronic contribution, much simpler although not necessarily less problematic from the standpoint of taming systematic uncertainties, has been proposed~\cite{passeraabbiendi}. It involves the calculation of the ratio of the cross section integrated in two separate ranges of $q^2$: one, a ``normalization region'' (NR), where the hadronic contribution is expected to be negligible; and another, a ``signal region'' (SR), where the wanted effect achieves its largest relative size. Having defined the boundaries of these two regions ($[q^2_{min,NR},q^2_{max,NR}]$ and $[q^2_{min,SR},q^2_{max,SR}]$), one may compute\par

\begin{equation*}
\begin{aligned}
N = N_{obs,SR}- f N_{obs,NR} \\
f = N_{EW,SR}^{th} / N_{EW,NR}^{th} = \frac {\int_{q^2_{min,SR}}^{q^2_{max,SR}} \frac{dN_{EW}}{dq^2} dq^2 }{\int_{q^2_{min,NR}}^{q^2_{max,NR}} \frac{dN_{EW}}{dq^2} dq^2 },\\
\end{aligned}
\end{equation*}

\noindent
from which one gets\par

\begin{equation}
\Delta \alpha_{had} =  N/(kL).
\end{equation}

\noindent
Above, $L$ is the integrated luminosity of the considered data, $k$ is a theoretical estimate of the fraction of the hadronic contribution in the signal region, and $N_{EW,SR}$ and $N_{EW,NR}$ are the predicted number of events expected from the electroweak contribution in the signal and normalization regions, while $N_{obs,SR}$ and $N_{obs,NR}$ are the observed event counts in the corresponding regions. An optimization of the normalization and signal region can be performed based on the amount of accumulated statistics. 
Such a calculation is easier to perform than a fit to the full differential shape of the measured cross section, but it is riddled by the same uncertainties, in particular those affecting our knowledge of the precision of the $q^2$ determination for each event. A systematic effect on the measured value of $\Delta \alpha_{had}$ also results from neglecting the hadronic contribution to the normalization region, although it is in principle easy to remove it by an iterative procedure, if the electroweak shape of the $q^2$ distribution is known with high precision. 

The precise impact on the final uncertainty on $\Delta \alpha_{had}$ of the theoretically modeled distributions is hard to assess. Equally hard is to foresee how well a real experiment may end up determining, after dedicated studies, the exact model of the resolution in measured $q^2$, and in particular its non-Gaussian tails, from the event kinematics: that function is a crucial input to any accurate fit to the cross-section shape. Because of this, we believe it is better in our study to stick with the intermediate goal of minimizing the uncertainty in the event $q^2$. A statistic correlated with that quantity is easy to define and determine directly, using the results of the global likelihood fit to the hit positions produced by simulated scattering events. Since the $q^2$ resolution is a function of $q^2$ itself, it is useful to try and be more specific. 
In the following we will focus both on the full-range RMS of the distribution of relative residuals $r_q = \frac{q^2_{meas}-q^2_{true}}{q^2_{true}}$, and on the RMS of residuals in the restricted region $q^2>0.1$ which is the most relevant for the measurement of the hadronic contribution to the cross section. While the RMS neglects to consider the non-Gaussian shape of residuals, its minimization should go quite far in the way of optimizing the measurement potential. Indeed, to make the investigated statistics even more robust, we have chosen to truncate the positive and negative $5\%$ tails of the distributions before computing their RMS and other quantities reported in the rest of this document, after verifying that the residuals have in all cases very close to Gaussian behaviour. This choice allows to focus on the properties of the bulk of the data, as the truncated RMS is less dependent on the occasional large residual which can always occur in pathological configurations.

In an attempt at capturing more precisely the effect on the measurement process of design variations, during our studies we tried insuccessfully to define, in addition to the above two, several alternative optimization measures related to an appraisal of the distinguishability of the hadronic component from the electroweak differential cross section curve. While the statistics we studied appear good choices in general, they proved to be insufficiently sensitive to the relative variations of functional shapes caused by the design variations that are the focus of this work. In the rest of this document we only occasionally show results of the use of two of them, which are discussed in Appendix A.

\section {A look at the main choice: concentrated versus distributed target \label{s:geometries}}

\subsection {The baseline geometry \label{s:baselinegeometry}}

The geometry we consider as our baseline option for a muon-electron scattering detector is the one originally proposed by the MUonE experiment, as our goal is to determine how much one may gain (in an appropriate metric such as one of those discussed in the previous Section) by choosing the most proficuous arrangement of detection and target layers.  The exact foreseen positioning of the detection modules and concentrated target within each station of MUonE is not precisely stated in public documents, but an approximated layout can be extracted from the figures in~\cite{muonedoc}. We model it by fixing the following parameters in our simulation (also see Sec.~\ref{s:detector}):\par

\begin{itemize}
\item total number of target layers per station: 1;
\item total width of target layer: $W_{Be}=1.5$cm;
\item position of the left edge of the three tracking modules in each station: $z_{mod \,0} = 5$cm; $z_{mod \,1}=50$cm, $z_{mod \,2}=95$cm;~\footnote{ Shortly before submission G. Venanzoni indicated that the space between target and first tracking module of the proposed MUonE detector is actually of 15cm This difference has only a minor impact; we provide some results for the different configuration {\em infra}. }
\item position of the left edge of the target layer in each station: $z_{Be \,0} =1.5$cm;
\item stereo angle in middle tracking module: $\theta_{stereo}=\pi/4$ rad;
\item spacing between silicon layers in double-sided sensors: $\Delta z_{Si}=0.18$cm;
\item transverse staggering between strips on the two sides of a double-sided sensor: $\Delta h_{stag}=0 \mu$m.
\end{itemize}

\begin{figure}[h!]
\begin{center}
\includegraphics[scale=0.8]{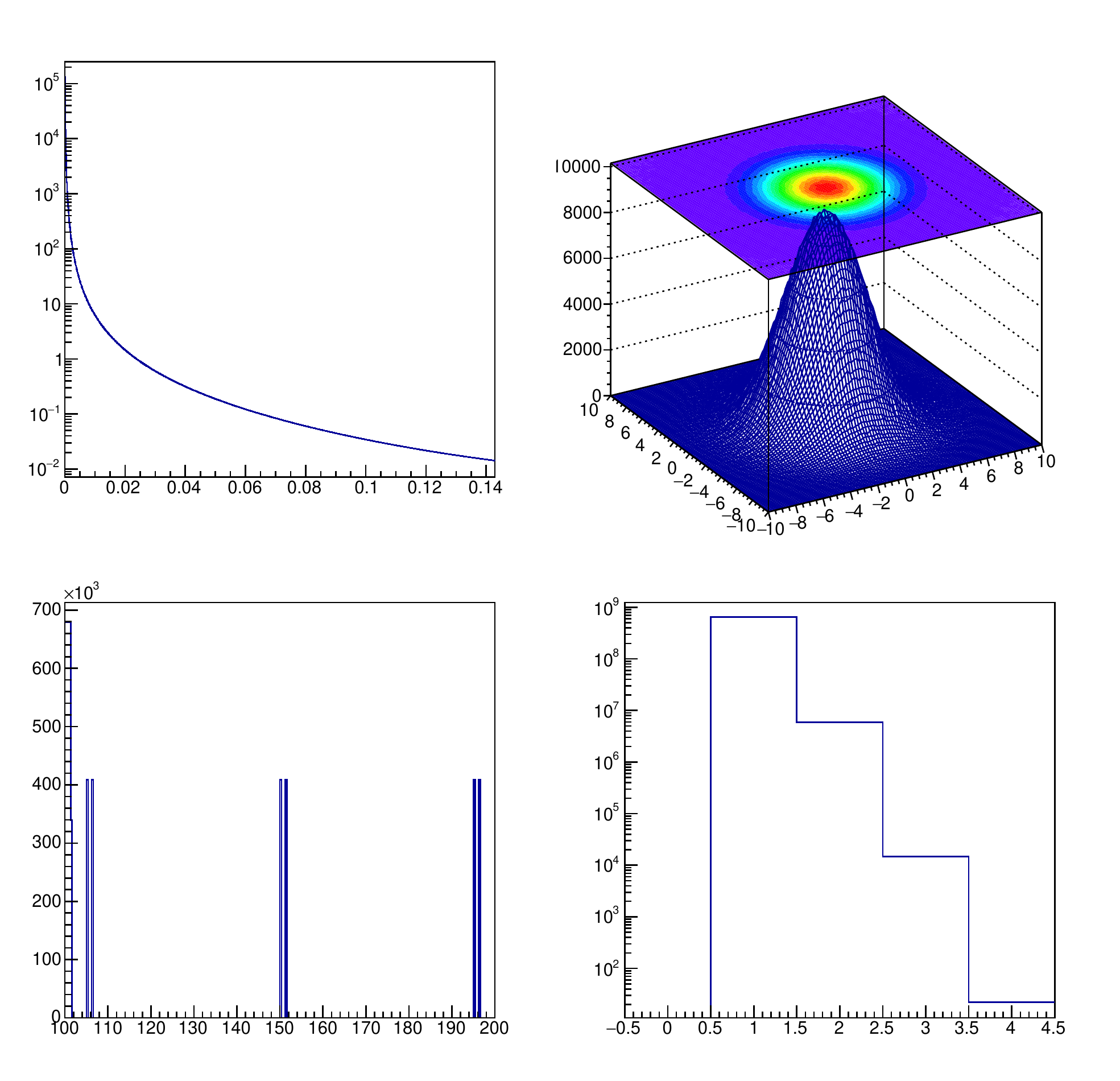} 
\caption {\em Top left: Elastic scattering cross section as a function of $q^2$. Top right: distribution of incoming muon coordinates in the transverse plane. Bottom left: distribution of the scattering position along the z axis. The coordinate is in centimeters from the left edge of the first station; one can see the contribution from the target material as well as from each double-sided sensor in the three tracking modules, for the baseline geometry. Bottom right: strip multiplicity of the clusters.} 
\label{f:baseline1}
\end{center}
\end{figure}

\begin{figure}[h!]
\begin{center}
\includegraphics[scale=0.8]{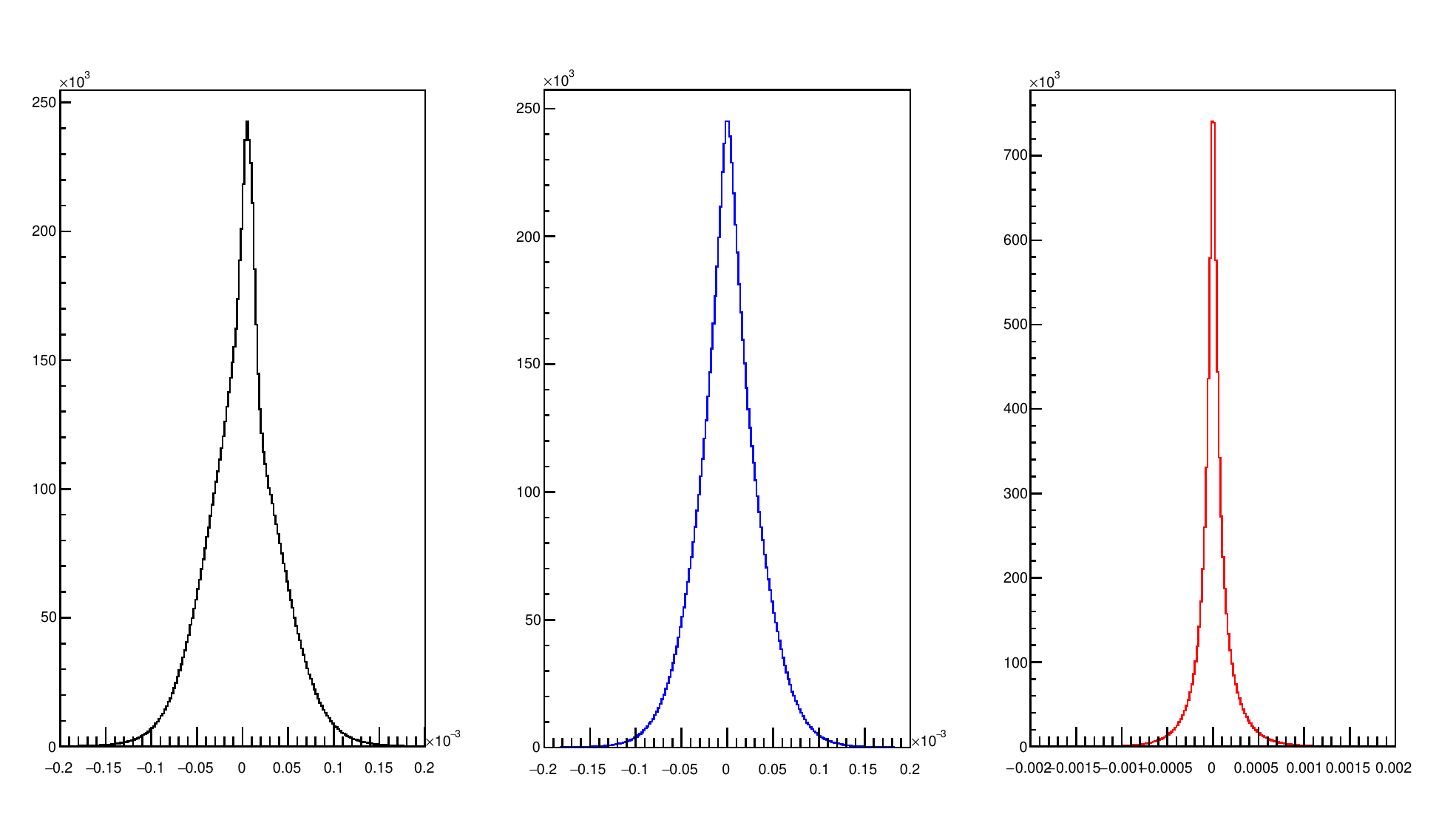}
\includegraphics[scale=0.6]{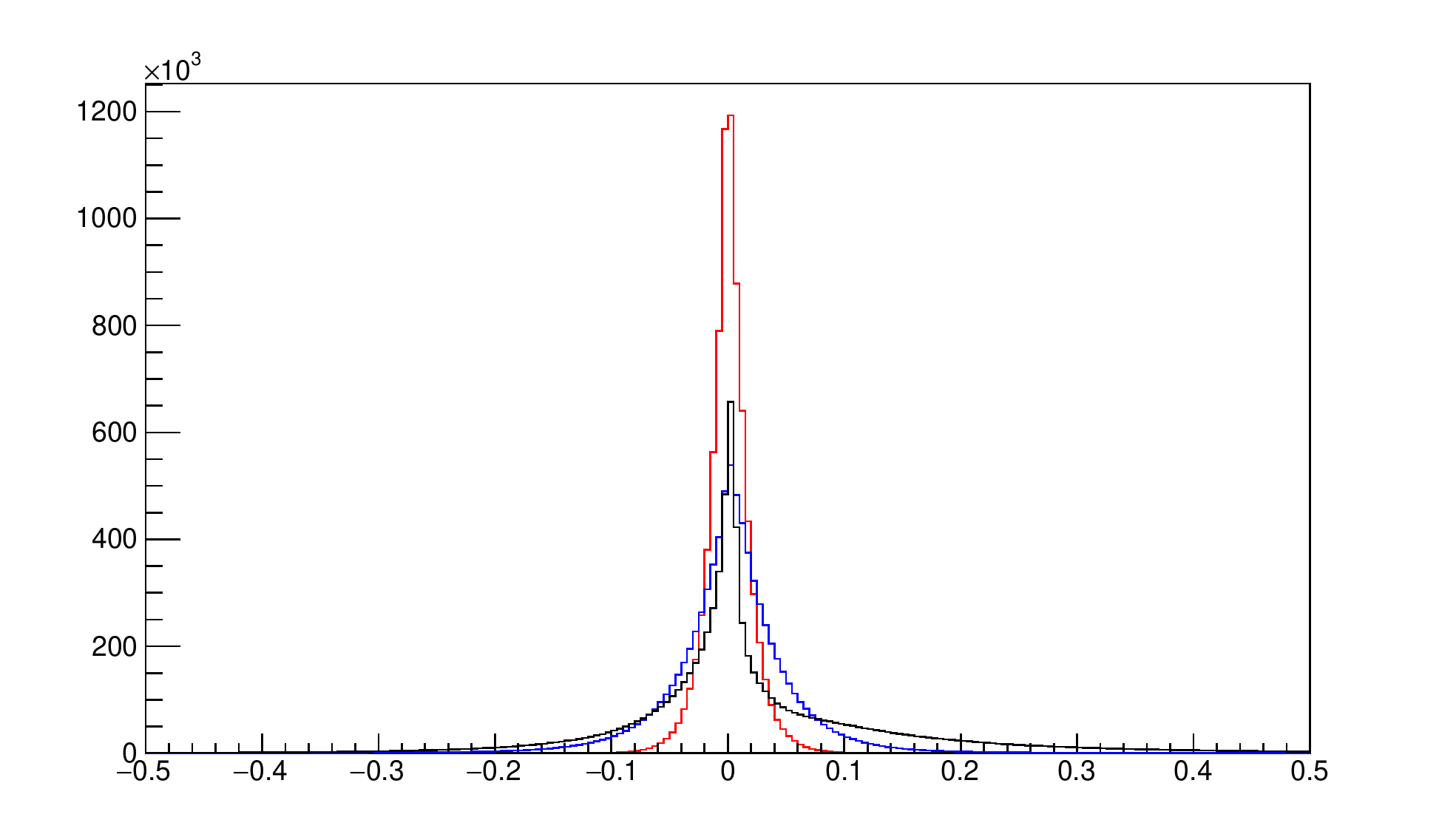}
\caption {\em Top: Difference between estimated and true $\theta$ angles in the laboratory frame for incoming muons (left), outgoing muons (center), and outgoing electrons (right) as returned by the likelihood fit, for the baseline geometry. Bottom: Difference between estimated and true $\phi$ angles in the laboratory frame for incoming muons (black), outgoing muons (blue), and outgoing electrons (red) as returned by the likelihood fit, for the baseline geometry.} 
\label{f:baseline2}
\end{center}
\end{figure}

\begin{figure}[h!]
\begin{center}
\includegraphics[scale=0.7]{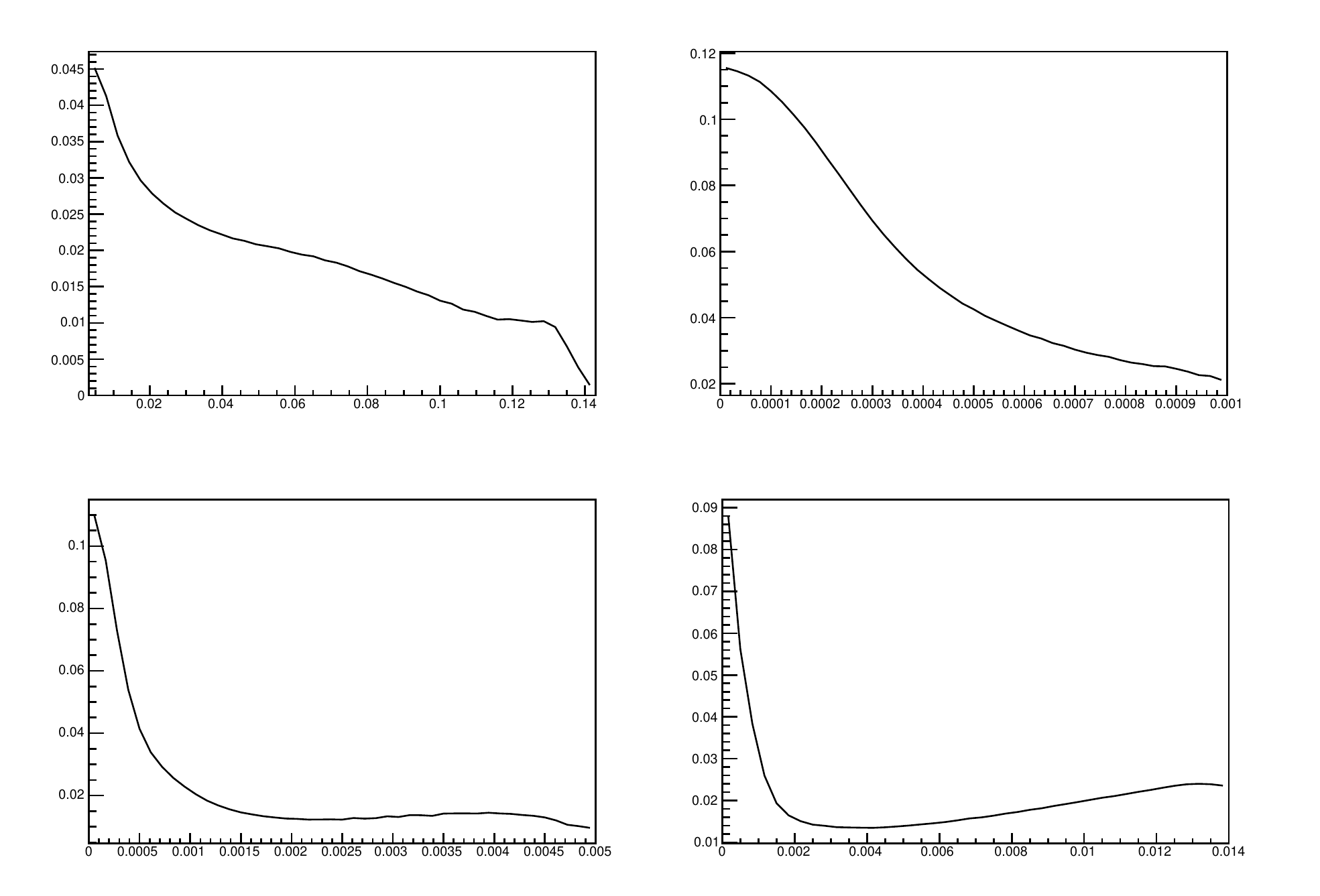} 
\caption {\em Relative resolution on kinematic quantities for the baseline geometry, obtained from $10^7$ elastic scattering events. Top left: Relative resolution $\sigma(q^2)/q^2$ as a function of $q^2$; top right: relative resolution $\sigma(\theta_{in})/\theta_{in}$ on incoming muon divergence as a function of incoming muon divergence; bottom left: relative resolution $\sigma(\theta_{\mu})/\theta_{\mu}$ in outgoing muon divergence as a function of outgoing muon divergence; bottom right: relative resolution $\sigma(\theta_e)/\theta_e$ in electron divergence as a function of electron divergence.} 
\label{f:baseline3}
\end{center}
\end{figure}

\noindent
We show {\em infra} (Figs.~\ref{f:baseline1}, \ref{f:baseline2}, \ref{f:baseline3}) some of the results of the simulation of $10^7$ muon interactions in the second station, when the detector is arranged as detailed above. Not all simulated interactions result in a well-reconstructed scattering event, as the divergence of the beam and the limited extension of tracking modules (in particular, the effect of the transverse rotation of the stereo module with respect to the others) reduce the acceptance. A further minor reduction comes from enforcing that each of the three particles (incoming muon, outgoing muon and electron) produces valid two-hit stubs in each of three consecutive modules. The total reduction of statistics amounts to about 21\% and is practically independent on the alternative geometry choices we discuss in the remainder of this article. A further fraction of less than 0.1\% of the events is removed because the likelihood maximization fails to converge. The failures are concentrated  in the low-$q^2$ region of phase space; we do not consider them further in this study.

\noindent
The simulation of ten million scattering events with the baseline geometry of the MUonE apparatus~\footnote {For the larger $\Delta z_{BeSi}=15 cm$ spacing of the target and the first station downstream mentioned {\em supra} (with positions $z_{mod0}=18$cm, $z_{mod1}=56.5$cm, $z_{mod2}=95$cm) the all-range ($q^2>0.1 GeV^2$) $q^2$ resolution results instead equal to $1.9891(5)$ ($0.6514(12)$).} results in the estimates reported in Table~\ref{t:baseline} for the figures of merit discussed in Sec.~\ref{s:optimize}.

\begin{table}[h!]
\begin{center}
\begin{tabular}{l|rrrrr} 
Configuration & $\overline{\sigma(\theta_{\in})_{rel}}$ & $\overline{\sigma(\theta_{\mu})_{rel}}$   & $\overline{\sigma(\theta_{e})_{rel}}$ & $\overline{\sigma(q^2_{rel})}$   & $\overline{\sigma(q^2_{rel})_{0.1}}$ \\ 
        & \%   & \% & \% & \% & \% \\ 
\hline
Baseline & $5.745(16)$ & $2.1143(5)$ & $1.2764(3)$ & $2.1222(5)$ & $0.6637(13)$ \\
\hline
\end{tabular}
\caption{\em Average relative resolution on the particles divergences from the z axis, on $q^2$ in the full investigated range ($[0.0057:0.143]$) and in the restricted range $[0.1:0.143]$, for the baseline geometry. See the text for details.}
\label{t:baseline}
\end{center}
\end{table}

\subsection{Distributed target options \label{s:distributed}}

We compare the baseline geometry with a set of alternative arrangements where the $1.5$cm-thick beryllium target of each section is divided into a number of thin layers. To gauge the effect of different possible choices we perform several studies, all based on a uniform placement of the tracking modules along each station ( $z_{mod \,0} = 31.9$cm; $z_{mod \,1}=65.2$cm, $z_{mod \,2}=98.5$cm); other parameters not mentioned below are for now kept at their default values listed in Sec.~\ref{s:defaultpars}.\par

\begin{figure}[h!]
\begin{center}
\includegraphics[scale=0.75]{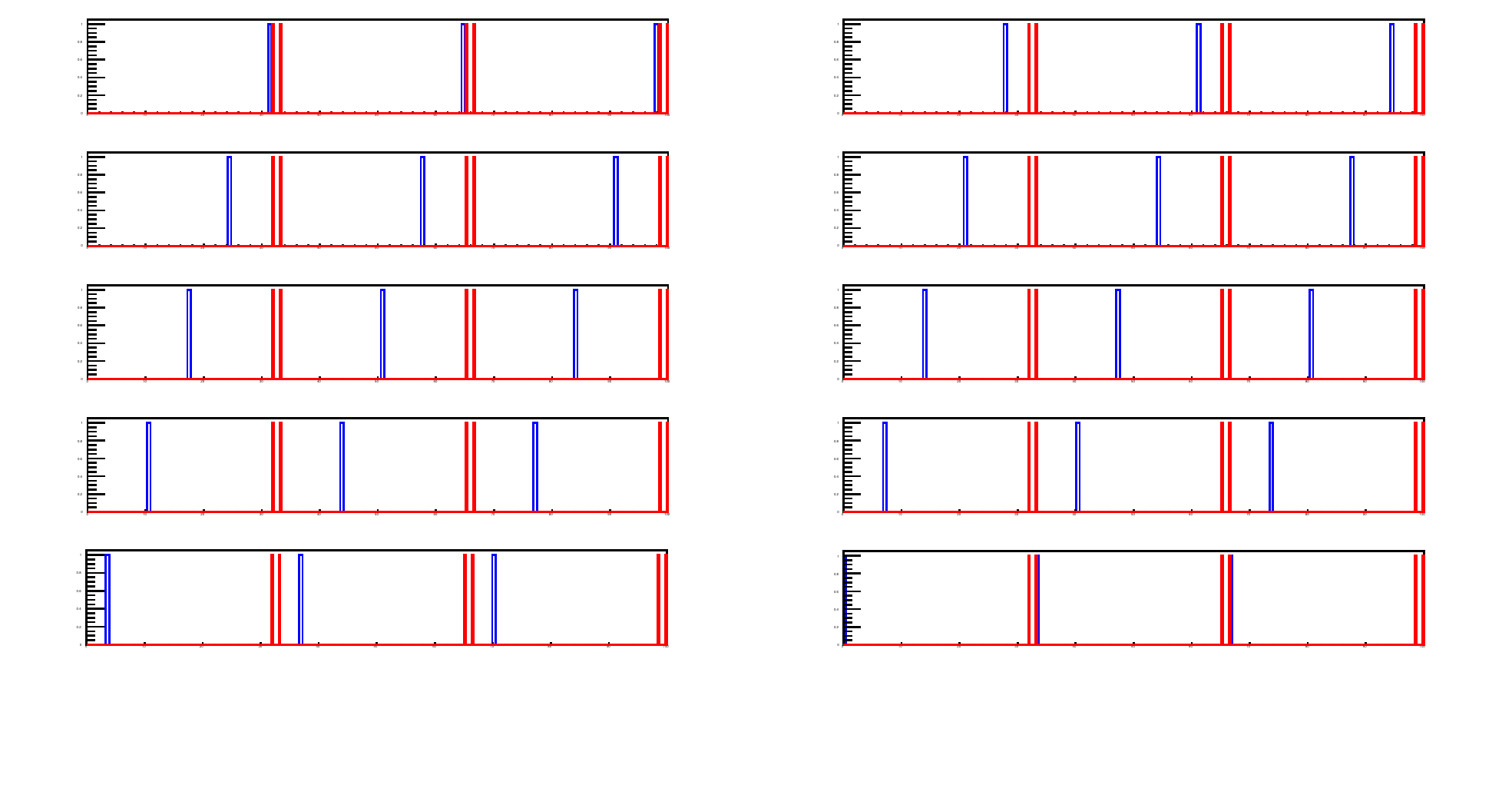}
\caption {\em Ten configurations of three target layers in a section, considered for the study of their positioning. 
In blue are shown the target layers, and in red are indicated the tracking modules.}
\label{f:geom_6_3tl}
\end{center}
\end{figure}

\begin{enumerate}

\item We consider a geometry with one target layer of $W_{Be}=0.5$cm width placed between each pair of tracking modules. We then vary the distance of each target layer to its downstream tracking module $\Delta z_{BeSi}$ from $0$ to its maximum value of $31.0$cm in ten equal intervals, to see what effect this simple rearrangement has on resolutions (see Fig.~\ref{f:geom_6_3tl}). In this setup we observe that the placement of the target with respect to the tracking modules upstream and downstream must be chosen with care, as one may get modifications of five percent or higher in the relative resolutions of reconstructed tracks by simply changing that construction choice.

\item We consider twelve target layers per station, forming three sections with four layers positioned between each pair of tracking modules. Each layer has initially a width $W_{Be}=0.125$cm; the layers are positioned such that the spacing between the tracking module to the left and right of the closest target layer is half of the spacing between two adjacent layers. We then iteratively double the number of layers, retaining the above requirement; as much as the spacings, the width of each target layer is of course halved at each iteration, keeping the material budget constant. 
This exercise shows the effect of increasing the number of layers alone, in a symmetric configuration. The advantage of a distributed target is significant, as for the high-$q^2$ events most important for the determination of $\Delta \alpha_{had}$, the resolution increases by about five percent in going from the baseline geometry to one with 384 layers per station.

\item We repeat the study of the effect of varying $\Delta z_{BeSi}$, for the case of $N=300$ target layers per station, divided in three 1.49cm-long stacks of 100 layers, each of width $W_{Be}=0.005$cm and spaced by 0.01cm from its neighbors. The issue of where to place the stacks of targets is a complex problem, as the resolution on particle trajectories is influenced by several factors; one of them is the interplay between the length of the extrapolation arms from silicon hits to scattering vertex and the constraint coming from the positioning of the target layers; another is the discreteness of the silicon readout, which may cause a periodicity in the precision of the silicon hits positions (as tracks incident on the sensors in na favourable position will have their position measured in two adjacent strips, with a much smaller resulting uncertainty) and a dependence of that parameter on the track incidence angle. Furthermore, these effects have an opposite valence for incoming and outgoing particles.
The general conclusions seen for the case of only one layer per section appear to be robust with respect to the increase of number of target layers: relatively packed stacks as those considered here appear to be most proficuously positioned closer to their upstream tracking module than otherwise.

\item We again consider $N=300$ target layers per station, each of width $W_{Be}=0.005$cm, again divided in three 100-layer stacks placed between the pairs of tracking modules (the first one has at its left the rightmost module of the previous station). We now vary the $\Delta z_{Be}$ spacing between adjacent layers from $0.0$cm to $0.3$cm in regular intervals. Once $\Delta z_{Be}$ is defined, the positioning of the stacks of target layers is determined by the smallest gap between the rightmost target layer and the silicon tracking module to its right, $\Delta z_{BeSi}$; we fix this parameter to $0.5$cm here. We observe that a wider spacing of target layers produces a significant improvement (about five percent) in the relative resolution of particle divergences. This is understood to be due to the higher effectiveness of the z constraint, as more widely spaced thin targets help the fit converge to the correct solution.

\end{enumerate}

\begin{figure}[h!]
\begin{center}
\includegraphics[scale=0.7]{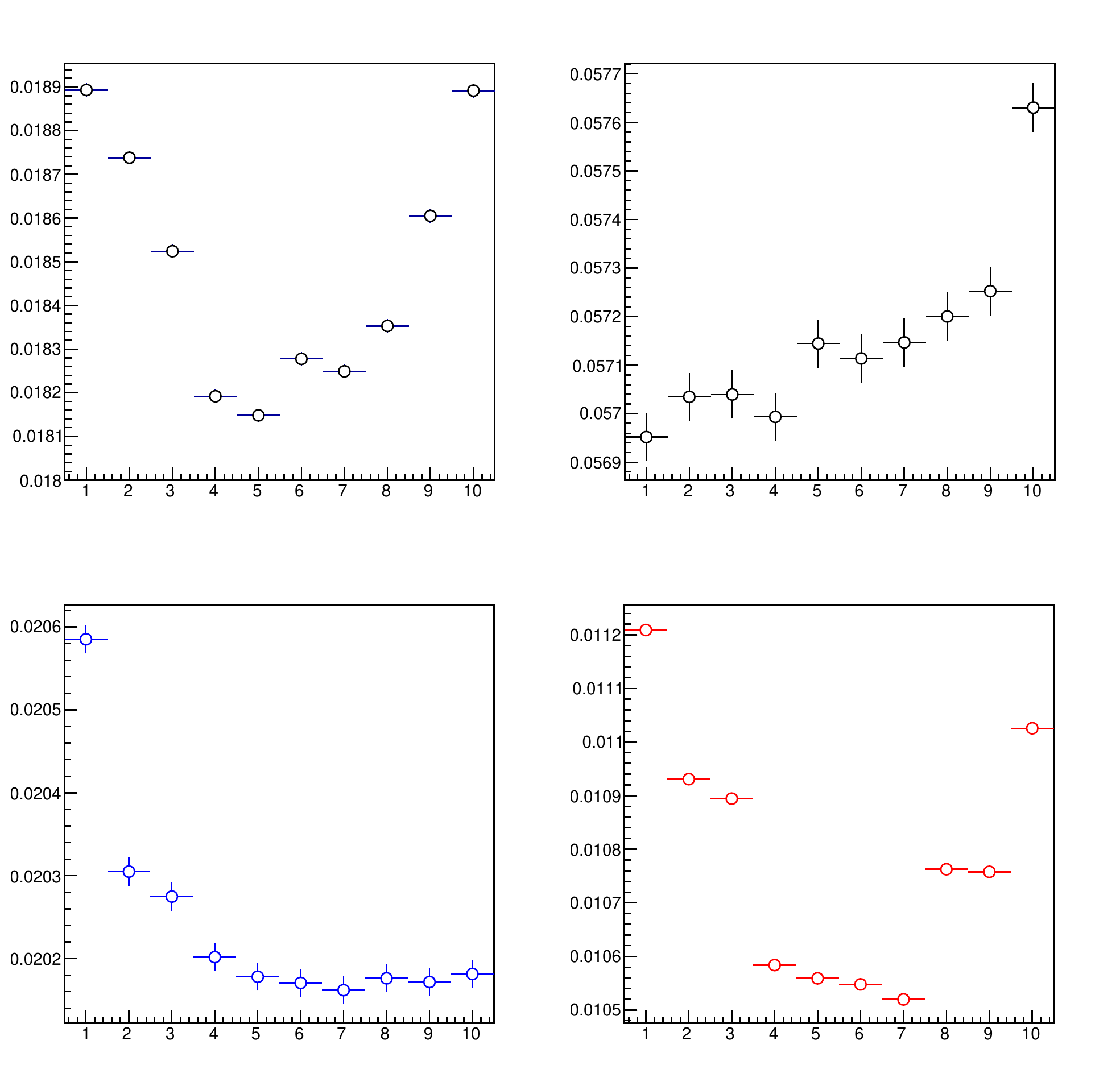} 
\caption {\em Resolution on relevant quantities for ten configurations of three target layers discussed in point (1) (supra). Top left: RMS of the $q^2$ measurement as a function of the layer positioning in each of the three section of each station. Top right: relative RMS of the incoming muon divergence, $\theta_{in}$. Bottom left: relative RMS of the outgoing muon divergence, $\theta_{\mu}$. Bottom right: relative RMS of the electron divergence, $\theta_{e}$.}
\label{f:relres_6_3tl}
\end{center}
\end{figure}

\begin{figure}[h!]
\begin{center}
\includegraphics[scale=0.8]{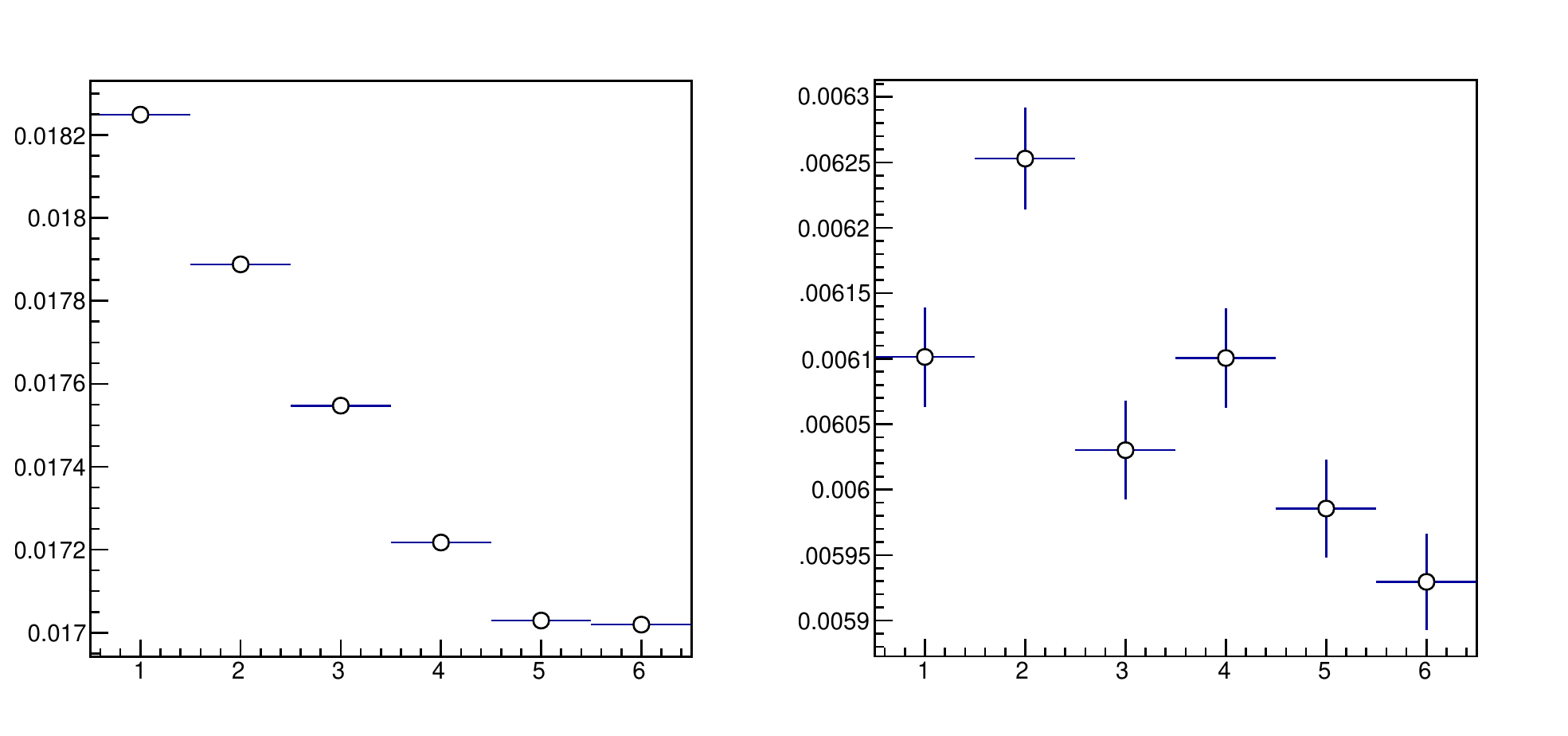} 
\includegraphics[scale=0.8]{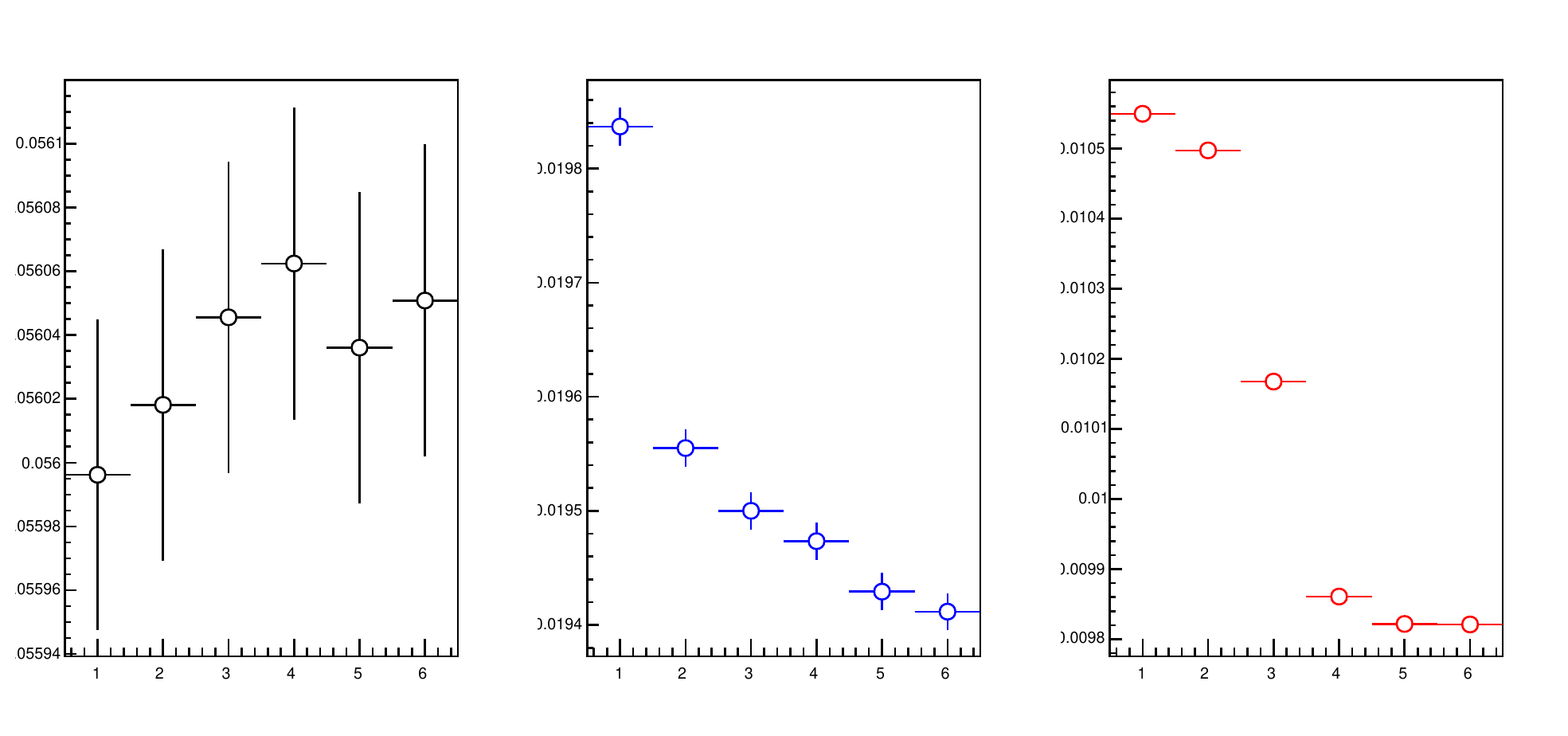} 
\caption{\em Top: average RMS of the relative resolution on $q^2$ in the full range (left) and in the restricted $[0.1:0.143]$ range, as a function of the distance between target layers, for a configuration with 100 $50\mu$m layers in each section of a station. The spacing varies between $0.0$cm and $0.3$cm in regular intervals. Bottom: average RMS of the relative resolution on the incoming muon, outgoing muon, and electron divergences from the z axis. See the text for more detail.}
\label{f:distr5} 
\end{center}
\end{figure}

\noindent
The results of these comparisons, performed with the simulation of $10^6$ elastic scatterings in the second station, repeated per each configuration with the random number generation recipe discussed {\em supra}, are shown in Table~\ref{t:distributed} below, and in Figs.~\ref{f:relres_6_3tl},~\ref{f:distr1},~\ref{f:distr3}, and~\ref{f:distr5}.

\begin{table}[h!]
\begin{center}
\begin{tabular}{r|rrrrr} 
Configuration & $\overline{\sigma(\theta_{\in})_{rel}}$ & $\overline{\sigma(\theta_{\mu})_{rel}}$   & $\overline{\sigma(\theta_{e})_{rel}}$ & $\overline{\sigma(q^2_{rel})}$   & $\overline{\sigma(q^2_{rel})_{0.1}}$ \\ 
\hline
   $\Delta z_{BeSi}$ (cm)     & \%   & \% & \% & \% & \% \\ 
  0.0   & $5.6952(49)$ & $2.0585(17)$ & $1.1209(9)$ & $1.8893(15)$ & $0.6780(42)$ \\ 
  3.48  & $5.7034(49)$ & $2.0304(16)$ & $1.0930(9)$ & $1.8738(15)$ & $0.6542(40)$  \\ 
  6.96  & $5.7039(49)$ & $2.0274(16)$ & $1.0894(9)$ & $1.8523(15)$ & $0.6487(40)$  \\ 
  10.43 & $5.6993(49)$ & $2.0201(16)$ & $1.0583(8)$ & $1.8191(15)$ & $0.6203(38)$  \\ 
  13.91 & $5.7144(49)$ & $2.0177(16)$ & $1.0558(8)$ & $1.8148(15)$ & $0.6278(39)$  \\ 
  17.39 & $5.7113(49)$ & $2.0170(16)$ & $1.0547(8)$ & $1.8277(15)$ & $0.6237(38)$  \\ 
  20.87 & $5.7146(50)$ & $2.0161(16)$ & $1.0519(8)$ & $1.8248(15)$ & $0.6263(38)$  \\ 
  24.34 & $5.7200(50)$ & $2.0176(16)$ & $1.0762(8)$ & $1.8352(15)$ & $0.6065(37)$  \\ 
  27.82 & $5.7252(50)$ & $2.0171(16)$ & $1.0757(8)$ & $1.8604(15)$ & $0.6085(37)$  \\ 
  31.30 & $5.7630(50)$ & $2.0181(16)$ & $1.1025(9)$ & $1.8891(15)$ & $0.6171(38)$  \\ 
\hline
 $N_{layers}$   & \%   & \% & \% & \% & \%  \\ 
  12  & $5.6852(49)$ & $1.9822(16)$ & $1.0579(8)$ & $1.7922(15)$ & $0.6254(38)$  \\ 
  24  & $5.6836(49)$ & $1.9769(16)$ & $1.0266(8)$ & $1.7732(14)$ & $0.5995(37)$  \\ 
  48  & $5.6714(49)$ & $1.9748(16)$ & $1.0218(8)$ & $1.7515(14)$ & $0.6244(38)$  \\ 
  96  & $5.6219(49)$ & $1.9462(16)$ & $0.9891(8)$ & $1.7431(14)$ & $0.6072(37)$ \\ 
  192 & $5.6492(49)$ & $1.9463(16)$ & $0.9841(8)$ & $1.7053(14)$ & $0.5889(36)$ \\ 
  384 & $5.6037(48)$ & $1.9423(16)$ & $0.9790(8)$ & $1.6823(14)$ & $0.5923(36)$ \\ 
\hline
$\Delta z_{BeSi}$ (cm)  & \%   & \% & \% & \% & \%  \\ 
  0.0   & $5.5980(48)$ & $1.9855(16)$ & $1.0564(8)$ & $1.8272(15)$ & $0.6452(40)$ \\ 
  7.58  & $5.6109(48)$ & $1.9527(16)$ & $1.0176(8)$ & $1.7563(14)$ & $0.6167(38)$ \\ 
  15.16 & $5.6465(49)$ & $1.9443(16)$ & $0.9845(8)$ & $1.7055(14)$ & $0.5875(36)$ \\ 
  22.73 & $5.6444(49)$ & $1.9417(16)$ & $0.9815(8)$ & $1.7020(14)$ & $0.5840(36)$ \\   
  30.31 & $5.6886(49)$ & $1.9431(16)$ & $1.0062(8)$ & $1.7287(14)$ & $0.5817(36)$ \\ 
\hline
$\Delta z_{Be}$ (cm)   & \%   & \% & \% & \% & \%  \\ 
  0.00 & $5.5996(48)$ & $1.9836(16)$ & $1.0549(8)$ & $1.8249(15)$ & $0.6101(38)$  \\ 
  0.06 & $5.6018(48)$ & $1.9555(16)$ & $1.0497(8)$ & $1.7888(14)$ & $0.6252(38)$  \\ 
  0.12 & $5.6045(48)$ & $1.9500(16)$ & $1.0167(8)$ & $1.7547(14)$ & $0.6030(37)$ \\ 
  0.18 & $5.6062(48)$ & $1.9473(16)$ & $0.9860(8)$ & $1.7217(14)$ & $0.6100(37)$ \\ 
  0.24 & $5.6036(48)$ & $1.9429(16)$ & $0.9821(8)$ & $1.7029(14)$ & $0.5985(37)$  \\ 
  0.30 & $5.6050(48)$ & $1.9411(16)$ & $0.9821(8)$ & $1.7019(14)$ & $0.5929(36)$ \\ 
\hline
\end{tabular}
\caption{\em Average relative resolutions for the studied configurations of distributed targets. The first block refers to studies labeled (1) in the text, and so on. In all cases, $10^6$ elastic scattering have been simulated for each geometry. See the text for more detail.}
\label{t:distributed}
\end{center}
\end{table}

\begin{figure}[h!]
\begin{center}
\includegraphics[scale=0.8]{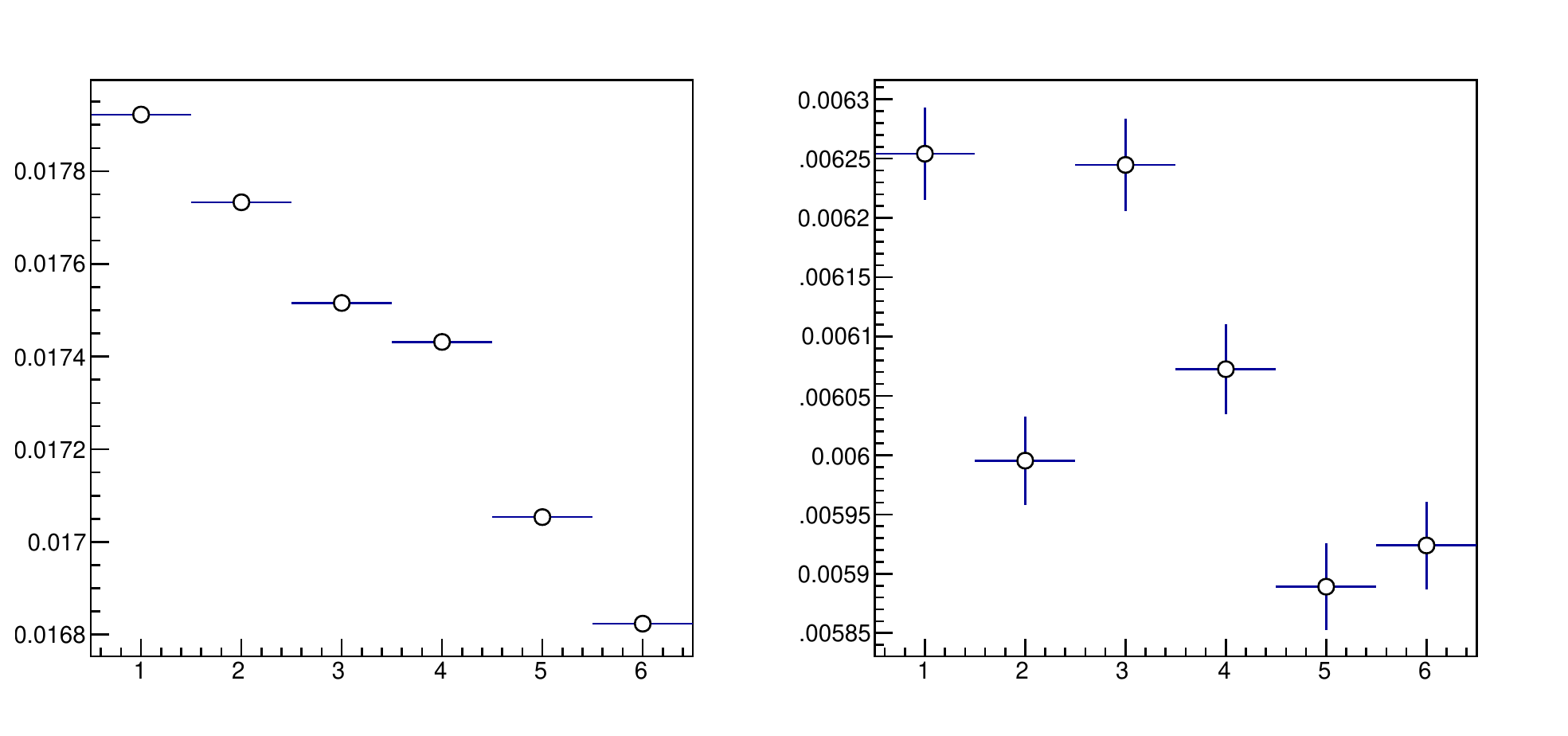} 
\caption{\em Average relative RMS of the $q^2$ determination (left: full range; right: average restricted to true $q^2$ values in the $[0.1:0.143]$ range) of scatterings as a function of the number of target layers considered for the configuration (2) above; from left to right, in each graph the six bins describe the results of having four, eight, sixteen, 32, 64, and 128 layers in each of the three sections of a station. }
\label{f:distr1} 
\end{center}
\end{figure}

\begin{figure}[h!]
\begin{center}
\includegraphics[scale=0.8]{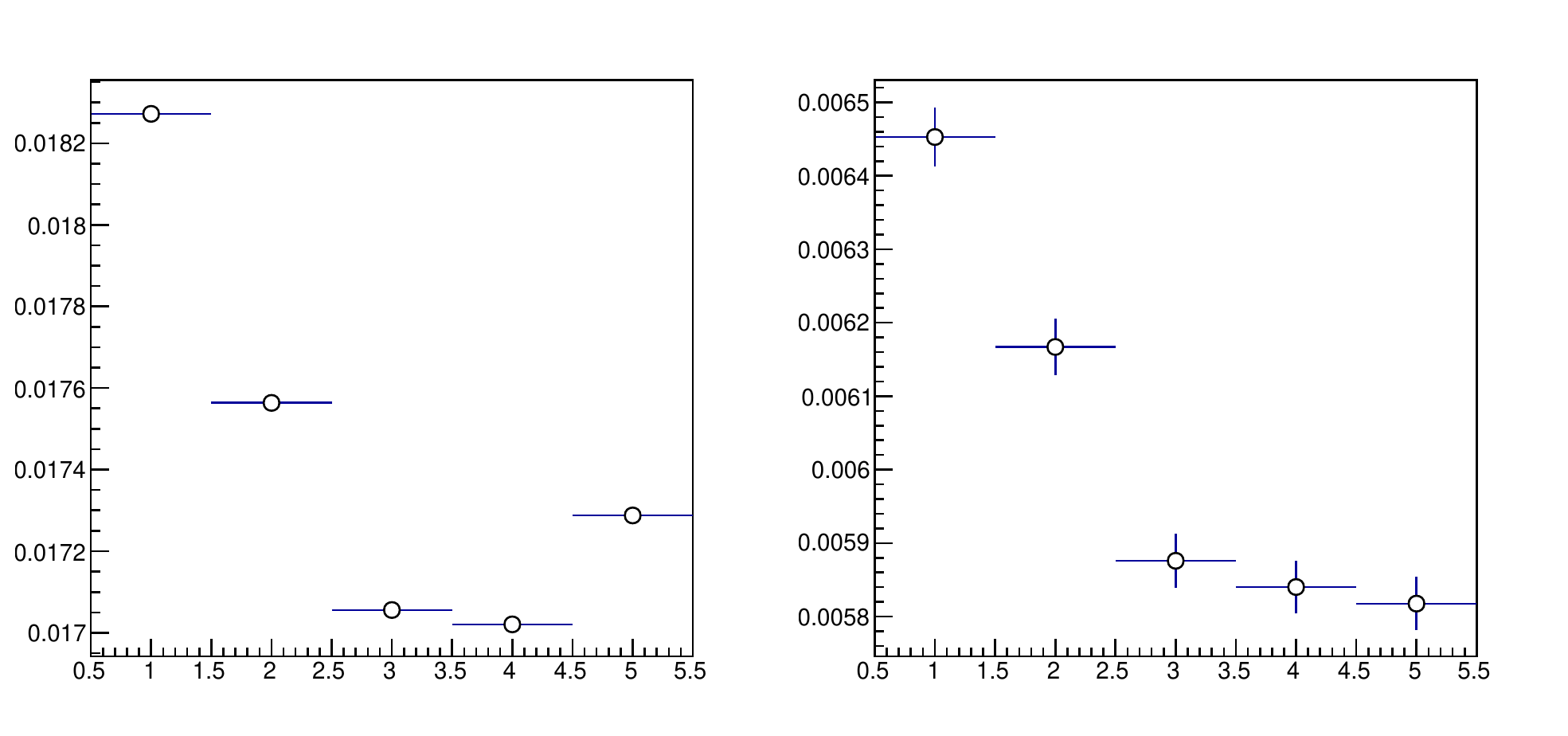} 
\caption{\em Average RMS of the relative resolution of $q^2$ in the full range (left) and in the restricted $[0.1:0.143]$ range, as a function of the distance between the stack of 100 target layers and the tracking module to its right. From left to right, the distance is sampled in five uniform intervals from 0 to 30.31cm. }
\label{f:distr3} 
\end{center}
\end{figure}

\begin{figure}[h!]
\begin{center}
\includegraphics[scale=0.9]{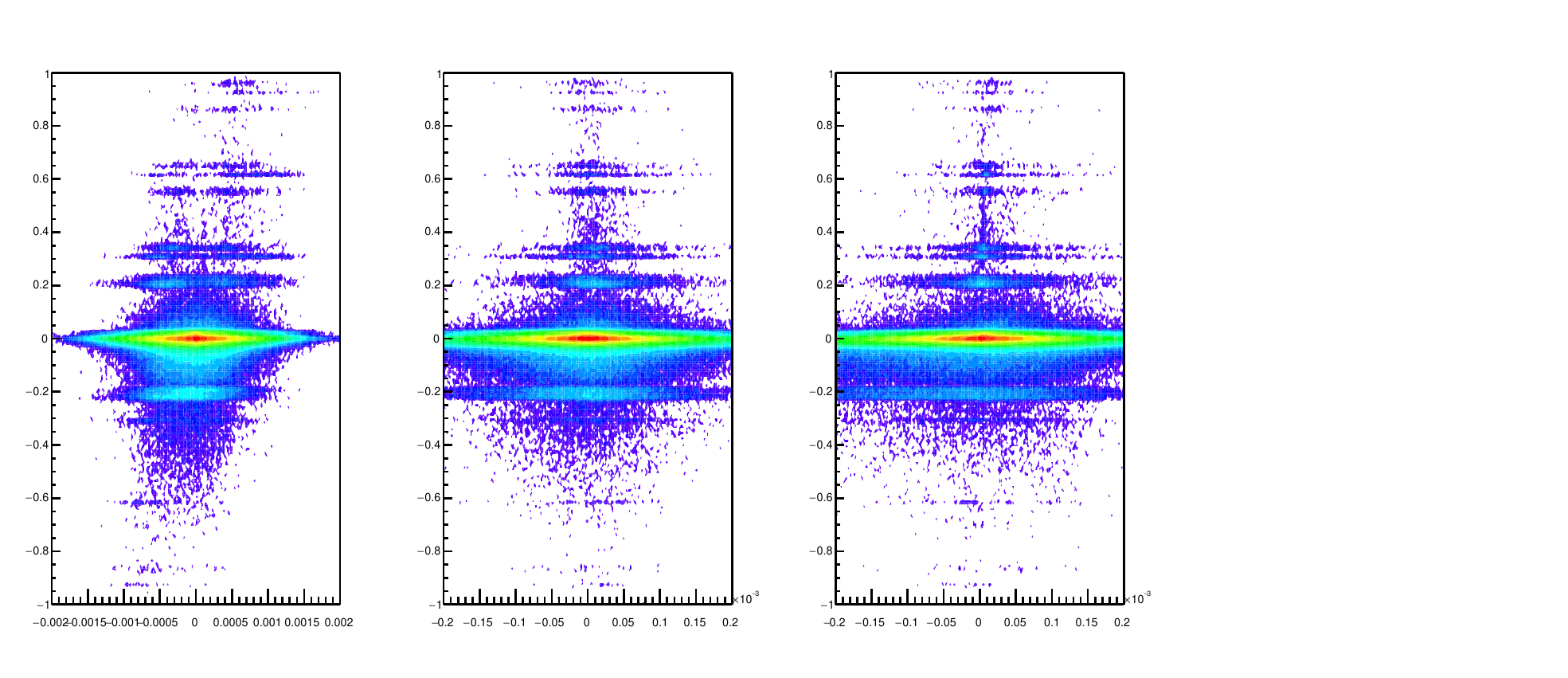}
\includegraphics[scale=0.9]{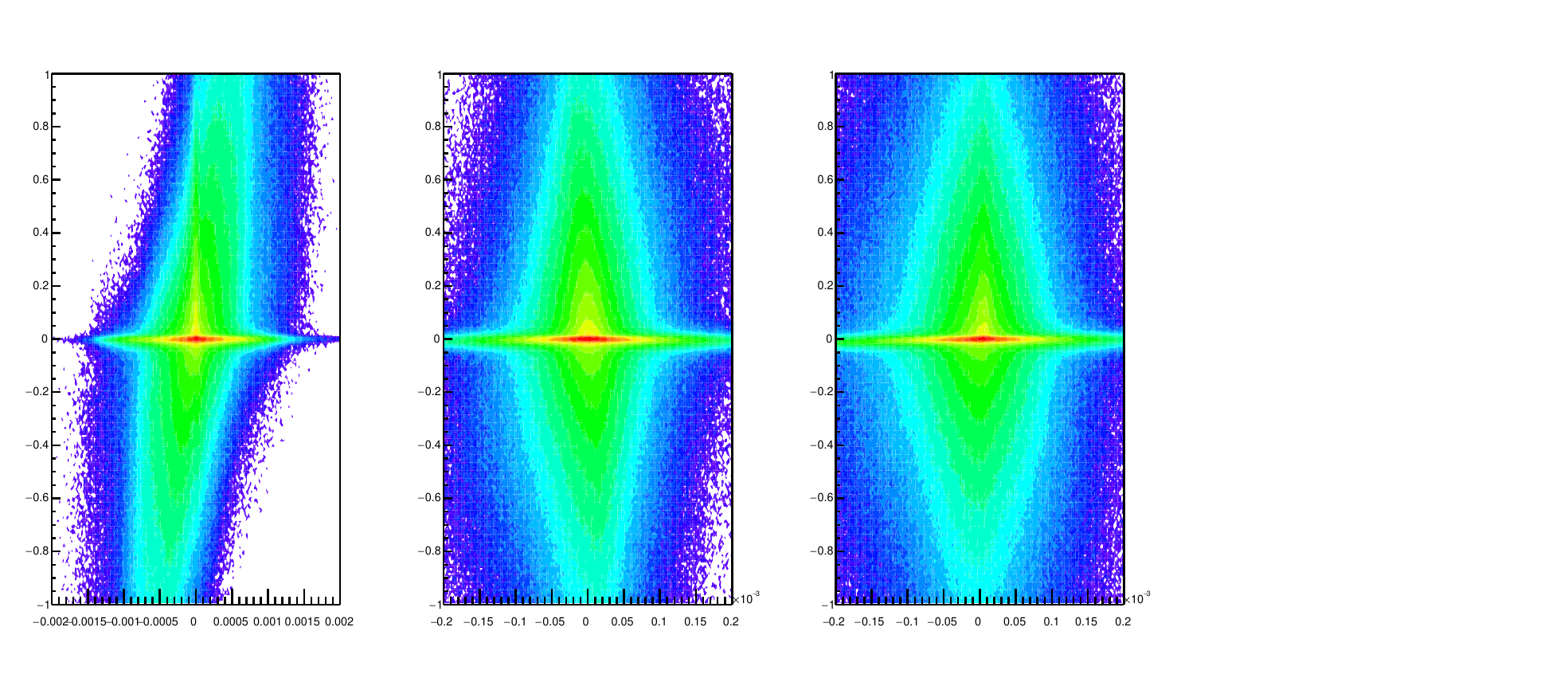}
\caption {\em Residuals in the measured z of the scattering vertex as a function of residuals in the divergence from the z axis of the electron track (left), muon track (center) and initial muon track (right), for a detector with 300 target layers (see text for details). In the top panel a z-vertex constraint has been used in the likelihood fit to the event kinematics; in the bottom panel the z-vertex constraint has been removed.}
\label{f:zvthetas}
\end{center}
\end{figure}

\section {Other geometry options}

In this section we summarize some of the studies performed on other parameters describing aspects of the detector geometry not previously discussed. Their modification produces minor effects on the measurement of $\Delta \alpha_{had}$, with the exception of a non-null vertical staggering of the strips on the two sides of the double-sided sensors.

The baseline layout of target elements and tracking modules in each station which we use for the studies discussed in this section results from the insight acquired in Sec.~5, and has 300 $50\mu$m layers divided into three $30.2$cm-long stacks of 100 layers, with a $0.3$cm gap between layers; the z coordinates of left edges of the stacks, as read from the left edge of the station, are $z_0=0.2$cm, $z_1=33.6$cm, $z_2=66.9$cm. The tracking modules, positioned at coordinates $z_{mod0}=31.8$cm, $z_{mod1}=65.2$cm, $z_{mod3}=98.5$cm, provide for an approximately symmetrical spacing of the three sections of each station, as in the studies of the previous section. The central module in each station has by default strips rotated by $\pi/4$ with respect to those of the other two modules, except when that parameter is varied to study the corresponding effect. The staggering of strips in the right-side element of double-sided sensors with respect to the corresponding strips in the left-side one is of 0. $\mu$m by default, except when that parameter is varied.

\subsection {Relative spacing of sensors in double-sided silicon modules \label{s:relspacingsisi}}

The nominal geometry of the silicon modules for the proposed MUonE experiment foresees that the two $320 \mu$m-thick silicon layers of a double-sided sensor be spaced by 0.18cm from one another, and that two such sensors be mounted together, forming a wafer of four detection elements, the two left ones reading one coordinate (here taken as the $x$ coordinate) and the two right ones reading the other coordinate orthogonal to the beam direction ($y$). The assembly is 1.5cm thick, hence in the original design there is a 1.012cm gap between the two double-sided sensors.

We study the effect of a wider spacing of the two sides in double-sided sensors on the angular and $q^2$ resolution of the scattering fit, by varying that parameter from $0.18$cm to $0.58$cm in $0.1$cm steps. All other parameters are kept to their baseline value as discussed earlier in this Section. The results of this exercise are summarized graphically in Fig.~\ref{f:2sspacing_relres} and listed in Table~\ref{t:zsisispacing} below. A wider space between each pair of coordinate measurements seems to help the determination of the particle directions, with gains of five to ten percent in the $q^2$ resolution. 

\begin{table}[h!]
\begin{center}
\begin{tabular}{l|rrrrr} 
Configuration  & $\overline{\sigma(\theta_{\in})_{rel}}$ & $\overline{\sigma(\theta_{\mu})_{rel}}$   & $\overline{\sigma(\theta_{e})_{rel}}$ & $\overline{\sigma(q^2_{rel})}$   & $\overline{\sigma(q^2_{rel})_{0.1}}$ \\ 
\hline
$\Delta z_{Si}$ (cm)       & \%   & \% & \% & \% & \%  \\ 
  0.18 & $5.6058(48)$ & $1.9418(16)$ & $0.9819(8)$ & $1.7011(14)$ & $0.5943(36)$ \\ 
  0.28 & $5.5948(48)$ & $1.9347(16)$ & $0.9741(8)$ & $1.6579(13)$ & $0.5862(36)$ \\ 
  0.38 & $5.5775(48)$ & $1.9266(16)$ & $0.9666(8)$ & $1.6337(13)$ & $0.5536(34)$ \\ 
  0.48 & $5.5641(48)$ & $1.9200(16)$ & $0.9351(7)$ & $1.6109(13)$ & $0.5558(34)$ \\ 
  0.58 & $5.5235(48)$ & $1.9150(15)$ & $0.9342(7)$ & $1.6072(13)$ & $0.5401(33)$ \\ 
\hline
\end{tabular}
\caption{\em Effect of the spacing $\Delta z_{Si}$ between the sensors in double-sided modules on relative resolutions. In each case $10^6$ elastic scattering events were simulated. See the text for detail.}
\label{t:zsisispacing}
\end{center}
\end{table}

\begin{figure}[h!]
\begin{center}
\includegraphics[scale=0.7]{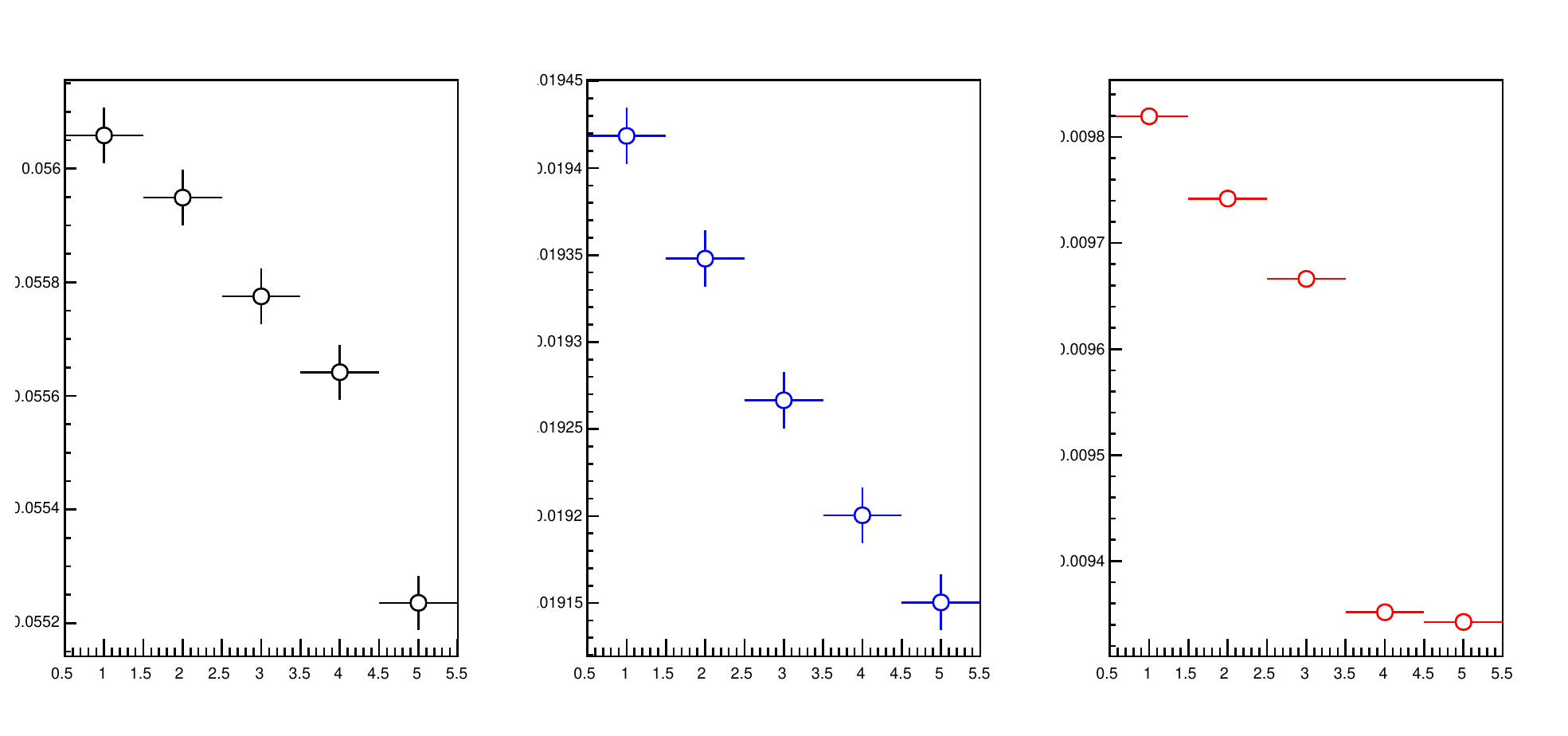} 
\includegraphics[scale=0.7]{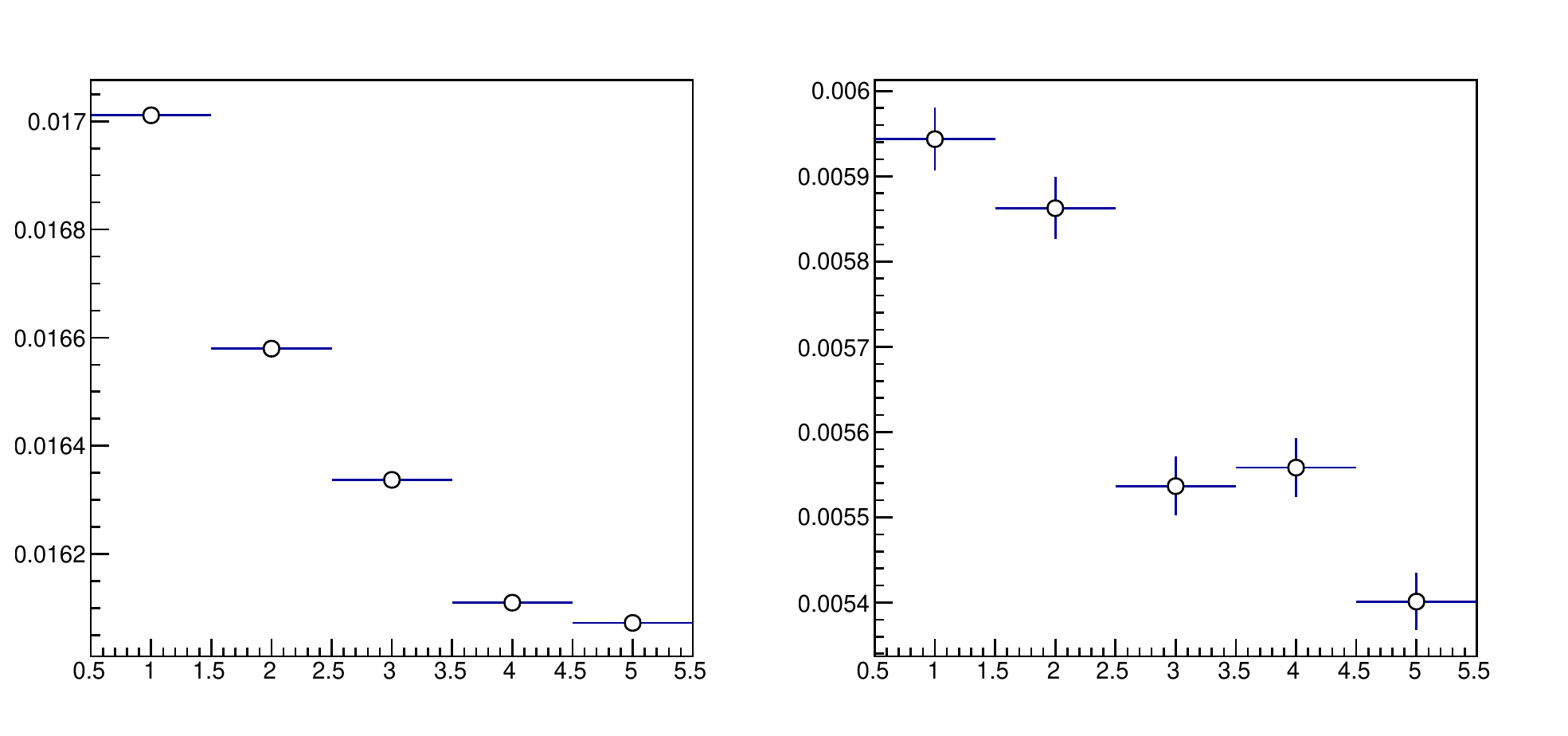} 
\caption {\em Top row: average relative resolution on the beam divergences of incoming (left), outgoing muon (center), and electron (right) for the five considered spacing configurations between the two sides of double-sided sensors in tracking modules. The five measurement points refer to a spacing of 0.18 to 0.58 cm. Bottom row: average relative resolution in event $q^2$ in the full studied range (left) and in the high-$q^2$ range $[0.1:0.143]$ (right). } 
\label{f:2sspacing_relres}
\end{center}
\end{figure}

\subsection {Offset in pitch position of strips on the two sides of double-sided silicon modules \label{staggering}}

In the application of precision tracking of ionizing particles for the CMS experiment there is no need for an optimization of the relative positioning of the strips in the two sides of double-sided sensors. There are two reasons for this: first, particles incide on the sensors surface with widely varied angles; second, the magnetic field in the tracker produces a transverse effect on the drift of the charge in the semiconductor. These two effects guarantee that a large fraction of the particles deposit ionization signal over more than a single strip, with considerable gains in the resulting position resolution along the coordinate orthogonal to the strips. The situation is quite different for MUonE, where there is no magnetic field providing a Lorentz force on the charges, and where particles incide on the sensors with typical angles of thousandths of a radian. We believe this calls for a modification of the relative positioning of the sensors for the double-sided modules which would be produced for MUonE, and we show evidence for the benefits that a staggering would provide.

\begin{figure}[h!]
\begin{center}
\includegraphics[scale=0.8]{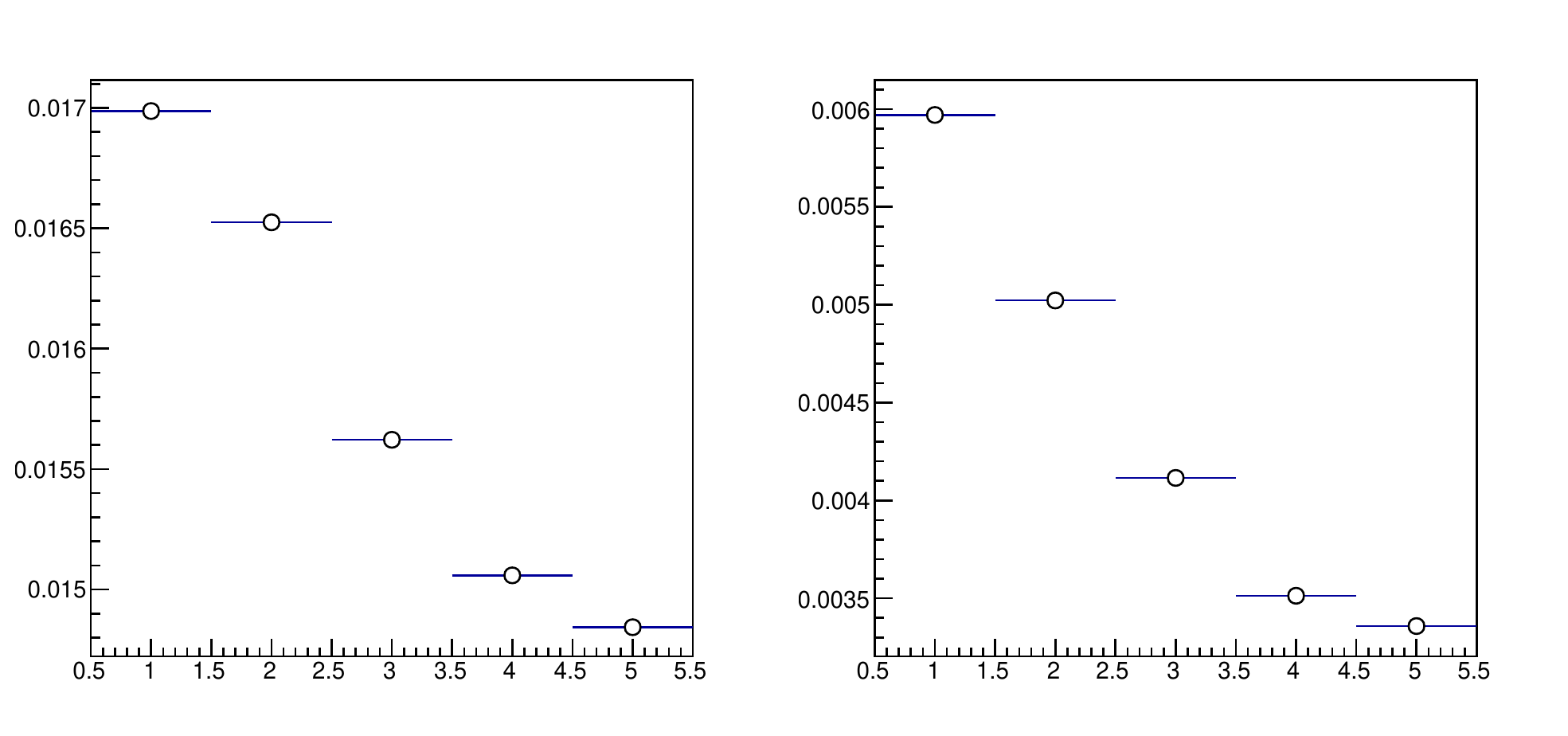} 
\includegraphics[scale=0.8]{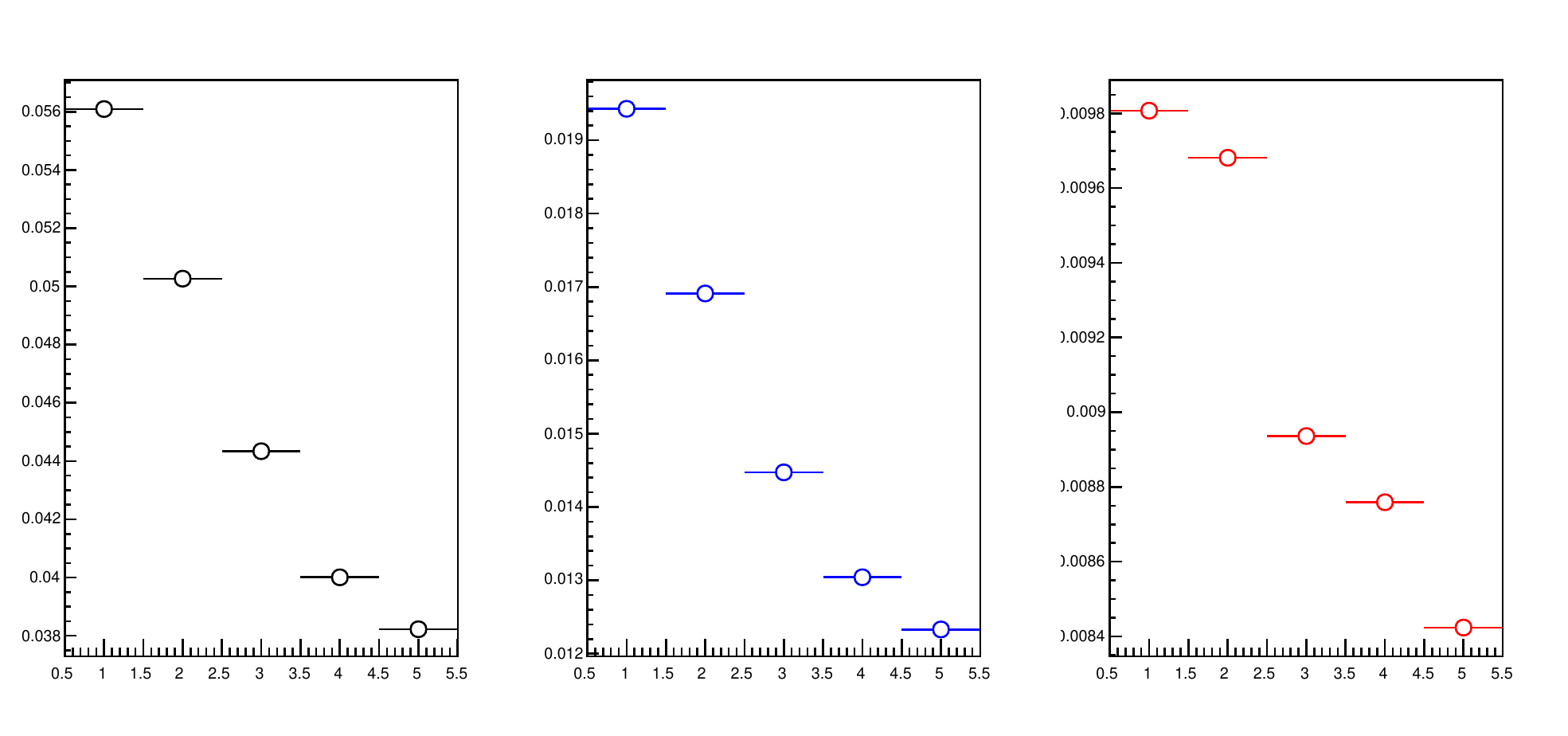} 
\caption {{\em Top: RMS of the relative precision in the measurement of event $q^2$ in the full investigated range (left) and in the restricted $[0.1:0.143]$GeV$^2$ range, as a function of the amount of staggering of strips in the right-sided element of a double-sided sensor. Bottom: RMS of the relative precision in the measurement of the tracks divergence from the z axis, as a function of the same quantity; the graphs refer to incoming muons (left), outgoing muon (center) and electron (right). In all graphs, the six bins correspond to staggerings varying from 0 to} $45\mu m$ {\em in 9}$\mu m$ {\em intervals.}}
\label{f:relativerms_staggering}
\end{center}
\end{figure}

Below it is possible to clearly see the effect of a half-pitch staggering of the right-side strips of a double-sided sensor. Such a choice maximizes the resolution in all the relevant variables, as shown in Fig.~\ref{f:relativerms_staggering}. The improvement in $q^2$ resolution when going from zero to 45 $\mu$m staggering in the most sensitive $q^2$ region amounts to a very significant $44.7 \%$ for the considered case of 300-layer stations detailed {\em supra}; similar gains however persist in different geometries and setups.

A makeshift alternative to the relative staggering of the strips, in case the construction of the double-sided modules could not provide the wanted relative positioning of the sensors, or in case the modules were already produced with no staggering, consists in positioning each double-sided sensor rotated with an opportune tilt angle with respect to the axis parallel to the strips, such that particles inciding on the sensor with no divergence from the z axis have a large chance of leaving ionization signal on two adjacent strips. However, such a setup is quite considerably more complex~\footnote{To be fully effective, the tilt angle $\phi_{tilt}$ should be large, {\em i.e.} of the order of the arctangent of the $p/w$ ratio discussed in Sec.~\ref{s:silicon}, and thus of about 15 degrees. Furthermore, the two double-sided sensors reading the x and y coordinate should be rotated along orthogonal directions, making the tracking modules look like pieces of modern art.}, and it has also the drawback of reducing the acceptance of the whole apparatus by a factor $(1-\cos{\phi_{tilt}})^2$. We did not study this possibility in detail, as we believe it is impractical to implement.

\begin{figure}[h!]
\begin{center}
\includegraphics[scale=0.8]{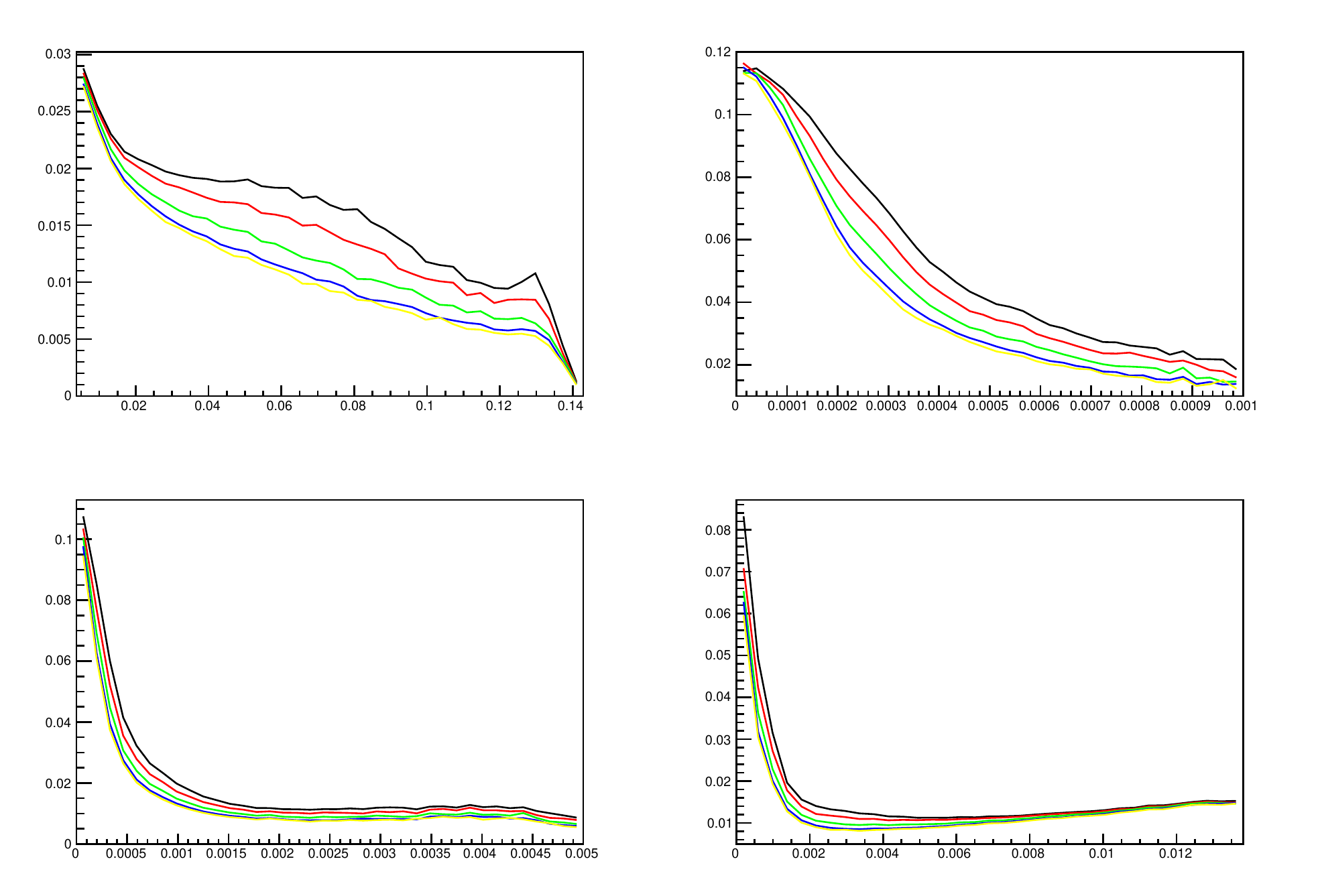} 
\caption {{\em Top left: Relative resolution in $q^2$ as a function of $q^2$; top right: relative resolution on the incoming muon divergence $\theta_{in}$ as a function of $\theta_{in}$; bottom left: relative resolution on outgoing muon divergence $\theta_{\mu}$ as a function of $\theta_{\mu}$; bottom right: relative resolution on electron divergence $\theta_e$ as a function of $\theta_e$. The six curves correspond to increasing staggering of strips in the right-sided element of double-sided sensors, from 0 to} $45\mu m$ {\em in 9}$\mu m$ {\em intervals.}}
\label{f:relativerms_staggering}
\end{center}
\end{figure}

\begin{table}[h!]
\begin{center}
\begin{tabular}{r|rrrrr} 
Configuration & $\overline{\sigma(\theta_{\in})_{rel}}$ & $\overline{\sigma(\theta_{\mu})_{rel}}$   & $\overline{\sigma(\theta_{e})_{rel}}$ & $\overline{\sigma(q^2_{rel})}$   & $\overline{\sigma(q^2_{rel})_{0.1}}$ \\ 
\hline
$\Delta h_{stag}$ ($\mu$m)     & \%   & \% & \% & \% & \% \\ 
   0.00 & $5.6091(48)$ & $1.9428(16)$ & $0.9807(8)$ & $1.6987(14)$ & $0.5969(37)$ \\ 
  11.25 & $5.0266(43)$ & $1.6911(14)$ & $0.9681(8)$ & $1.6525(13)$ & $0.5022(31)$ \\ 
  22.50 & $4.4339(37)$ & $1.4472(12)$ & $0.8936(7)$ & $1.5621(13)$ & $0.4115(25)$ \\ 
  33.75 & $4.0008(34)$ & $1.3042(10)$ & $0.8759(7)$ & $1.5058(12)$ & $0.3512(22)$ \\ 
  45.00 & $3.8224(32)$ & $1.2330(10)$ & $0.8423(7)$ & $1.4843(12)$ & $0.3358(21)$  \\ 
\hline
\end{tabular}
\caption{\em Effect on relative resolutions of the staggering $\Delta h_{stag}$ of strips in the right-side sensor, with respect to the left-sided one, in double-sided tracking modules. In each case $10^6$ elastic scattering events were simulated. }
\label{t:zprioreffect}
\end{center}
\end{table}

\subsection {Reprise: A further look at $\Delta z_{Si}$ for staggered sensors \label{s:dzsi_stag}}

\begin{figure}[h!]
\begin{center}
\includegraphics[scale=0.7]{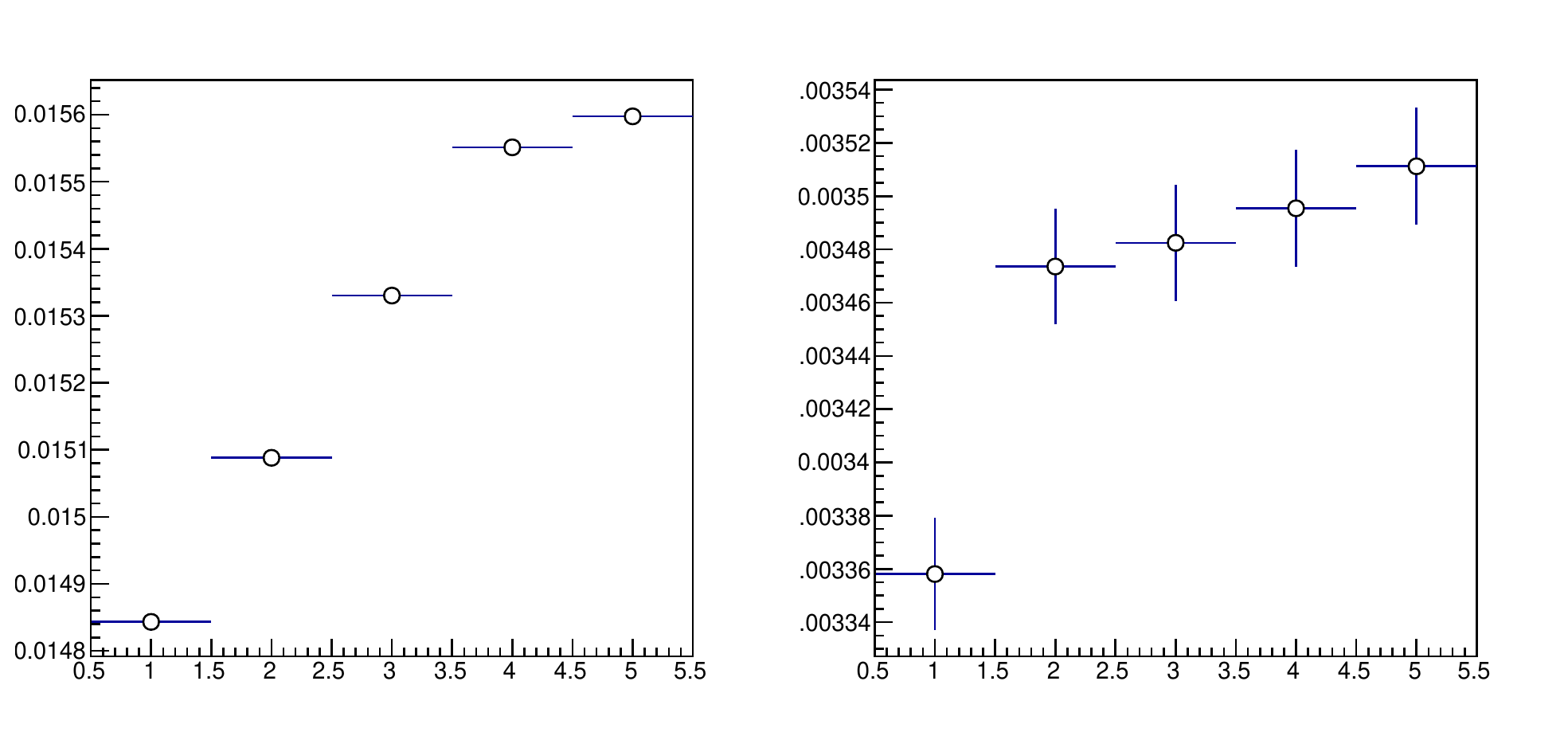} 
\caption {\em Average relative resolution in event $q^2$ in the full studied range (left) and in the high-$q^2$ range $[0.1:0.143]$ (right) as a function of the spacing of sensors in double-sided sensors when the right-hand strips have been staggered transversally by 45 $\mu$m. See the text for detail.} 
\label{f:dzsisi_stag}
\end{center}
\end{figure}

As a side effect, the application of a 45 $\mu$m staggering of the strips on the right-side sensor of double-sided modules turns out to completely change the conclusions we had reached on the advisability of a wider spacing between the sensors in Sec.~\ref{s:relspacingsisi}, which had been obtained with no staggering. Indeed, the two parameters play together in affecting the precise estimate of particle trajectories through the discreteness of the silicon strip measurements, and together they conjure a warning that a true optimization can only be achieved by considering all parameters together --something we of course cannot afford to do in a study of this kind. Below (Fig.~\ref{f:dzsisi_stag} and Table~\ref{t:dzsisi_stag}) we show how the $\Delta z_{Si}$ parameter affects resolutions in case strips are staggered by half the pitch width.

\begin{table}[h!]
\begin{center}
\begin{tabular}{r|rrrrr} 
Configuration & $\overline{\sigma(\theta_{\in})_{rel}}$ & $\overline{\sigma(\theta_{\mu})_{rel}}$   & $\overline{\sigma(\theta_{e})_{rel}}$ & $\overline{\sigma(q^2_{rel})}$   & $\overline{\sigma(q^2_{rel})_{0.1}}$ \\ 
\hline
$\Delta z_{Si}$ (cm)     & \%   & \% & \% & \% & \% \\ 
  0.18 & $3.8224(32)$ & $1.2330(10)$ & $0.8423(7)$ & $1.4843(12)$ & $0.3358(21)$ \\  
  0.28 & $3.8252(32)$ & $1.2365(10)$ & $0.8752(7)$ & $1.5088(12)$ & $0.3473(21)$ \\  
  0.38 & $3.8534(32)$ & $1.2392(10)$ & $0.8817(7)$ & $1.5330(12)$ & $0.3482(21)$ \\
  0.48 & $3.8323(32)$ & $1.2429(10)$ & $0.8866(7)$ & $1.5551(12)$ & $0.3495(21)$ \\
  0.58 & $3.8575(32)$ & $1.2456(10)$ & $0.8888(7)$ & $1.5597(13)$ & $0.3511(22)$ \\
\hline
\end{tabular}
\caption{\em Effect on relative resolutions of the changing of $\Delta z_{Si}$ parameter if a $\Delta h_{stag}=45 \mu$m has been applied to strips in the right-side sensor with respect to the left-sided one, in double-sided tracking modules. In each case $10^6$ elastic scattering events were simulated. }
\label{t:dzsisi_stag}
\end{center}
\end{table}

\subsection {Stereo angle variations}

\begin{figure}[h!]
\begin{center}
\includegraphics[scale=0.6]{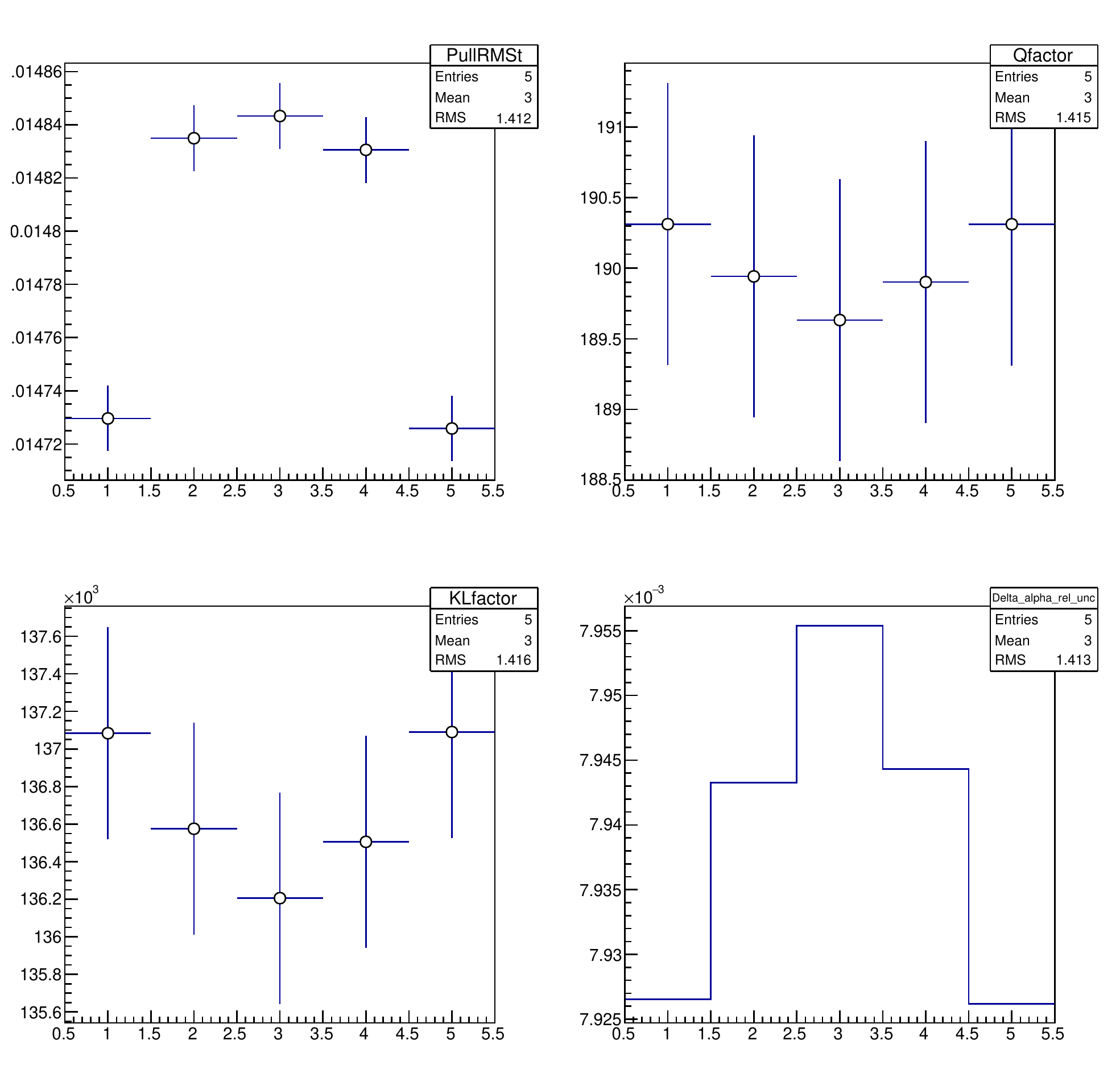}
\caption{\em Top left: relative RMS of the determination of event $q^2$ as a function of the stereo angle of rotation of middle tracking modules; bins 1 to 5 refer to rotation of 0 to $\pi/2$ in $\pi/8$ increments. Top right: values of Q test statistic (see Appendix A for details) as a function of the rotation angle. Bottom left: KL test statistic values as a function of the rotation angle. Bottom right: Relative uncertainty in the fitted $\Delta \alpha_{had}$ component from a two-component fit, where the hadronic component has been increased by a factor of 2000; see Appendix B for details.}
\label{f:stereo}
\end{center}
\end{figure}

The rotation by 45 degrees of the central module of each station, with respect to the orientation of the other two, provides measurements along the diagonal of the $xy$ plane. This arrangement does not improve the information offered by the hits, but it simplifies the tracking in some configurations, and it improves background discrimination. We study the effect of the rotation angle of the central module on the performance of the measurement, scanning the angle in 5-degree intervals from 0 to 45 degrees (larger angles produce the same effect, mirrored around the 45 degree point). Results are shown in Fig.~\ref{f:stereo} and Table~\ref{t:stereo}. We observe no appreciable variations in the relative resolution of the relevant kinematic quantities, although the considered figures of merit all indicate a very slight worsening of the measurement potential\footnote {We have chosen to show in this graph, as an example, the behavior of the two test statistics described in Appendix A, as well as the relative resolution on $\Delta \alpha_{had}$ achievable with template fits as discussed in Appendix B. The very limited changes of detector geometry produced by a rotation of the central tracking module in this case do not introduce elements of stochasticity, making those figures of merit effective in this particular case.}. One effect of to the rotation of the stereo modules which is easy to understand is the slight reduction in acceptance due to the reduction of nominal coverage (for tracks with zero divergence) from a square to an octagonal area as the stereo angle changes from zero to 45 degrees. The effect, for the considered definition of ``reconstructable'' events ({\em i.e.} ones where all three particles yield at least three x and three y stubs in consecutive modules), is a decrease by less than a percent, given the nominal beam parameters and incoming muon divergences considered in our study.

\begin{table}[h!]
\begin{center}
\begin{tabular}{r|rrrrr} 
Configuration & $\overline{\sigma(\theta_{in})_{rel}}$ & $\overline{\sigma(\theta_{\mu})_{rel}}$   & $\overline{\sigma(\theta_{e})_{rel}}$ & $\overline{\sigma(q^2)_{rel}}$   & $\overline{\sigma(q^2_{rel})_{0.1}}$ \\ 
\hline
$\phi_{stereo}$ ($\mu$m)       & \%   & \% & \% & \% & \% \\ 
  0.      & $3.8098(32)$ & $1.2291(10)$ & $0.8421(6)$ & $1.4729(12)$ & $0.3351(20)$ \\
  $\pi/8$ & $3.8115(32)$ & $1.2314(10)$ & $0.8680(7)$ & $1.4834(12)$ & $0.3434(21)$ \\
  $\pi/4$ & $3.8224(32)$ & $1.2330(10)$ & $0.8423(7)$ & $1.4843(12)$ & $0.3358(21)$ \\
  $3\pi/8$& $3.8165(32)$ & $1.2333(10)$ & $0.8685(7)$ & $1.4830(12)$ & $0.3482(21)$ \\
  $\pi/2$ & $3.8117(32)$ & $1.2292(10)$ & $0.8417(6)$ & $1.4725(12)$ & $0.3473(21)$ \\
\hline
\end{tabular}
\caption{\em Variation of angular and $q^2$ resolutions with the angle of rotation of the second tracking module in each station. In each case $10^6$ elastic scattering events were simulated. In addition to the other geometry options of results of this Section, a 45 $\mu$m staggering was considered for this simulation.}
\label{t:stereo}
\end{center}
\end{table}

\subsection {An additional option: square-mesh targets \label{s:mesh}}

One of the take-away points discussed in the previous sections is the advantage of dividing the target material up into many thin layers: one is then capable of acquiring information on the $z$ position of the scattering vertex, if this takes place in the target. The precise positioning along the beam axis of thin target layers is, we believe, a not so difficult problem to solve in practice, if rigid structures are produced with target layers spaced by appropriate frames: relatively small amounts of data are anyway sufficient to monitor it (see Sec.~\ref{s:systs}). 

\begin{figure}[h!]
\begin{center}
\includegraphics[scale=0.6]{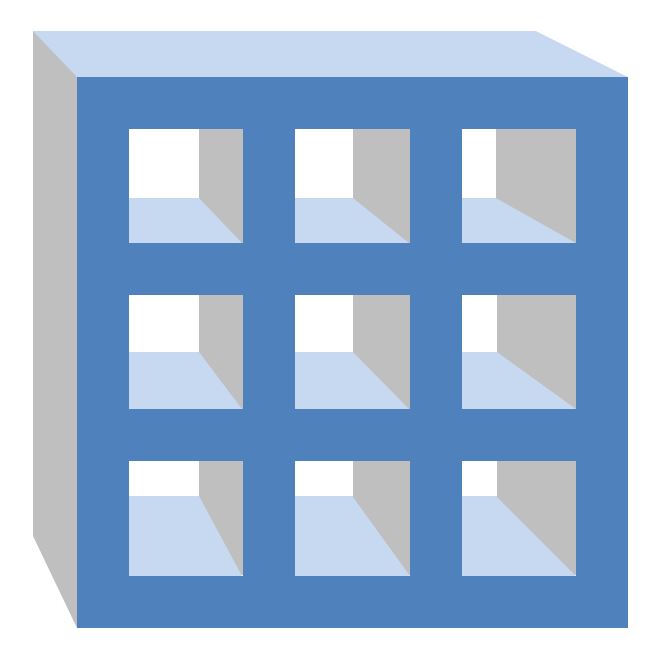}
\caption {\em Possible microscopic structure of a target layer. A periodic $xy$ lattice such as the one shown effectively provides one additional measurement point in either $x$ or $y$ if the scattering takes place in the layer material, except when it is located at the crossing of vertical and horizontal sheets.}
\label{f:lattice}
\end{center}
\end{figure}

The next logical step should then be obvious: provided that their accurate machining and production is practical, target layers built of a lattice of material alternated to holes should capture our attention. A square lattice such as the one shown in Fig.~\ref{f:lattice} might provide an additional useful constraint to the scattering vertex along the $x$ or $y$ coordinate, depending on where this is located --provided, of course, that the interaction does take place in the target and not in the silicon sensors.

The effect of lattice-shaped targets on the simulated interactions is CPU-consuming and cumbersome to model if one insists that the multiple scattering of each particle perfectly conforms to the effective amount of material crossed in target layers depending on the crossing position and angle; however, this is a minor effect which does not impact the optimization problem. For the limited purpose of a first appraisal of the beneficial effect on the precision of kinematic reconstruction we may ignore it. At simulation stage, we only insist that the scattering interaction does take place only if the position in the beryllium targets does not coincide with the assumed location of the square voids, by simply dropping from consideration events failing that criterion. We can then model the added piece of information in our fitting procedure by adding to the likelihood function a term accounting for the x and y probability of the scattering interaction in the layer.

We model the mesh as a repeated square pattern of holes distributed with an equal step $w_{gap}$ in x and y, taking a fraction $f$ of the step in both coordinates. We do not attempt an optimization of the $f$ parameter, allowing for once our intuition to pick a reasonable value for it: $f=\sqrt{2}/2$. With such a choice, $50\%$ of the target area is devoid of material. Of course this implies that the total effective thickness of 1.5 cm of Be is actually of 0.75cm per station, so one would then need to envision a doubled thickness of the layers to compensate for it. We ignore this detail, reasoning that a different choice of target material (of doubled radiation length) may retain the same target layers width; as explained {\em supra}, our simulation is oblivious of the removed target material except for the scattering position, so our results will remain consistent albeit approximate.

The probability of a scattering interaction occurring in a target layer at coordinates $x_{loc} = mod (x,w)$, $y_{loc} = mod (y,w)$ can be modeled as follows:\par

\vskip .2cm
\begin{tabular}{ll|l}
x range & y range & $P(y)$ \\
\hline
$x_{loc} < (1-f) w_{gap}$ & $[0,w_{gap}]$ & 1.0\\
\hline
$(1-f) w_{gap} \leq x_{loc} \leq w_{gap}$ & $y_{loc}<0.5 (1-f) w_{gap}$ & $0.5+0.5 Erf(\frac{y_{loc}}{\sigma_y})$\\
    & $0.5 (1-f) w_{gap} \leq y_{loc} < (1-f/2) w_{gap}$ & $0.5+0.5 Erf(\frac{(1-f) w_{gap}-y_{loc}}{\sigma_y})$ \\
    & $y_{loc} \geq (1-f/2) w_{gap}$ & $0.5+0.5 Erf(\frac{y_{loc}-w_{gap}}{\sigma_y})$ \\
\hline
y range & x range & $P(x)$ \\
\hline
$y_{loc} < (1-f) w_{gap}$ & $[0,w_{gap}]$ & 1.0 \\
\hline
$(1-f) w_{gap} \leq y_{loc} \leq w_{gap}$ & $x_{loc}<0.5 (1-f) w_{gap}$ & $0.5+0.5 Erf(\frac{x_{loc}}{\sigma_x})$ \\
   & $0.5 (1-f) w_{gap} \leq x_{loc} < (1-f/2) w_{gap}$ & $0.5+0.5 Erf(\frac{(1-f) w_{gap}-x_{loc}}{\sigma_x})$ \\
   & $x_{loc} \geq (1-f/2) w_{gap}$ & $0.5+0.5 Erf(\frac{x_{loc}-w_{gap}}{\sigma_x})$ \\
\hline
\end{tabular}
\vskip .2cm

\noindent
Once $P(x)$ and $P(y)$ are defined, they are combined by taking $P(x,y) = min[P(x),P(y)]$ except where 
$P(x)<0.5$ and $P(y)<0.5$, when we define $P(x,y) = max[P(x),P(y)]$.

We have not considered a more precise optimization of the grid design, as we believe that in this case construction issues and cost become the drivers of the available choices for geometry solutions; a full Pareto optimality can only be studied by factoring in those parameters, to which we have no access here. We thus limit ourselves to showing, in Table~\ref{t:summary}, whether a square mesh like the one of the figure above, with a spacing of 40 microns between holes, would improve the resolution on the scattering kinematics. As shown below, there is no real gain apparent from the use of such a configuration. The simple reason of this conclusion is that the $x,y$ position of the scattering vertex is already extremely well constrained by the combined fit of the three particle trajectories, with uncertainties in each coordinate of the order of 10 micrometers or less.


\subsection {Summary table}

Here we summarize the studies of this section in terms of relevant quantities of interest. As in the previous tables of this document, we consider beryllium as the target material, and a total of 1.5 cm of beryllium equivalent in the total material of the target in each station. For different target materials, one should consider the changes in layers width resulting from the different specific radiation lengths. The proposed alternative geometry we compare to the MUonE baseline here, labeled ``optimized'' in Table~\ref{t:summary} below, includes all optimal parameter values discussed in Sec.~5 and in this Section, except the etching of a lattice on the target layers. We list again the value of the improved parameters below for the benefit of the reader.\par

\begin{itemize}
\item Total number of target layers per station: 300;
\item width of each layer: $W_{Be}=0.005$cm;
\item Spacing of rightmost target layer and nearest silicon sensor: $\Delta z_{BeSi} = 0.5$cm;
\item interspacing of target layers: $\Delta z_{Be}=0.3$cm;
\item position of the left edge of the three tracking modules in each station:  $z_{mod0}=31.8$cm, $z_{mod1}=65.2$cm, $z_{mod3}=98.5$cm;
\item stereo angle in middle tracking module: $\theta_{stereo}=\pi/4$ rad;
\item spacing between silicon layers in double-sided sensors: $\Delta z_{Si}=0.18$cm;
\item transverse staggering between strips on the two sides of a double-sided sensor: $\Delta h_{stag}=45 \mu$m.
\end{itemize}

\noindent
The last line in Table~\ref{t:summary}, labeled ``etched target'', includes in addition to the above choices of geometry layout the option of etching $40 (\sqrt(2)/2) = 28.3 \mu$m square holes in the target layers, spaced by 40 $\mu$m in x and y, as discussed {\em supra}, Sec.\ref{s:mesh}. The relative improvement in $q^2$ resolution over the already optimized geometry is very small.

\begin{table}[h!]
\begin{center}
\begin{tabular}{r|rrrrr} 
Configuration & $\overline{\sigma(\theta_{\in})_{rel}}$ & $\overline{\sigma(\theta_{\mu})_{rel}}$   & $\overline{\sigma(\theta_{e})_{rel}}$ & $\overline{\sigma(q^2_{rel})}$   & $\overline{\sigma(q^2_{rel})_{0.1}}$ \\ 
        & \%   & \% & \% & \% & \% \\ 
\hline
baseline      & $5.7450(16)$ & $2.1143(05)$ & $1.2764(3)$ & $2.1222(05)$ & $0.6637(13)$ \\ 
optimized     & $3.8202(10)$ & $1.2331(03)$ & $0.8424(2)$ & $1.4833(03)$ & $0.3456(06)$ \\
etched target & $3.8220(32)$ & $1.2308(10)$ & $0.8394(6)$ & $1.4647(12)$ & $0.3366(21)$ \\
\end{tabular}
\caption{\em Figures of merit describing the separability of the hadronic contribution to $e \mu$ scattering for the baseline geometry of the MUonE detector, for a possible improved design with 300 target layers per station, and for a full optimization including target sheets with a 40 $\mu$m square lattice. The first two results are based on $10^7$ simulated elastic scattering events, the third is based on $10^6$ events. See the text for details.}
\label{t:summary}
\end{center}
\end{table}

\subsection {Effect of the identification of final state particles \label{s:switch}}

Here we study what degradation in the determination of the scattering parameters occurs if no identification of the outgoing muon and electron is provided by the experimental apparatus. In principle, a combined fit of the event kinematics should be able to determine in all cases the most likely configuration, and therefore provide indirectly an assignment of each outgoing track to a final state particle, because the functional dependence of $\theta_{\mu}$ and $\theta_{e}$ (see Fig.~\ref{f:thetas}, Sec.~\ref{s:scattering}) does not in general have a solution for the switch of those two parameters. The exception to this is the ``confusion'' region where the outgoing angles in the scattering frame are very similar, $\theta_{\mu} \simeq \theta_e$. This region corresponds to angles of about 2.5 milliradians, for which the nominal angular resolution on the outgoing particles provided by the multi-layer baseline detector layout discussed {\em supra} sits in the one to two percent range~\footnote{ In passing, we note that the functional relationship depends on the incoming muon energy, which has a $3.5\%$ spread; as noted {\em supra}, the effect of this nuisance parameter has not been accounted in the studies presented here).}.

In the likelihood definition of Sec.~\ref{s:likelihood} we have implicitly assumed that we could distinguish the hits due to the final state electron and muon tracks. This can be done with the help of a calorimeter or some other particle ID detector (not studied here), or by comparing the ionization deposits in the sensors (also not discussed in this article), or (if tracks are measured in many silicon layers) by comparing their multiple scattering angles from the fit residuals (this option should also be considered in a future study, and has been neglected here). For the discussion offered in this Section we instead drop the assumption altogether. The likelihood we use here is one which determines the maximum of the $\log L$ function for each of the two competing hypotheses (track 1 is the muon and track 2 is the electron, or vice-versa), and picks the best of the two solutions to the overall topology. The worsening effect of allowing both hypotheses to compete can be assessed by comparing, in the region of confusion, the resolution on particle angles and event $q^2$ for the configuration we have identified as the most promising {\em supra}~\footnote{ Of course a still larger number of target layers would offer further small gains, but we do find that 300 layers per station constitute an excellent practical compromise.}, {\em i.e.} \par

\begin{itemize}
\item $N_{Be,0} = N_{Be,1} = N_{Be,2} = 100$ layers, $N_{Be, tot}=300$ per station;
\item $\Delta z_{BeSi} = 0.5$cm;
\item interspacing of target layers: $\Delta z_{Be}=0.3$cm
\item position of the left edge of the three tracking modules in each station:  $z_{mod0}=31.8$cm, $z_{mod1}=65.2$cm, $z_{mod3}=98.5$cm
\item $\Delta z_{Si}=0.18$cm;
\item $\Phi_{stereo}=\pi /4$;
\item $\Delta h_{stag}=45 \mu$m.
\end{itemize}

\begin{figure}[h!]
\begin{center}
\includegraphics[scale=0.6]{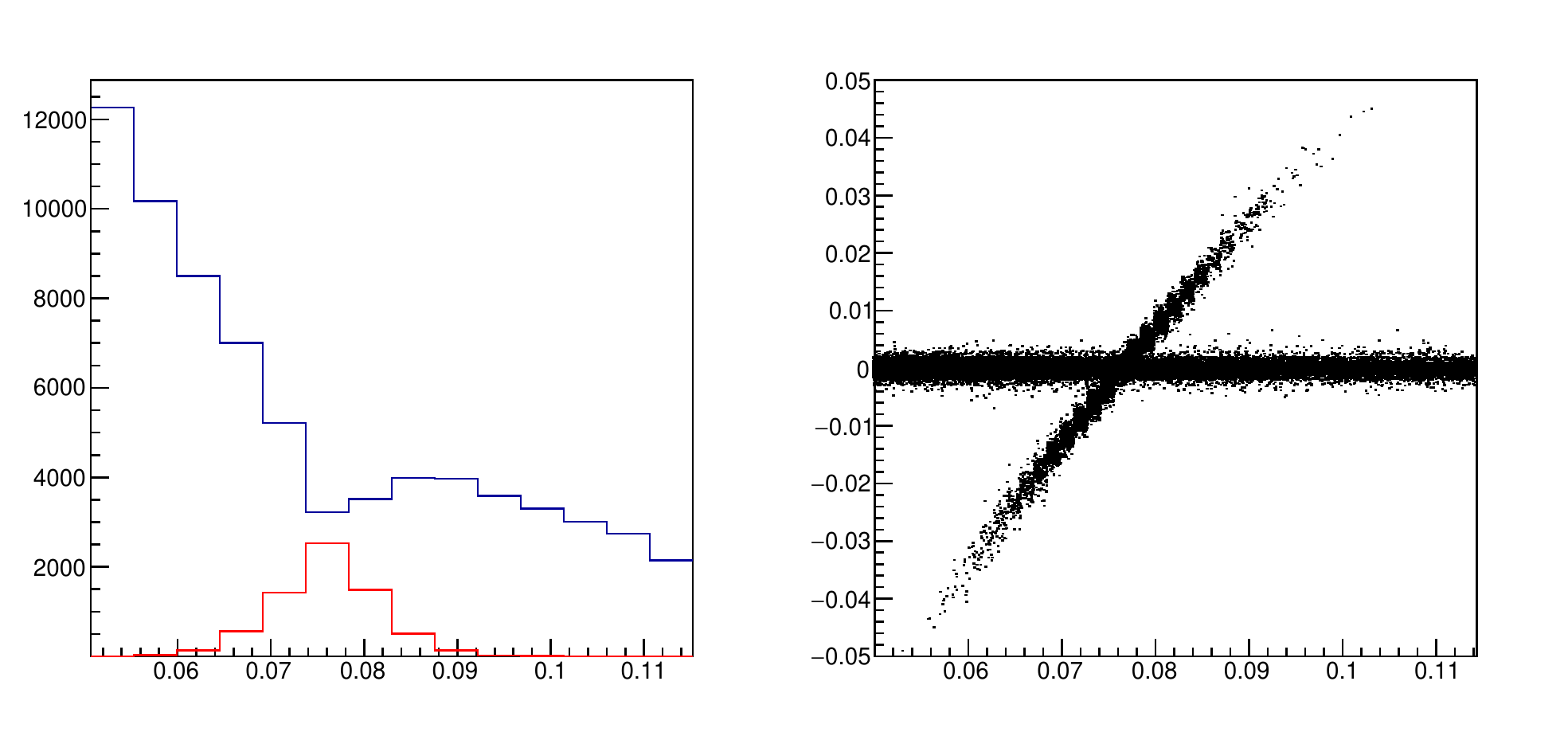} 
\caption {\em Left: True $q^2$ distribution of events for which the fit preferred the ``switched'' solution (red) to the true one (blue). See the text for details. Right: Difference between true and measured $q^2$ as a function of true $q^2$ for reconstructed scattering events where the solution of highest likelihood is chosen between the two possible assignments of final state particles to hit trajectories. A minority population of events for which the measured $q^2$ is biased toward values close to $0.08$ extends from true $q^2$ values of 0.06 to 0.1. }
\label{f:switch_q2vsq2}
\end{center}
\end{figure}

\begin{figure}[h!]
\begin{center}
\includegraphics[scale=0.6]{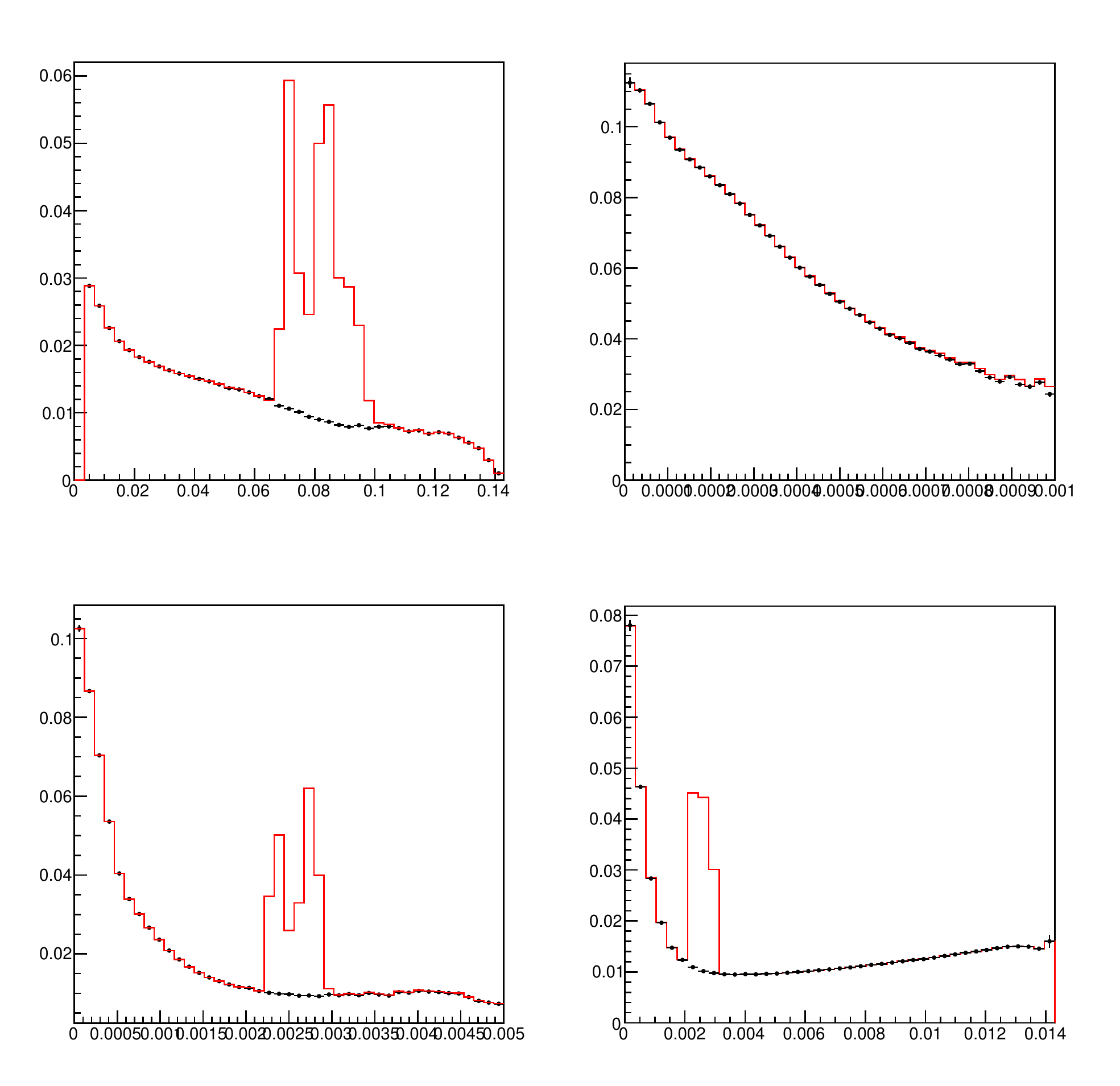}
\caption {\em Comparison of the relative resolution on $q^2$ (top left) and particle divergences from the z axis $\theta_{in}$ (top right), $\theta_{\mu}$ (bottom left), and $\theta_e$ (bottom right) when knowledge of the ID of the final state particles is assumed (black) or ignored (red). }
\label{f:switch_relres}
\end{center}
\end{figure}

\noindent
Figure~\ref{f:switch_relres} shows that the resolutions on $q^2$ and incoming muon, outgoing muon, and electron divergences from the z axis do suffer a large worsening in the region of confusion. Further, Fig.~\ref{f:switch_q2vsq2} highlights the ``attractive'' behavior of true $q^2$ values around 0.08. Away from that region, however, no significant effect is apparent. The region of highest sensitivity in the $q^2$ distribution --the one of high four-momentum transfer-- seems to be largely unaffected by the lack of particle ID.

\clearpage

\section { Constraints on the relative positioning of detector components \label{s:systs} \label{s:biases}}

In the proposal of the MUonE experiment~\cite{muonedoc} a demonstration is provided of how the average momentum of the muon beam can be determined with high accuracy by considering elastic scattering events for which the final state particles emerge with equal angles $\theta_{\mu}^{SC}$, $\theta_e^{SC}$ from the incoming muon direction. The method would suffer from a deteriorating systematic bias if the position of the detection layers along the $z$ axis were subjected to an offset from their nominal value. That appears to be motivation for an effort to secure a precision to better than 10 $\mu$m on the positioning of the sensors, along with a precise placement of the sensors orthogonally to the beam: in fact, a tilt of the sensors by angles of the order of a few milliradians could also worsen significantly the precision of the beam momentum measurement. For those reasons, a holographic system is under development, which would use laser interferometry to extract a precise determination of the relative placement of the sensors along with their tilts along the $x$ or $y$ axes. However, we argue in this section that a fully-software-based alternative is available, as discussed {\em infra}.

\subsection {Precision of the determination of the $z$ positioning of a module \label{s:zbiasconstraint}}

Here we show how a data-driven measurement of the placement of silicon sensor layers is easy to carry out, by studying the profile likelihood of the fits to the full scattering kinematics, as a function of the assumed position of a detection layer. We consider al large set of scattering events ($O(10^6)$), and for each event, whose measurements are indicated by $\vec{x}$ below, for simplicity, we \par

\begin{enumerate}
\item reconstruct the kinematics from a full kinematic fit to the elastic $\mu e \to \mu e$ hypothesis, using Eq.~\ref{eq:likelihood} and the method outlined in Sec.~\ref{s:likelihood};
\item store the maximum of the likelihood, $L_{max}(\vec{x},\hat{p})$, where $\hat{p}$ is the vector of parameters maximizing $L$ given data $\vec{x}$;
\item vary the position of the layer under study by $\Delta z$, and maximize again the likelihood, $L_{max}(\vec{x},\hat{p'})$; note that in general, this will happen at parameter values $\hat{p'}$ differing from the previous ones;
\item store the difference between likelihood maxima as a function of $\Delta z$,

\end{enumerate}
\begin{equation*}
\Delta \chi^2 (\Delta z) = -2[ \log L_{max}(\hat{p'}|\Delta z) - \log L_{max}(\hat{p}|0)]
\end{equation*}

\noindent
The distribution of total $\Delta \chi^2$, summed over all events, will approximately take the shape of a parabola with positive quadratic coefficient, passing by a point of coordinates $(\Delta z=0, \Delta \chi^2=0)$. It is not guaranteed that the parabola will have a minimum at $(0,0)$ even when using simulated data with nominal coordinate values of the considered tracking module, as the details of the detector geometry ({\em e.g.} discreteness of the hit position determination and z-vertex constraint) do introduce small biases to the estimated layer position~\footnote {We have observed that biases of the order of 10-20 microns indeed arise due to the causes mentioned above, for tracks with very small divergence which are the majority in our case.}; however, a robust extraction of those biases, which are driven by geometry configurations more than physical details, is possible by using a detailed simulation. Indeed, the minimum of the $\Delta \chi^2$ distribution calculated as above in a sample of real data represents a data-driven determination of the true $\Delta z$ value, {\em i.e.} the offset from the nominal position of the considered tracking module, after the reconstruction bias for nominal position (obtained from simulation) is subtracted.

\begin{figure}[h!]
\begin{center}
\includegraphics[scale=0.6]{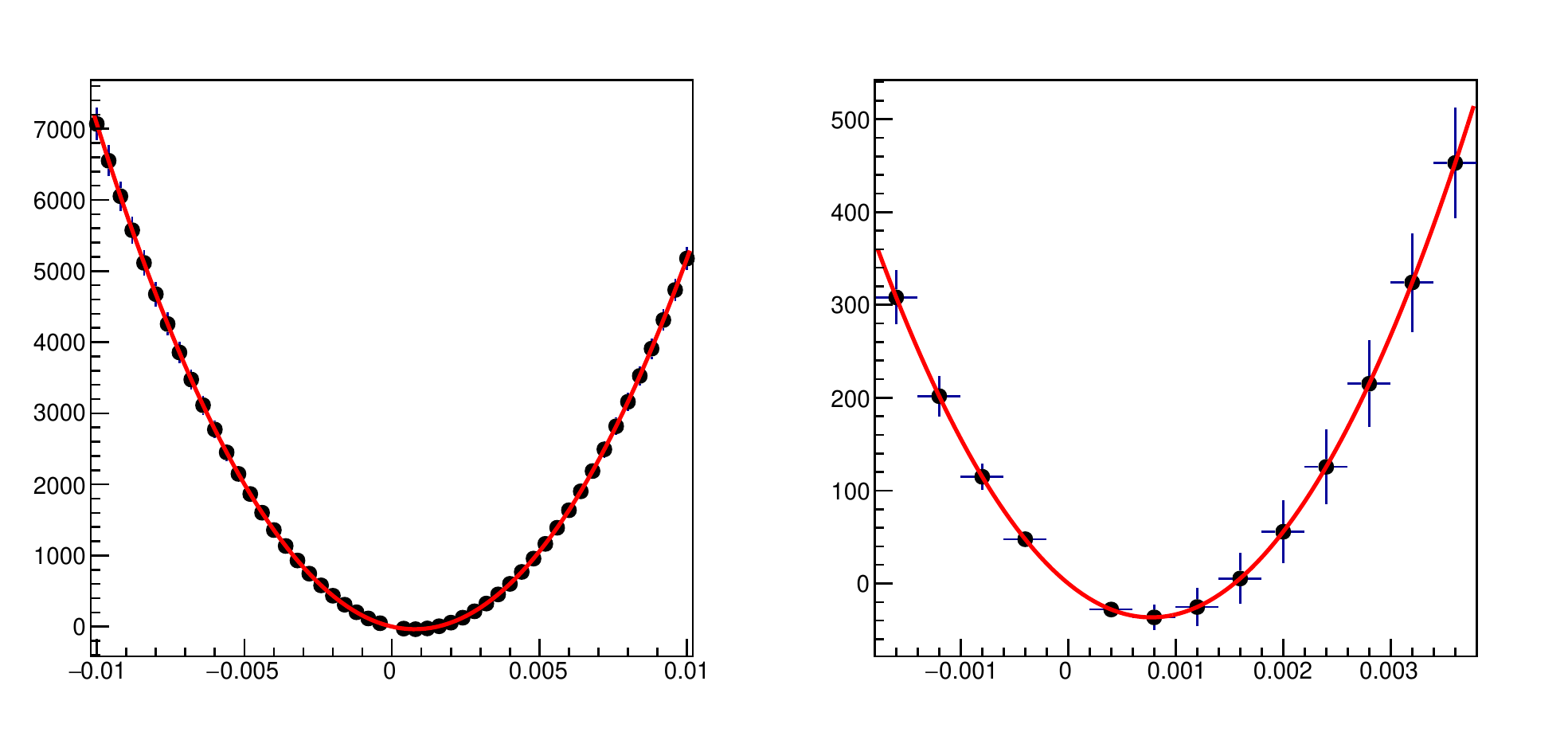} 
\caption {\em $\Delta \chi^2$ profile as a function of a shift in the positioning of tracking module 2 in the second station, for a statistics of $10^7$ generated interactions in the baseline geometry. On the left a full scan is shown from -0.01cm to 0.01cm, on the right a zoom near the minimum of the $\Delta \chi^2$ curve.  }
\label{f:offset_baseline}
\end{center}
\end{figure}

\begin{figure}[h!]
\begin{center}
\includegraphics[scale=0.6]{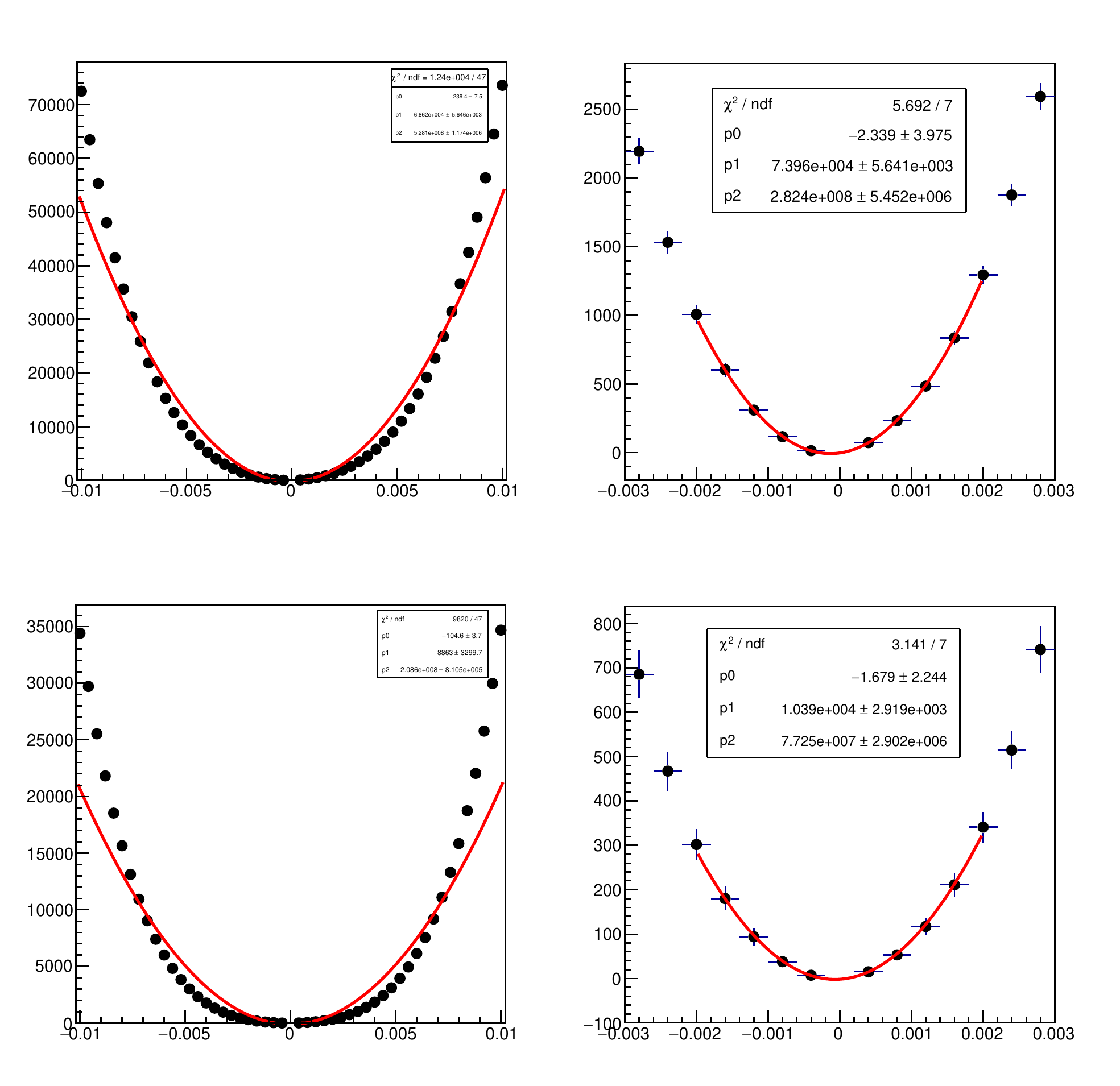} 
\caption {\em Top: $\Delta \chi^2$ profile as a function of a x tilt in the positioning of tracking module two in the second station, for a statistics of $10^7$ generated interactions in the baseline geometry. On the left a full scan is shown from -0.01 rad to 0.01 rad, on the right a zoom near the minimum of the $\Delta \chi^2$ curve. Bottom: same, for tilts along the y direction. As the likelihood profile is not well approximated by a parabola in the full considered range (left), the parabolic fit that extract the tilts are performed in a reduced range, as shown in the right panels.}
\label{f:xtilt_baseline}
\label{f:ytilt_baseline}
\end{center}
\end{figure}

\begin{figure}[h!]
\begin{center}
\includegraphics[scale=0.3]{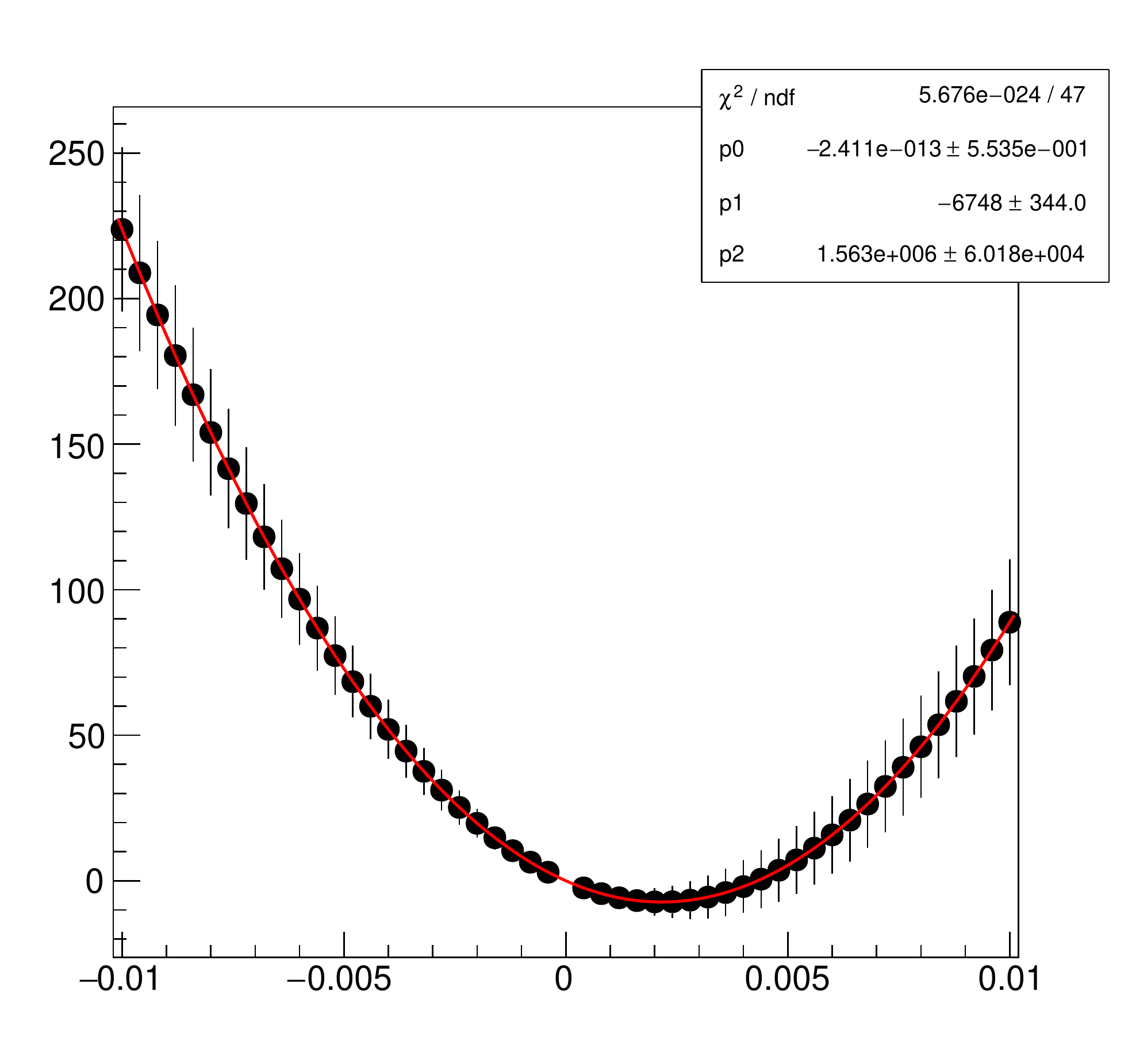} 
\caption {\em $\Delta \chi^2$ profile as a function of the bow (measured as a displacement along z) at sensor center in tracking module two in the second station, for a statistics of $10^7$ generated interactions in the baseline geometry. }
\label{f:xbow_baseline}
\end{center}
\end{figure}

\begin{table}[h!]
\begin{center}
\begin{tabular}{l|r}
Geometry & Bias in z position of tracking layer \\
\hline
Baseline        & $ 7.73 \pm 1.28 \mu$m \\
Optimized       & $-33.06 \pm 1.20 \mu$m \\
\end{tabular}
\caption {\em Estimated value of z-positioning bias parameter of the central tracking module of the second section in the considered four-stations assembly, from parabolic fits to the $\Delta \chi^2$ values corresponding to the maximum likelihood returned by fits to $10^7$ generated scattering events in the baseline and optimized geometries discussed in the text.}
\label{t:off_chi2}
\end{center}
\end{table}

\subsection {Precision in determination of the orthogonality of modules to $z$ axis}

Quite similarly to the above determination of $z$ offsets, one may obtain a determination of possible biases in the orthogonality of the sensors. A rotation of a detection layer around the axis parallel to the silicon strips --say, the $y$ axis for the left double-sided sensor of a module, which measures the $x$ coordinate-- and located on the $z$ axis will have an impact on the measurement of the track hits along $x$, while it will to first order produce no bias on the $y$ measurements. Since the modules can be constructed in rigid structures, both $x$ and $y$ rotations from the nominal orthogonal position can be detected by studying the distribution of the total $\chi^2$ of the scattering fits as a function of those angles. In Figs.~\ref{f:xtilt_baseline} we show a typical determination of the effect, for detectors that are supposed to be positioned perfectly,  {\em i.e.} with no tilts. One observes that the minima of the parabolas do not exactly correspond to zero tilt, yet their displacement, mostly due to the discreteness of the detection system, are smaller than in the case of linear offsets~\footnote{For a 10-cm-wide sensor, a rotation of $3.46 \times 10^{-4}$ radians around its center produces, for uniform illumination of its width, the same systematic effect of an offset of $10 \mu$m along $z$.}; in any case, when estimating the tilt in real data, a simulation of the apparatus is still necessary to determine the expectation value for no sensor misplacement, such that an observed departure from those values can be used to correct the track fits.

\begin{table}[h!]
\begin{center}
\begin{tabular}{l|r}
\hline
Geometry & Est. bias in x tilt \\
\hline
baseline & $-0.000131 \pm 0.000009$ rad \\   
optimized& $ 0.000020 \pm 0.000200$ rad \\
\hline
Geometry & Est. bias in y tilt \\
\hline
baseline & $ -0.000067 \pm 0.000019$ rad \\ 
optimized& $  0.000206 \pm 0.000211$ rad \\
\hline
\end{tabular}
\caption {\em Estimated value of x- and y-tilt bias parameter of the central tracking module of the second section in the considered four-stations assembly, from parabolic fits to the $\Delta \chi^2$ values corresponding to the maximum likelihood returned by fits to $10^7$ generated scattering events.}
\label{t:tilt_chi2}
\end{center}
\end{table}

\subsection {Precision determination of the bow of silicon sensors }

Another subtle effect comes from possible bows of the thin silicon layers of a detection module. A bow may result from compressive or tensile effects of the mounting supports on the sensor. Here we consider a bow along one of the orthogonal directions (the $x$ axis) of the third module of station 2 as an example, and study as a function of the maximum displacement (which results at the detector center for the modeled symmetric effect, {\em i.e.} on the $z$ axis) the total $\Delta \chi^2$ of a collection of scattering fits~\footnote{The deformation is modeled as a quadratic displacement of the true z position of the sensor with respect to its nominal one.}. Similarly to what is found for $z$ offsets, a non-zero bow estimate may result from the discreteness of the measurements of track coordinates --the effect is of the same order of magnitude.

\begin{table}[h!]
\begin{center}
\begin{tabular}{l|r}
Geometry & bow at layer center \\
\hline
baseline  & $  21.6 \pm 8.0 \mu$m \\
optimized & $ -87.2 \pm 6.3 \mu$m \\
\end{tabular}
\caption {\em Estimated value of bow along z of the central tracking module of the second section in the considered four-stations assembly, from parabolic fits to the $\Delta \chi^2$ values corresponding to the maximum likelihood returned by fits to $10^7$ generated scattering events.}
\label{t:bow_chi2}
\end{center}
\end{table}

\subsection {A note on the positioning of target layers}

The discussions outlined {\em supra} focused on the possible biases in the positioning of silicon sensors, and on data-driven ways to constrain them. The demonstration that precise fits to the scattering kinematics have the potential of tightly constraining those effects should come as a relief; on the other hand, throughout this document we have argued in favour of preferring distributed target geometries, with very thin layers precisely positioned along the $z$ axis acting as an additional measurement constraint on the vertex position. The question then arises of what is the constraining power of the data on the positioning of those detector elements, too: indeed, an imprecise placement of the target material may jeopardize all the gains of its distributed geometry.

We have not attempted in our studies to demonstrate that the same method discussed earlier in this section can be successfully applied to constrain the positioning of target layers. On the other hand, the technique is very similar. The information reported {\em supra}, Sec.\ref{s:zbiasconstraint} should be sufficient proof that the constraining power of the direct parametrization of the scattering vertex position in the likelihood function (Eq.~\ref{eq:likelihood}) can be a solution to the problem. The question to us is thus not whether this approach is viable, as much as how doable is the machining of a precise, rigid structure --a ``distributed target block'' wherein the relative position of each layer will remain the same throughout a lengthy data taking campaign, provided that the detector area is kept at constant temperature, so that the original determination of possible deviations from the design positions, obtained from data, will remain valid on the time scale of several hours, without the need to be redetermined with high frequency. A second issue is the availability of a sufficiently precise simulation to estimate the ``bias for no offset'' subtraction terms that allow to correct for the misplacement of the detector elements; we have no means of gauging whether a full GEANT simulation suffices for the task, but on the other hand the simulation can in turn be tuned with real data before these effects can be modeled. Hence it appears that a software solution to the issue of positioning errors is available, albeit maybe not as straightforward as it seems in our idealized setup.

\clearpage
\section {Conclusions and design recommendations}

In this concluding section we summarize our recommendations on some design aspects and construction choices for the detector of a muon-electron scattering experiment targeting the measurement of the $\Delta \alpha_{had}$ parameter, as well as on future studies which we believe are potentially fruitful. Our results are of course approximate and the resulting conclusions should be verified with the help of a full simulation; however, we believe they are still useful in guiding those more refined and detailed investigations: in a word, they show the direction that should be taken in furthering the optimization of design choices.

We have approached this study with the preconception, partly derived from previous discussions with colleagues, that the design of a detector built to measure such a subtle physical effect as the running of $\alpha$ due to hadronic loop effects should be as simple and robust to systematic effects as possible. Indeed, this appears as the driving consideration, given that hadronic loop contributions never exceed a part in a hundred of the total cross section, even in the most favourable regions of high $q^2$. However, the precise study of the scattering kinematics, and of the constraining power of the various relations between measurable quantities, leaves us at the end of this investigation with the opposite cognitive bias: the very high statistics of collectable scattering events, combined with the redundant measurement of an overconstrained system, allow for a data-driven {\em in situ} reduction of the most worrysome systematic sources of uncertainty. The results of the previous section demonstrate how even the ``small'' statistics of O($10^7$) fully reconstructed scattering interactions, collectable in a few minutes of data taking, allow to detect misplacements, tilts, and bows of the detection modules by microscopic amounts. After obtaining those results, we are left with the clear impression that {\em all} of the detector parameters bearing some relevance for the reconstruction of the events kinematics may be determined with very high precision and with dataset sizes as small as those collectable on the time scale of minutes of data taking. If true, this is very good news for the experiment, as the beam instability during data taking will require to frequently re-calculate the mean beam energy. The MUonE collaboration has shown how this can be done with the study of events where electron and muon emerge with equal divergence from the scattering, provided that the $z$ position of detection modules is very precisely known. For that purpose, they proposed to endow the stations with built-in laser interferometers. We believe those devices are useful but not strictly needed in principle, and we trust that their calibration method can be carried out without being affected by large systematic uncertainties from the relative positioning of sensors and target layers.

\subsection {Recommendations}

Below we list the main take-home points we obtained from our study of the geometry of detection and target components for the proposed muon scattering experiment.

\begin{enumerate}

\item The advantages of an independence of the stations making up the detector (ease of construction and assembly, reduction of trigger logic) should be considered with care and compared to the advantages brought by the alternative designs proposed in this work, to be reassessed with a full simulation of the detector and interactions.
 
\item The option of dividing up the target material into as many thin layers as it is practical to assemble in rigid structures should be investigated in detail, taking into account material choices, production costs, and machining issues. The single choice of dividing the 1.5 cm of beryllium envisioned for each station into 300 $50 \mu$m thick layers, stacked into 31-cm-long structures where each layer is spaced by 3mm from its neighbors, wins a considerable amount of constraining power on the parameter of interest.

\item If shorter stacks of thin target layers are built, which thus do not occupy the full longitudinal space between two consecutive modules in a section, their placement with respect to the tracking modules should be studied with care, as considerable variations in the $q^2$ resolution may result from the variation of that parameter. From our studies it appears advantageous to position the target stacks closer to the module on their left, as this increases the precision of the measurement of outgoing particles at the price of a less crucial decrease of the precision with which the incoming muon is traced.

\item The construction of double-sided silicon strip detection elements should be customized such that one side has strips staggered by half the pitch width ({\em i.e.} $45 \mu$m for the CMS Phase-2 sensors) with respect to the position of the strips on the other side. While the advantage of such setup depends on the details of the charge collection in the sensors, which has been simulated with a quite crude model, we believe that the peculiar kinematics of MUonE, with almost all tracks traveling with almost null divergence from the beam axis and thus will typically create single-strip hits, makes this conclusion robust.

\item The option of rotating by a 45-degree stereo angle the middle module of each station should be studied with a full simulation which could appraise the relative merits of such a setup with respect to other effects. In the absence of backgrounds and in the idealized setup we have considered here, a slight worsening of the considered figures of merit is apparent from the rotated setup. In addition, a small loss in acceptance results from the misalignment of the sensitive area of the second module with respect to the other two in each station.

\item If practical, the separation of sensors in double-sided modules by a larger amount than the $0.18$cm default of the CMS phase 2 detection elements appears to improve the resolution for the reconstruction of elastic muon-electron scattering events, in modules mounted with no staggering of the strips; however, if a 45 $\mu$m staggering is used in mounting the two sensors back-to-back in a double-sided module, the spacing should be kept at its minimum value.

\item Although in principle advantageous, the production of thin layers machined in a square lattice by etching away material in a grid of narrow holes does not appear to provide a sufficient additional constraining power to be worth pursuing, as the resolution on the transverse position of the scattering vertex achievable with a combined fit to the three tracks is already quite good.

\item We suggest that the positioning, the tilt, and the bow of each of the detection elements can be determined with very high accuracy by studying the distribution of the profile likelihood of the scattering fits to a large number of interactions as a function of the considered biased parameter. In order to evidence those shifts and constrain the parameters in an optimal way, the fits should handle the scattering as a whole as is done in this study, rather than consider independently the trajectory of each track. 

\item In case a distributed target is chosen for the detector, the option of sealing the volume external to the tracking modules in bags filled with low-pressure helium should be considered (as argued {\em supra}, at the end of Sec.~\ref{s:generation}). While we did not compare the resolution provided by such an arrangement to that obtained by ignoring the effect of scatterings in air, a small gain is clearly predictable by reducing the scattering with non-constrainable vertex z.

\item We suggest that a global likelihood fit to the track hit information, which included the hit position determination in the likelihood calculation, would improve the sensitivity of the determination of the event $q^2$ over other choices. The benefits of a simultaneous fit to all available information comes from avoiding first-order Taylor approximations to the covariance of the individual determinations, as well as, in the case of hit finding, from the interplay of the center-of-gravity determination and the track incident angle on the silicon sensors. We are however aware that the presently envisioned readout of the CMS phase-2 silicon modules does not include analogic charge readout capabilities, so this option may not be implemented in practice.
\end{enumerate}

\noindent
While we believe that the above conclusions are robust enough to be largely independent of the approximations we used to derive them, we suggest that the main differences between the considered design choices be studied by modeling the relative geometries in a full simulation of the device, which may correctly account for non-Gaussianity of multiple scattering, delta rays, background hits, non-elastic scattering events, and a full model of charge deposition in the silicon sensors. 

\section { Acknowledgements}

We are indebted to several colleagues for interesting and insightful discussions, and for providing information on detector detail and theoretical inputs; in particular we thank Giovanni Abbiendi, Nicola Bacchetta, Carlo Carloni Calame, Enrico Conti, Umberto Marconi, Clara Matteuzzi, Massimo Passera, Roberto Tenchini, and Graziano Venanzoni for insightful input.

\section*{Appendix A: Two figures of merit \label{s:fom}}

We define here two different figures of merit which may be useful to quantify the information content of the reconstructed differential cross section, with respect to the extraction of its hadronic contribution. The first one is a simple pseudo-significance measure, summed over all considered bins of the cross section distribution: \par

\begin{equation}
Z_{SB}= \sqrt{ \sum_{i=i_{min}}^{i=i_{max}} \left[ 2 (\sqrt{N^{SB}}-\sqrt{N^{B}}) \right] ^{2} }
\end{equation}

\noindent
Above, $N_{i}^{SB}$ is the expectation value of the number of event counts in bin $i$, assuming the nominal value of $\Delta \alpha_{had}$ and $N_{i}^{B}$ is the corresponding expected number of events if the hadronic contribution (as modeled in Eq.~\ref{eq:hadcontrib}) is neglected; $i_{min}$ and $i_{max}$ are the indices of the first and last bin considered in the histograms comparison. The construction requires the specification of a predicted distribution of the electroweak part of the elastic cross section, which is the theoretical curve folded with the experimental acceptance and resolution effects. These are determined separately, as discussed {\em infra}. 

The acceptance factor is extracted from a binned ratio between reconstructed and generated events in each bin of true $q^2$: because the same generated events are used in the calculation, this factor is thus the true one affecting the measured distribution, hence no uncertainty from imperfect acceptance affects the comparison of measured and expected electroweak spectrum. As for the resolution effects, we tried unsuccessfully to extract them from mean and RMS values of the $q^2$ residuals (measured minus generated) from the generation and reconstruction of simulated elastic scattering events at different reference values of $q^2$. Imperfections in the resulting model --which crucially requires the independent determination of parameters for each of the different studied geometries-- consistently out-weigh the effect that the $q^2$ resolution alone has on the test statistic defined above. We therefore also in this case decided to use the ``true'' $q^2$ resolution, as determined event per event by comparing fitted and generated $q^2$. At the price of some throwing up of our hands, we gain some more power for the defined statistics.

Since each $2(\sqrt{N_{i}^{SB}}-\sqrt{N_{i}^{B}})$ factor is an approximate measure of the significance of a departure of the observed rate from its Poisson mean $N_{i}^{B}$, and $Z_{SB}$ is a quadrature sum of those factors, it results in a pseudo-significance measure of the shape difference between the two distributions. 

A second, well-known measure of the information content of the difference between two distributions is provided by the Kullbach-Leibler divergence~\cite{kullbackleibler}. It is defined as \par

\begin{equation}
D_{KL} = \sum_i{p_i^{SB} \log \left( \frac{p_i^{SB}}{p_i^{B}} \right)}
\end{equation}

\noindent
where here $p_i^{SB}$ and $p_i^{B}$ are here the two probability distributions under discussion (respectively, the expected density function of the differential cross section which includes the hadronic contribution, and the expected density function of the same, considering the electroweak part alone; in both cases, the true resolution and acceptance are considered, as discussed {\em supra}); index $i$ runs on the bins from $i_{min}$ to $i_{max}$ as for the $Z_{SB}$ statistic defined {\em supra}. $D_{KL}$ is a measure of entropy --in other words, it is an estimate of the information content provided by the difference of the two compared densities, and is thus arguably very well suited to capture the quality of the reconstruction, when the hadronic contribution to elastic scattering is the focus of the measurement.

As mentioned in the body of this document, the above test statistics are insufficiently sensitive to the effect we are trying to put in evidence, for simulated datasets of the size reachable by our computing power. They can only be useful when there is no stochasticity involved in passing from one studied geometry to the next; this happens, {\em e.g.}, when we study effects that do not modify the propagation of the particles in the material, such as variations of the stereo angle of the central tracking module (particles crossing the rotated module encounter the same amount of material as particles crossing an unrotated module). The residual statistical variations due to different acceptance of the configurations at different stereo angles are small enough that the two test statistics discussed here produce a coherent picture, as shown in Fig.~\ref{f:stereo}.

\section*{Appendix B: $\Delta \alpha_{had}$ extraction by cross section fits \label{s:switch}}

Here we describe the extraction of the $\Delta \alpha_{had}$ parameter from a shape fit to the observed $q^2$ distribution of elastic muon-electron scattering events, used for some checks described in this document. Originally our intent was to extract directly the uncertainty on the parameter of interest with this method. However, we realized that it was not practical to do so; in fact, the crucial input of the fit is a precise model of the $q^2$ resolution, as of course this directly affects the precision on the extractable value of $\Delta \alpha_{had}$. We tried to model the resolution with parametric forms of considerable complexity, but the comparison of the merits of different geometries showed to be too dependent on unwanted differences in the precision of the resolution models produced for each of them.  

\begin{figure}[h!]
\begin{center}
\includegraphics[scale=0.8]{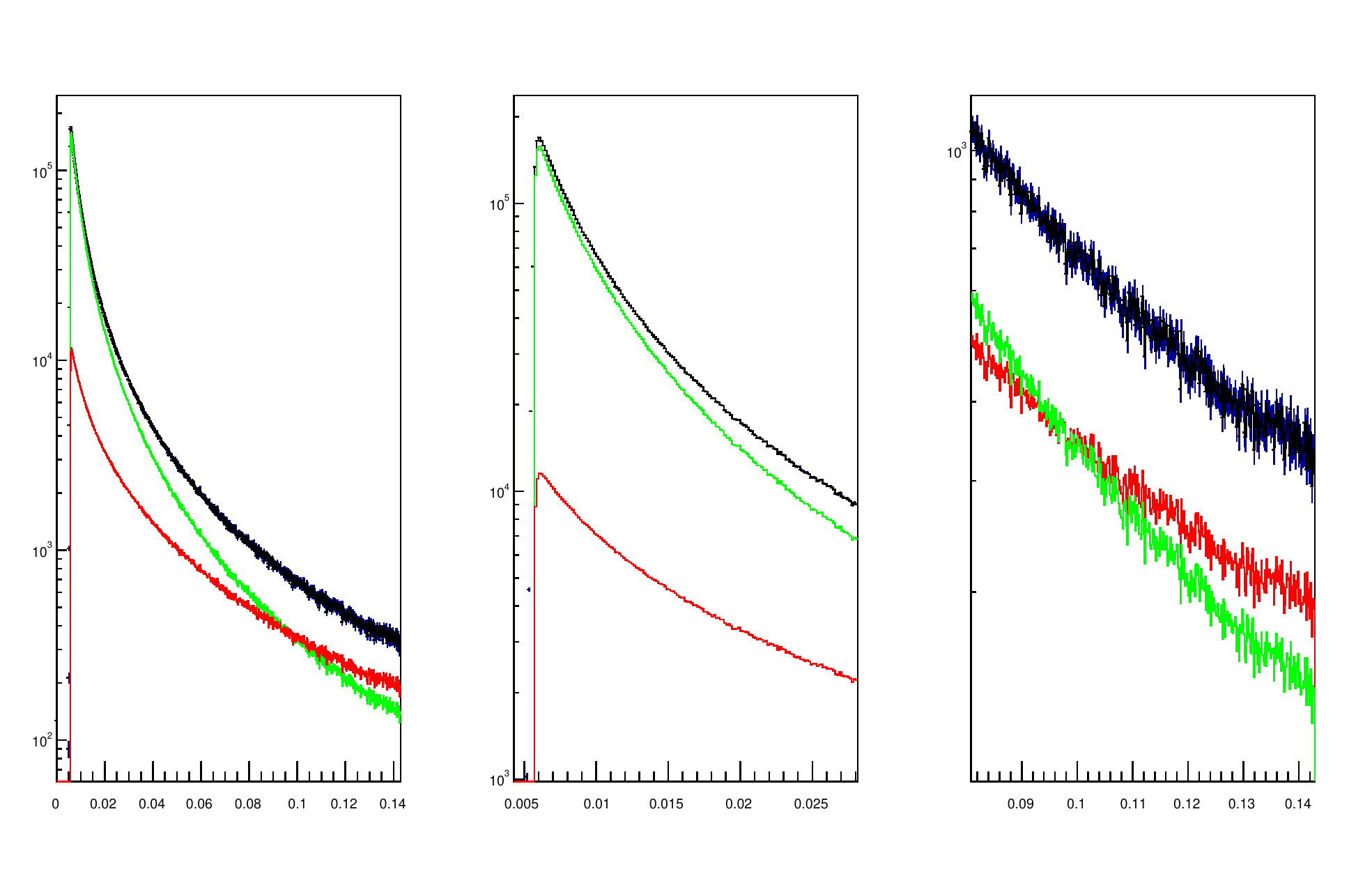}
\caption{\em Example of a shape fit to the differential cross section of elastic muon-electron scattering as a function of $q^2$. The blue points show simulated data, the black curve is the result of the fit, and the green and red curves show the electroweak and hadronic contributions which constitute the two fitted components. In this simulation the value of the hadronic contribution has been artificially increased by a factor of 2000 to better study the fit performance in case of a significant hadronic contribution, given the sample size ($10^7$ generated interactions, of which $79\%$ are fully reconstructed). The left panel shows the full spectrum, the middle one shows a closeup of its low end, and the right one shows its high-$q^2$ tail. Note that the data extends to values where the templates are null at the small end. This is an innocuous feature of the template generation, which includes resolution and efficiency effects. The fit is performed by ignoring the first four non-zero bins of the templates as well as the last two bins.}
\label{f:fitexample}
\end{center}
\end{figure}

Because of the above, we use simulation information: rather than modeling the $q^2$ resolution, for each event we reconstruct we read off the difference between true and estimated $q^2$. This allows to construct perfect templates of ``signal'' (the hadronic contribution to the differential cross section) and ``background'' for each studied geometry configuration. Again, this is the absolute optimum one may ever obtain with a perfect modeling of the resolution map; its application to a two-component fit thus allows to extract information on the relative merits of the different geometries which is oblivious of the issues connected with that part of the analysis problem.

Since this technique has not been used for the results reported in the article, here we only exemplify how a shape fit performs in the baseline geometry, with a value of $\Delta \alpha_{had}$ increased by a factor sufficient to be estimated precisely with a statistics of $10^7$ produced interactions. Figure~\ref{f:fitexample} shows the appearance of the data and its interpretation as the sum of background and signal templates with a likelihood fit, where the signal fraction is the only free parameter. The signal in this case has been increased by a factor of 2000 from its true value, to evidence its different shape and to allow a convergence of the fit with a limited number of fitted data events.

\clearpage


\end{document}